# Gravity Control by means of *Electromagnetic Field* through *Gas* or *Plasma* at Ultra-Low Pressure

**Fran De Aquino**
Maranhao State University, Physics Department, S.Luis/MA, Brazil.

It is shown that the gravity acceleration just above a chamber filled with *gas* or *plasma* at ultra-low pressure can be strongly reduced by applying an Extra Low-Frequency (ELF) electromagnetic field across the gas or the plasma. This Gravitational Shielding Effect is related to recent discovery of *quantum correlation* between gravitational mass and inertial mass. According to the theory samples hung above the gas or the plasma should exhibit a weight decrease when the frequency of the electromagnetic field is decreased or when the intensity of the electromagnetic field is increased. This Gravitational Shielding Effect is unprecedented in the literature and can not be understood in the framework of the General Relativity. From the technical point of view, there are several applications for this discovery; possibly it will change the paradigms of *energy* generation, *transportation* and *telecommunications*.

**CONTENTS**

# I. INTRODUCTION

It will be shown that the local gravity acceleration can be controlled by means of a device called Gravity Control Cell (GCC) which is basically a recipient filled with gas or plasma where is applied an *electromagnetic field*. According to the theory samples hung above the gas or plasma should exhibit a weight decrease when the frequency of the electromagnetic field is decreased or when the intensity of the electromagnetic field is increased. The electrical *conductivity* and the *density* of the gas or plasma are also highly relevant in this process.

With a GCC it is possible to convert the gravitational energy into rotational mechanical energy by means of the *Gravitational Motor*. In addition, a new concept of spacecraft (the *Gravitational Spacecraft*) and aerospace flight is presented here based on the possibility of gravity control. We will also see that the gravity control will be very important to *Telecommunication*.

# II. THEORY

It was shown [1] that the relativistic *gravitational mass* $M_g = m_g \big/ \sqrt{1 - V^2/c^2}$ and the relativistic *inertial mass* $M_i = m_{i0} \big/ \sqrt{1 - V^2/c^2}$ are *quantized*, and given by $M_g = n_g^2 m_{i0(min)}$ , $M_i = n_i^2 m_{i0(min)}$ where $n_g$ and $n_i$ are respectively, the *gravitational quantum number* and the *inertial quantum number* ; $m_{i0(min)} = \pm 3.9 \times 10^{-73} kg$ is the elementary *quantum* of inertial mass. The masses $m_g$ and $m_{i0}$ are correlated by means of the following expression:

$$m_g = m_{i0} - 2\left[\sqrt{1 + \left(\frac{\Delta p}{m_i c}\right)^2} - 1\right] m_{i0}. \qquad (1)$$

Where $\Delta p$ is the *momentum* variation on the particle and $m_{i0}$ is the inertial mass at rest.

In general, the *momentum* variation $\Delta p$ is expressed by $\Delta p = F\Delta t$ where $F$ is the applied force during a time interval $\Delta t$. Note that there is no restriction concerning the *nature* of the force $F$, i.e., it can be mechanical, electromagnetic, etc.

For example, we can look on the *momentum* variation $\Delta p$ as due to absorption or emission of *electromagnetic energy* by the particle.

In the case of radiation, $\Delta p$ can be obtained as follows: It is known that the *radiation pressure*, $dP$, upon an area $dA = dxdy$ of a volume $dV = dxdydz$ of a particle ( the incident radiation normal to the surface $dA$ )is equal to the energy $dU$ absorbed per unit volume $(dU/dV)$.i.e.,

$$dP = \frac{dU}{dV} = \frac{dU}{dxdydz} = \frac{dU}{dAdz} \qquad (2)$$

Substitution of $dz = vdt$ ($v$ is the speed of radiation) into the equation above gives

$$dP = \frac{dU}{dV} = \frac{(dU/dAdt)}{v} = \frac{dD}{v} \qquad (3)$$

Since $dPdA = dF$ we can write:

$$dFdt = \frac{dU}{v} \qquad (4)$$

However we know that $dF = dp/dt$, then

$$dp = \frac{dU}{v} \qquad (5)$$

From this equation it follows that

$$\Delta p = \frac{U}{v}\left(\frac{c}{c}\right) = \frac{U}{c} n_r$$

Substitution into Eq. (1) yields

$$m_g = \left\{1 - 2\left[\sqrt{1 + \left(\frac{U}{m_{i0}c^2} n_r\right)^2} - 1\right]\right\} m_{i0} \qquad (6)$$

Where $U$, is the electromagnetic energy absorbed by the particle; $n_r$ is the index of refraction.

Equation (6) can be rewritten in the following form

$$m_g = \left\{1 - 2\left[\sqrt{1 + \left(\frac{W}{\rho c^2} n_r\right)^2} - 1\right]\right\} m_{i0} \qquad (7)$$

Where $W = U/V$ is the *density of electromagnetic energy* and $\rho = m_{i0}/V$ is the density of inertial mass.

The Eq. (7) is the expression of the *quantum correlation* between the *gravitational mass* and the *inertial mass* as a function of the *density of electromagnetic energy*. This is also the expression of correlation between gravitation and electromagnetism.

The density of electromagnetic energy in an *electromagnetic* field can be deduced from Maxwell's equations [2] and has the following expression

$$W = \tfrac{1}{2}\varepsilon E^2 + \tfrac{1}{2}\mu H^2 \qquad (8)$$

It is known that $B = \mu H$, $E/B = \omega/k_r$ [3] and

$$v = \frac{dz}{dt} = \frac{\omega}{\kappa_r} = \frac{c}{\sqrt{\frac{\varepsilon_r \mu_r}{2}\left(\sqrt{1 + (\sigma/\omega\varepsilon)^2} + 1\right)}} \qquad (9)$$

Where $k_r$ is the real part of the *propagation vector* $\vec{k}$ (also called *phase constant* [4]); $k = |\vec{k}| = k_r + ik_i$ ; $\varepsilon$, $\mu$ and $\sigma$, are the electromagnetic characteristics of the medium in which the incident (or emitted) radiation is propagating ($\varepsilon = \varepsilon_r \varepsilon_0$ where $\varepsilon_r$ is the *relative dielectric permittivity* and $\varepsilon_0 = 8.854 \times 10^{-12} F/m$ ; $\mu = \mu_r \mu_0$ where $\mu_r$ is the *relative magnetic permeability* and $\mu_0 = 4\pi \times 10^{-7} H/m$; $\sigma$ is the *electrical conductivity*). It is known that for *free-space* $\sigma = 0$ and $\varepsilon_r = \mu_r = 1$ then Eq. (9) gives

$$v = c \qquad (10)$$

From (9) we see that the *index of refraction* $n_r = c/v$ will be given by

$$n_r = \frac{c}{v} = \sqrt{\frac{\varepsilon_r \mu_r}{2}\left(\sqrt{1 + (\sigma/\omega\varepsilon)^2} + 1\right)} \qquad (11)$$

Equation (9) shows that $\omega/\kappa_r = v$. Thus, $E/B = \omega/k_r = v$, i.e., $E = vB = v\mu H$ . Then, Eq. (8) can be rewritten in the following form:

$$W = \tfrac{1}{2}\left(\varepsilon v^2 \mu\right)\mu H^2 + \tfrac{1}{2}\mu H^2 \qquad (12)$$

For $\sigma << \omega\varepsilon$, Eq. (9) reduces to

$$v = \frac{c}{\sqrt{\varepsilon_r \mu_r}}$$

Then, Eq. (12) gives

$$W = \tfrac{1}{2}\left[\varepsilon\left(\frac{c^2}{\varepsilon_r \mu_r}\right)\mu\right]\mu H^2 + \tfrac{1}{2}\mu H^2 = \mu H^2 \qquad (13)$$

This equation can be rewritten in the following forms:

$$W = \frac{B^2}{\mu} \qquad (14)$$

or

$$W = \varepsilon E^2 \qquad (15)$$

For $\sigma >> \omega\varepsilon$, Eq. (9) gives

$$v = \sqrt{\frac{2\omega}{\mu\sigma}} \qquad (16)$$

Then, from Eq. (12) we get

$$W = \tfrac{1}{2}\left[\varepsilon\left(\frac{2\omega}{\mu\sigma}\right)\mu\right]\mu H^2 + \tfrac{1}{2}\mu H^2 = \left(\frac{\omega\varepsilon}{\sigma}\right)\mu H^2 + \tfrac{1}{2}\mu H^2 \cong$$

$$\cong \tfrac{1}{2}\mu H^2 \qquad (17)$$

Since $E = vB = v\mu H$ , we can rewrite (17) in the following forms:

$$W \cong \frac{B^2}{2\mu} \qquad (18)$$

or

$$W \cong \left(\frac{\sigma}{4\omega}\right)E^2 \qquad (19)$$

By comparing equations (14) (15) (18) and (19) we see that Eq. (19) shows that the better way to obtain a strong value of $W$ *in practice* is by applying an *Extra Low-Frequency* (ELF) *electric field* $(w = 2\pi f << 1Hz)$ through a *mean with high electrical conductivity*.

Substitution of Eq. (19) into Eq. (7), gives

$$m_g = \left\{1 - 2\left[\sqrt{1 + \frac{\mu}{4c^2}\left(\frac{\sigma}{4\pi f}\right)^3 \frac{E^4}{\rho^2}} - 1\right]\right\} m_{i0} \qquad (20)$$

This equation shows clearly that if an

electrical *conductor mean* has $\rho << 1\ Kg.m^{-3}$ and $\sigma >> 1$, then it is possible obtain strong changes in its gravitational mass, with a relatively small ELF *electric field*. An electrical *conductor mean* with $\rho << 1\ Kg.m^{-3}$ is obviously a *plasma*.

There is a very simple way to test Eq. (20). It is known that inside a fluorescent lamp *lit* there is *low-pressure Mercury plasma.* Consider a 20W T-12 *fluorescent lamp* (80044–F20T12/C50/ECO GE, Ecolux® T12), whose characteristics and dimensions are well-known [5]. At around $T \cong 318.15^0 K$, an optimum mercury vapor pressure of $P = 6\times10^{-3}Torr = 0.8N.m^{-2}$ is obtained, which is required for maintenance of high luminous efficacy throughout life. Under these conditions, the mass density of the Hg plasma can be calculated by means of the well-known *Equation of State*

$$\rho = \frac{PM_0}{ZRT} \qquad (21)$$

Where $M_0 = 0.2006kg.mol^{-1}$ is the molecular mass of the Hg; $Z \cong 1$ is the *compressibility factor* for the Hg plasma; $R = 8.314 joule.mol^{-1}.^0K^{-1}$ is the *gases universal constant*. Thus we get

$$\rho_{Hg\ plasma} \cong 6.067\times10^{-5} kg.m^{-3} \qquad (22)$$

The electrical conductivity of the Hg plasma can be deduced from the *continuum form of Ohm's Law* $\vec{j} = \sigma\vec{E}$, since the *operating current* through the lamp and the *current density* are well-known and respectively given by $i = 0.35A$ [5] and $j_{lamp} = i/S = i/\frac{\pi}{4}\phi_{int}^2$, where $\phi_{int} = 36.1mm$ is the inner diameter of the lamp. The voltage drop across the electrodes of the lamp is $57V$ [5] and the distance between them $l = 570mm$. Then the electrical field *along* the lamp $E_{lamp}$ is given by $E_{lamp} = 57V/0.570m = 100\ V.m^{-1}$. Thus, we have

$$\sigma_{Hg\ plasma} = \frac{j_{lamp}}{E_{lamp}} = 3.419\ \ S.m^{-1} \qquad (23)$$

Substitution of (22) and (23) into (20) yields

$$\frac{m_{g(Hg\ plasma)}}{m_{i(Hg\ plasma)}} = \left\{1 - 2\left[\sqrt{1 + 1.909\times10^{-17}\frac{E^4}{f^3}} - 1\right]\right\} \qquad (24)$$

Thus, if an *Extra Low-Frequency electric field* $E_{ELF}$ with the following characteristics: $E_{ELF} \approx 100V.m^{-1}$ and $f < 1mHZ$ is applied through the Mercury plasma then a strong *decrease in the gravitational mass of the Hg plasma* will be produced.

It was shown [1] that there is an additional effect of *gravitational shielding* produced by a substance under these conditions. Above the substance the gravity acceleration $g_1$ is reduced at the same ratio $\chi = m_g/m_{i0}$, i.e., $g_1 = \chi\ g$, ($g$ is the gravity acceleration *under* the *substance*). Therefore, due to the *gravitational shielding effect* produced by the decrease of $m_{g(Hg\ plasma)}$ *in the region where the ELF electric field* $E_{ELF}$ *is applied*, the gravity acceleration just *above* this region will be given by

$$g_1 = \chi_{(Hg\ plasma)}g = \frac{m_{g(Hg\ plasma)}}{m_{i(Hg\ plasma)}}g =$$

$$= \left\{1 - 2\left[\sqrt{1 + 1.909\times10^{-17}\frac{E_{ELF}^4}{f_{ELF}^3}} - 1\right]\right\}g \qquad (25)$$

The trajectories of the electrons/ions through the lamp are determined by the electric field $E_{lamp}$ *along* the lamp. If the ELF electric field *across* the lamp $E_{ELF}$ is much greater than $E_{lamp}$, the current through the lamp can be interrupted. However, if $E_{ELF} << E_{lamp}$, these trajectories will be only slightly modified. Since here $E_{lamp} = 100\ V.m^{-1}$, then we can arbitrarily choose $E_{ELF}^{max} \cong 33\ \ V.m^{-1}$. This means that the *maximum* voltage drop, which can be applied across the metallic plates, placed at distance $d$, is equal to the outer diameter (max *) of the bulb $\phi_{lamp}^{max}$ of the 20W T-12 Fluorescent lamp, is given by

$$V_{max} = E_{ELF}^{max}\phi_{lamp}^{max} \cong 1.5\ V$$

Since $\phi_{lamp}^{max} = 40.3mm$ [5].

Substitution of $E_{ELF}^{max} \cong 33\ V.m^{-1}$ into (25) yields

$$g_1 = \chi_{(Hg\ plasma)}g = \frac{m_{g(Hg\ plasma)}}{m_{i(Hg\ plasma)}}g =$$

$$= \left\{1 - 2\left[\sqrt{1 + \frac{2.264\times10^{-11}}{f_{ELF}^3}} - 1\right]\right\}g \qquad (26)$$

Note that, for $f < 1mHz = 10^{-3}Hz$, the gravity acceleration can be strongly reduced. These conclusions show that the ELF Voltage Source of the set-up shown in Fig.1 should have the following characteristics:

- Voltage range: 0 – 1.5 V
- Frequency range: $10^{-4}$Hz – $10^{-3}$Hz

In the experimental arrangement shown in Fig.1, an ELF electric field with intensity $E_{ELF} = V/d$ crosses the fluorescent lamp; $V$ is the voltage drop across the metallic plates of the capacitor and $d = \phi_{lamp}^{max} = 40.3mm$. When the ELF electric field is applied, the gravity acceleration just above the lamp (inside the dotted box) decreases according to (25) and the changes can be measured by means of the system balance/sphere presented on the top of Figure 1.

In Fig. 2 is presented an experimental arrangement with *two* fluorescent lamps in order to test the gravity acceleration above the *second* lamp. Since gravity acceleration above the *first* lamp is given by $\vec{g}_1 = \chi_{1(Hg\ plasma)}\vec{g}$, where

$$\chi_{1(Hg\ plasma)} = \frac{m_{g1(Hg\ plasma)}}{m_{i1(Hg\ plasma)}} =$$

$$= \left\{1 - 2\left[\sqrt{1 + 1.909\times10^{-17}\frac{E_{ELF(1)}^4}{f_{ELF(1)}^3}} - 1\right]\right\} \qquad (27)$$

Then, above the *second* lamp, the gravity acceleration becomes

$$\vec{g}_2 = \chi_{2(Hg\ plasma)}\vec{g}_1 = \chi_{2(Hg\ plasma)}\chi_{1(Hg\ plasma)}\vec{g} \qquad (28)$$

where

$$\chi_{2(Hg\ plasma)} = \frac{m_{g2(Hg\ plasma)}}{m_{i2(Hg\ plasma)}} =$$

$$= \left\{1 - 2\left[\sqrt{1 + 1.909\times10^{-17}\frac{E_{ELF(2)}^4}{f_{ELF(2)}^3}} - 1\right]\right\} \qquad (29)$$

Then, results

$$\frac{g_2}{g} = \left\{1 - 2\left[\sqrt{1 + 1.909\times10^{-17}\frac{E_{ELF(1)}^4}{f_{ELF(1)}^3}} - 1\right]\right\}\times$$

$$\times\left\{1 - 2\left[\sqrt{1 + 1.909\times10^{-17}\frac{E_{ELF(2)}^4}{f_{ELF(2)}^3}} - 1\right]\right\} \qquad (30)$$

From Eq. (28), we then conclude that if $\chi_{1(Hg\ plasma)} < 0$ and also $\chi_{2(Hg\ plasma)} < 0$, then $g_2$ will have the *same direction* of $g$. This way it is possible to intensify several times the gravity in the direction of $\vec{g}$. On the other hand, if $\chi_{1(Hg\ plasma)} < 0$ and $\chi_{2(Hg\ plasma)} > 0$ the direction of $\vec{g}_2$ will be contrary to direction of $\vec{g}$. In this case will be possible to *intensify* and become $\vec{g}_2$ *repulsive* in respect to $\vec{g}$.

If we put a lamp above the *second* lamp, the gravity acceleration above the *third* lamp becomes

$$\vec{g}_3 = \chi_{3(Hg\ plasma)}\vec{g}_2 =$$

$$= \chi_{3(Hg\ plasma)}\chi_{2(Hg\ plasma)}\chi_{1(Hg\ plasma)}\vec{g} \qquad (31)$$

or

* After heating.

$$\frac{g_3}{g}=\left\{1-2\left[\sqrt{1+1.909\times10^{-17}\frac{E^4_{ELF(1)}}{f^3_{ELF(1)}}}-1\right]\right\}\times$$
$$\times\left\{1-2\left[\sqrt{1+1.909\times10^{-17}\frac{E^4_{ELF(2)}}{f^3_{ELF(2)}}}-1\right]\right\}\times$$
$$\times\left\{1-2\left[\sqrt{1+1.909\times10^{-17}\frac{E^4_{ELF(3)}}{f^3_{ELF(3)}}}-1\right]\right\} \qquad (32)$$

If $f_{ELF(1)} = f_{ELF(2)} = f_{ELF(3)} = f$ and

$$E_{ELF(1)} = E_{ELF(2)} = E_{ELF(3)} = V/\phi =$$
$$= V_0 \sin \omega t / 40.3mm =$$
$$= 24.814V_0 \sin 2\pi ft.$$

Then, for $t = T/4$ we get

$$E_{ELF(1)} = E_{ELF(2)} = E_{ELF(3)} = 24.814V_0.$$

Thus, Eq. (32) gives

$$\frac{g_3}{g}=\left\{1-2\left[\sqrt{1+7.237\times10^{-12}\frac{V_0^4}{f^3}}-1\right]\right\}^3 \qquad (33)$$

For $V_0 = 1.5V$ and $f = 0.2mHz$ $(t = T/4 = 1250s = 20.83\min)$ the gravity acceleration $\vec{g}_3$ above the *third* lamp will be given by

$$\vec{g}_3 = -5.126\vec{g}$$

Above the *second* lamp, the gravity acceleration given by (30), is

$$\vec{g}_2 = +2.972\vec{g} \qquad .$$

According to (27) the gravity acceleration above the *first* lamp is

$$\vec{g}_1 = \text{-1,724}\vec{g}$$

Note that, by this process an acceleration $\vec{g}$ can be increased several times in the direction of $\vec{g}$ or in the opposite direction.

In the experiment proposed in Fig. 1, we can start with ELF voltage sinusoidal wave of amplitude $V_0 = 1.0V$ and frequency $1mHz$. Next, the frequency will be progressively decreased down to $0.8mHz$, $0.6mHz$, $0.4mHz$ and $0.2mHz$. Afterwards, the amplitude of the voltage wave must be increased to $V_0 = 1.5V$ and the frequency decreased in the above mentioned sequence.

Table1 presents the *theoretical* values for $g_1$ and $g_2$, calculated respectively by means of (25) and (30).They are also plotted on Figures 5, 6 and 7 as a function of the frequency $f_{ELF}$.

Now consider a chamber filled with *Air* at $3\times10^{-12}torr$ and 300K as shown in Figure 8 (a). Under these circumstances, the mass density of the *air* inside the chamber, according to Eq. (21) is $\rho_{air} \cong 4.94\times10^{-15}kg.m^{-3}$.

If the frequency of the *magnetic* field, $B$, through the *air* is $f = 60Hz$ then $\omega\varepsilon = 2\pi f\varepsilon \cong 3\times10^{-9}S/m$. Assuming that the electric conductivity of the *air* inside the chamber, $\sigma_{(air)}$ is much less than $\omega\varepsilon$, i.e., $\sigma_{(air)} << \omega\varepsilon$ (The atmospheric air conductivity is of the order of $2-100\times10^{-15}S.m^{-1}$ [6, 7]) then we can rewritten the Eq. (11) as follows

$$n_{r(air)} \cong \sqrt{\varepsilon_r\mu_r} \cong 1 \qquad (34)$$

From Eqs. (7), (14) and (34) we thus obtain

$$m_{g(air)}=\left\{1-2\left[\sqrt{1+\left(\frac{B^2}{\mu_{air}\rho_{air}c^2}n_{r(air)}\right)^2}-1\right]\right\}m_{i(air)}=$$
$$=\left\{1-2\left[\sqrt{1+3.2\times10^6B^4}-1\right]\right\}m_{i(air)} \qquad (35)$$

Therefore, due to the *gravitational shielding effect* produced by the decreasing of $m_{g(air)}$, the gravity acceleration *above* the *air* inside the chamber will be given by

$$g' = \chi_{air}g = \frac{m_{g(air)}}{m_{i(air)}}g =$$
$$= \left\{1-2\left[\sqrt{1+3.2\times10^6B^4}-1\right]\right\}g$$

Note that the gravity acceleration above the *air* becomes *negative* for $B > 2.5\times10^{-2}T$.

For $B = 0.1T$ the gravity acceleration above the air becomes

$$g' \cong -32.8g$$

Therefore the ultra-low pressure air inside the chamber, such as the Hg plasma inside the fluorescent lamp, works like a Gravitational Shield that in practice, may be used to build *Gravity Control Cells* (GCC) for several practical applications.

Consider for example the GCCs of Plasma presented in Fig.3. The ionization of the plasma can be made of several manners. For example, by means of an electric field between the electrodes (Fig. 3(a)) or by means of a RF signal (Fig. 3(b)). In the first case the ELF electric field and the ionizing electric field can be the same.

Figure 3(c) shows a GCC filled with *air* (at ambient temperature and 1 atm) strongly ionized by means of alpha particles emitted from 36 radioactive ions sources (a very small quantity of *Americium* 241†). The radioactive element Americium has a half-life of 432 years, and emits *alpha particles* and low energy gamma rays $(\approx 60KeV)$. In order to shield the *alpha* particles and *gamma* rays emitted from the Americium 241 it is sufficient to encapsulate the GCC with *epoxy*. The alpha particles generated by the americium ionize the oxygen and nitrogen atoms of the air in the *ionization chamber* (See Fig. 3(c)) increasing the *electrical conductivity* of the air inside the chamber. The high-speed alpha particles hit molecules in the air and knock off electrons to form ions, according to the following expressions

$$O_2 + H_e^{++} \rightarrow O_2^+ + e^- + H_e^{++}$$
$$N_2 + H_e^{++} \rightarrow N_2^+ + e^- + H_e^{++}$$

It is known that the electrical conductivity is proportional to both the concentration and the mobility of the *ions* and the *free electrons*, and is expressed by

$$\sigma = \rho_e\mu_e + \rho_i\mu_i$$

Where $\rho_e$ and $\rho_i$ express respectively the concentrations $\left(C/m^3\right)$ of electrons and ions; $\mu_e$ and $\mu_i$ are respectively the mobilities of the electrons and the ions.

In order to calculate the electrical conductivity of the air inside the ionization chamber, we first need to calculate the concentrations $\rho_e$ and $\rho_i$. We start calculating the *disintegration constant*, $\lambda$, for the Am 241 :

$$\lambda = \frac{0.693}{T^{\frac{1}{2}}} = \frac{0.693}{432\left(3.15\times10^7 s\right)} = 5.1\times10^{-11}s^{-1}$$

Where $T^{\frac{1}{2}} = 432 years$ is the *half-life* of the Am 241.

One $kmole$ of an isotope has mass equal to atomic mass of the isotope expressed in kilograms. Therefore, $1g$ of Am 241 has

$$\frac{10^{-3}kg}{241\ \ kg/kmole} = 4.15\times10^{-6}kmoles$$

One $kmole$ of any isotope contains the Avogadro's number of atoms. Therefore $1g$ of Am 241 has

$$N = 4.15\times10^{-6}kmoles\times$$
$$\times 6.025\times10^{26}\ atoms/kmole = 2.50\times10^{21}atoms$$

Thus, the *activity* [8] of the sample is

† The radioactive element *Americium* (Am-241) is widely used in *ionization smoke detectors*. This type of smoke detector is more common because it is inexpensive and better at detecting the smaller amounts of smoke produced by flaming fires. Inside an ionization detector there is a small amount (perhaps 1/5000th of a gram) of americium-241. The Americium is present in oxide form ($AmO_2$) in the detector. The cost of the $AmO_2$ is US$ 1,500 per gram. The amount of radiation in a smoke detector is extremely small. It is also predominantly alpha radiation. Alpha radiation cannot penetrate a sheet of paper, and it is blocked by several centimeters of air. The americium in the smoke detector could only pose a danger if inhaled.

$R = \lambda N = 1.3\times10^{11}$ disintegrations/s.

However, we will use 36 ionization sources each one with 1/5000th of a gram of Am 241. Therefore we will only use $7.2\times10^{-3}\,g$ of Am 241. Thus, $R$ reduces to:

$$R = \lambda N \cong 10^{9}\text{ disintegrations/s}$$

This means that at *one* second, about $10^{9}\,\alpha$ *particles* hit molecules in the air and knock off electrons to form ions $O_2^+$ and $N_2^+$ inside the ionization chamber. Assuming that *each* alpha particle yields *one* ion at each $1/10^{9}$ second then the total number of ions produced in one second will be $N_i \cong 10^{18}\,ions$. This corresponds to an ions concentration

$$\rho_i = eN_i/V \approx 0.1/V \quad \left(C/m^3\right)$$

Where $V$ is the volume of the ionization chamber. Obviously, the concentration of electrons will be the same, i.e., $\rho_e = \rho_i$. For $d = 2cm$ and $\phi = 20cm$ (See Fig.3(c)) we obtain

$V = \frac{\pi}{4}(0.20)^2\left(2\times10^{-2}\right) = 6.28\times10^{-4}\,m^3$ Then we get:

$$\rho_e = \rho_i \approx 10^{2}\ C/m^3$$

This corresponds to the *minimum* concentration level in the case of *conducting materials*. For these materials, at temperature of 300K, the mobilities $\mu_e$ and $\mu_i$ vary from $10$ up to $100\ m^2V^{-1}s^{-1}$ [9]. Then we can assume that $\mu_e = \mu_i \approx 10\ m^2V^{-1}s^{-1}$. (*minimum* mobility level for conducting materials). Under these conditions, the electrical conductivity of the air inside the ionization chamber is

$$\sigma_{air} = \rho_e\mu_e + \rho_i\mu_i \approx 10^{3}\,S.m^{-1}$$

At temperature of 300K, the air *density* inside the GCC, is $\rho_{air} = 1.1452\,kg.m^{-3}$. Thus, for $d = 2cm$, $\sigma_{air} \approx 10^{3}\,S.m^{-1}$ and $f = 60Hz$ Eq. (20) gives

$$\chi_{air} = \frac{m_{g(air)}}{m_{i(air)}} = \left\{1 - 2\left[\sqrt{1 + \frac{\mu}{4c^2}\left(\frac{\sigma_{air}}{4\pi f}\right)^3 \frac{V_{rms}^4}{d^4\rho_{air}^2}} - 1\right]\right\} = \left\{1 - 2\left[\sqrt{1 + 3.10\times10^{-16}V_{rms}^4} - 1\right]\right\}$$

Note that, for $V_{rms} \cong 7.96KV$, we obtain: $\chi_{(air)} \cong 0$. Therefore, if the voltages range of this GCC is: $0 - 10KV$ then it is possible to reach $\chi_{air} \cong -1$ when $V_{rms} \cong 10KV$.

It is interesting to note that $\sigma_{air}$ can be strongly increased by increasing the amount of Am 241. For example, by using $0.1g$ of Am 241 the value of $R$ increases to:

$$R = \lambda N \cong 10^{10}\text{ disintegrations/s}$$

This means $N_i \cong 10^{20}\,ions$ that yield

$$\rho_i = eN_i/V \approx 10/V \quad \left(C/m^3\right)$$

Then, by reducing, $d$ and $\phi$ respectively, to 5mm and to 11.5cm, the volume of the ionization chamber reduces to:

$V = \frac{\pi}{4}(0.115)^2\left(5\times10^{-3}\right) = 5.19\times10^{-5}\,m^3$

Consequently, we get:

$$\rho_e = \rho_i \approx 10^{5}\ C/m^3$$

Assuming that $\mu_e = \mu_i \approx 10\ m^2V^{-1}s^{-1}$, then the electrical conductivity of the air inside the ionization chamber becomes

$$\sigma_{air} = \rho_e\mu_e + \rho_i\mu_i \approx 10^{6}\,S.m^{-1}$$

This reduces for $V_{rms} \cong 18.8V$ the voltage necessary to yield $\chi_{(air)} \cong 0$ and reduces to $V_{rms} \cong 23.5V$ the voltage necessary to reach $\chi_{air} \cong -1$.

If the outer surface of a metallic sphere with radius $a$ is covered with a radioactive element (for example Am 241), then the electrical conductivity of the air (very close to the sphere) can be strongly increased (for example up to $\sigma_{air} \cong 10^6 s.m^{-1}$). By applying a low-frequency electrical potential $V_{rms}$ to the sphere, in order to produce an electric field $E_{rms}$ starting from the outer surface of the sphere, then very close to the sphere the low-frequency electromagnetic field is $E_{rms} = V_{rms}/a$, and according to Eq. (20), the *gravitational mass* of the air in this region expressed by

$$m_{g(air)} = \left\{1 - 2\left[\sqrt{1 + \frac{\mu_0}{4c^2}\left(\frac{\sigma_{air}}{4\pi f}\right)^3 \frac{V_{rms}^4}{a^4 \rho_{air}^2}} - 1\right]\right\} m_{i0(air)},$$

can be easily reduced, making possible to produce a controlled *Gravitational Shielding* (similar to a GCC) surround the sphere.

This becomes possible to build a spacecraft to work with a gravitational shielding as shown in Fig. 4.

The *gravity accelerations* on the spacecraft (due to the rest of the Universe. See Fig.4) is given by

$$g'_i = \chi_{air} g_i \qquad i = 1, 2, 3 \ldots n$$

Where $\chi_{air} = m_{g(air)}/m_{i0(air)}$. Thus, the *gravitational forces* acting on the spacecraft are given by

$$F_{is} = M_g g'_i = M_g \left(\chi_{air} g_i\right)$$

By reducing the value of $\chi_{air}$, these forces can be reduced.

According to the *Mach's principle*;

"The *local inertial forces* are determined by the *gravitational interactions* of the local system with the distribution of the cosmic masses".

Thus, the local inertia is just the gravitational influence of the rest of matter existing in the Universe. Consequently, if we reduce the gravitational interactions between a spacecraft and the rest of the Universe, then *the inertial properties of the spacecraft* will be also reduced. This effect leads to a new concept of spacecraft and space flight.

Since $\chi_{air}$ is given by

$$\chi_{air} = \frac{m_{g(air)}}{m_{i0(air)}} = \left\{1 - 2\left[\sqrt{1 + \frac{\mu_0}{4c^2}\left(\frac{\sigma_{air}}{4\pi f}\right)^3 \frac{V_{rms}^4}{a^4 \rho_{air}^2}} - 1\right]\right\}$$

Then, for $\sigma_{air} \cong 10^6 s.m^{-1}$, $f = 6Hz$, $a = 5m$, $\rho_{air} \cong 1Kg.m^{-3}$ and $V_{rms} = 3.35KV$ we get

$$\chi_{air} \cong 0$$

Under these conditions, the gravitational forces upon the spacecraft become approximately nulls and consequently, the spacecraft practically *loses its inertial properties.*

Out of the terrestrial atmosphere, the gravity acceleration upon the spacecraft is negligible and therefore the gravitational shielding is not necessary. However, if the spacecraft is in the outer space and we want to use the gravitational shielding then, $\chi_{air}$ must be replaced by $\chi_{vac}$ where

$$\chi_{vac} = \frac{m_{g(vac)}}{m_{i0(vac)}} = \left\{1 - 2\left[\sqrt{1 + \frac{\mu_0}{4c^2}\left(\frac{\sigma_{vac}}{4\pi f}\right)^3 \frac{V_{rms}^4}{a^4 \rho_{vac}^2}} - 1\right]\right\}$$

The electrical conductivity of the ionized outer space (very close to the spacecraft) is small; however, its density is remarkably small $\left(<< 10^{-16} Kg.m^{-3}\right)$, in such a manner that the smaller value of the factor $\sigma_{vac}^3/\rho_{vac}^2$ can be easily compensated by the increase of $V_{rms}$.

It was shown that, when the gravitational mass of a particle is reduced to ranging between $+0.159M_i$ to $-0.159M_i$, it becomes *imaginary* [1], i.e., the gravitational and the inertial masses of the particle become *imaginary*. Consequently, the particle disappears from our ordinary space-time. However, the factor $\chi = M_{g(imaginary)}/M_{i(imaginary)}$ remains *real* because

$$\chi = \frac{M_{g(imaginary)}}{M_{i(imaginary)}} = \frac{M_g i}{M_i i} = \frac{M_g}{M_i} = real$$

Thus, if the gravitational mass of the particle is reduced by means of absorption of an amount of electromagnetic energy $U$, for example, we have

$$\chi = \frac{M_g}{M_i} = \left\{1 - 2\left[\sqrt{1+\left(U/m_{i0}c^2\right)^2} - 1\right]\right\}$$

This shows that the energy $U$ of the electromagnetic field *remains acting on* the imaginary particle. In practice, this means that *electromagnetic fields act on imaginary particles*. Therefore, the electromagnetic field of a GCC remains acting on the particles inside the GCC even when their gravitational masses reach the gravitational mass ranging between $+0.159M_i$ to $-0.159M_i$ and they become imaginary particles. This is very important because it means that the GCCs of a gravitational spacecraft keep on working when the spacecraft becomes imaginary.

Under these conditions, the gravity accelerations on the *imaginary* spacecraft particle (due to the rest of the imaginary Universe) are given by

$$g'_j = \chi\ g_j \qquad j = 1,2,3,\ldots,n.$$

Where $\chi = M_{g(imaginary)}/M_{i(imaginary)}$ and $g_j = -Gm_{gj(imaginary)}/r_j^2$. Thus, the gravitational forces acting on the spacecraft are given by

$$F_{gj} = M_{g(imaginary)}g'_j =$$
$$= M_{g(imaginary)}\left(-\chi Gm_{gj(imaginary)}/r_j^2\right) =$$
$$= M_g i\left(-\chi Gm_{gj} i/r_j^2\right) = +\chi GM_g m_{gj}/r_j^2 .$$

Note that these forces are *real*. Remind that, the Mach's principle says that the *inertial effects* upon a particle are consequence of the gravitational interaction of the particle with the rest of the Universe. Then we can conclude that the *inertial forces* upon an *imaginary* spacecraft are also *real*. Consequently, it can travel in the imaginary space-time using its thrusters.

It was shown that, *imaginary particles* can have *infinite speed* in the *imaginary space-time* [1] . Therefore, this is also the speed upper limit for the spacecraft in the imaginary space-time.

Since the gravitational spacecraft can use its thrusters after to becoming an imaginary body, then if the thrusters produce a total thrust $F = 1000kN$ and the gravitational mass of the spacecraft is reduced from $M_g = M_i = 10^5 kg$ down to $M_g \cong 10^{-6} kg$, the acceleration of the spacecraft will be, $a = F/M_g \cong 10^{12} m.s^{-2}$. With this acceleration the spacecraft crosses the "visible" Universe ($diameter = d \approx 10^{26} m$) in a time interval $\Delta t = \sqrt{2d/a} \cong 1.4\times 10^7 m.s^{-1} \cong 5.5\ months$

Since the inertial effects upon the spacecraft are reduced by $M_g/M_i \cong 10^{-11}$ then, in spite of the effective spacecraft acceleration be $a = 10^{12} m.s^{-1}$, the effects for the crew and for the spacecraft will be equivalent to an acceleration $a'$ given by

$$a' = \frac{M_g}{M_i} a \approx 10 m.s^{-1}$$

This is the order of magnitude of the acceleration upon of a commercial jet aircraft.

On the other hand, the travel in the *imaginary* space-time can be very safe, because there won't any material body along the trajectory of the spacecraft.

Now consider the GCCs presented in Fig. 8 (a). Note that below and above the *air* are the bottom and the top of the chamber. Therefore the choice of the material of the chamber is highly relevant. If the chamber is made of steel, for example, and the gravity acceleration below the chamber is $g$ then at the bottom of the chamber, the gravity becomes $g' = \chi_{steel}\, g$ ; in the air, the gravity is $g'' = \chi_{air} g' = \chi_{air}\chi_{steel} g$ . At the top of the chamber, $g''' = \chi_{steel} g'' = (\chi_{steel})^2 \chi_{air} g$. Thus, out of the chamber (close to the top) the gravity acceleration becomes $g'''$. (See Fig. 8 (a)). However, for the steel at $B < 300T$ and $f = 1\times 10^{-6}\,Hz$ , we have

$$\chi_{steel} = \frac{m_{g(steel)}}{m_{i(steel)}} = \left\{1 - 2\left[\sqrt{1 + \frac{\sigma_{(steel)} B^4}{4\pi f \mu \rho^2_{(steel)} c^2}} - 1\right]\right\} \cong 1$$

Since $\rho_{steel} = 1.1\times 10^6\, S.m^{-1}$, $\mu_r = 300$ and $\rho_{(steel)} = 7800 k.m^{-3}$.

Thus, due to $\chi_{steel} \cong 1$ it follows that

$$g''' \cong g'' = \chi_{air}\, g' \cong \chi_{air}\, g$$

If instead of one GCC we have *three* GCC, all with steel box (Fig. 8(b)), then the gravity acceleration above the *second* GCC, $g_2$ will be given by

$$g_2 \cong \chi_{air}\, g_1 \cong \chi_{air}\chi_{air}\, g$$

and the gravity acceleration above the *third* GCC, $g_3$ will be expressed by

$$g_3 \cong \chi_{air} g'' \cong \chi_{air}^{\;3} g$$

## III. CONSEQUENCES

These results point to the possibility to convert gravitational energy into rotational mechanical energy. Consider for example the system presented in Fig. 9. Basically it is a motor with massive iron rotor and a box filled with gas or plasma at ultra-low pressure (Gravity Control Cell-GCC) as shown in Fig. 9. The GCC is placed below the rotor in order to become *negative* the acceleration of gravity inside *half* of the rotor $\left(g' = (\chi_{steel})^2 \chi_{air}\, g \cong \chi_{air}\, g = -ng\right)$. Obviously this causes a torque $T = (-F' + F)r$ and the rotor spins with angular velocity $\omega$. The average power, $P$ , of the motor is given by

$$P = T\omega = [(-F' + F)r]\omega \qquad (36)$$

Where

$$F' = \tfrac{1}{2} m_g\, g' \qquad\qquad F = \tfrac{1}{2} m_g\, g$$

and $m_g \cong m_i$ ( mass of the rotor ). Thus, Eq. (36) gives

$$P = (n+1)\frac{m_i g\omega\, r}{2} \qquad (37)$$

On the other hand, we have that

$$-g' + g = \omega^2 r \qquad (38)$$

Therefore the angular speed of the rotor is given by

$$\omega = \sqrt{\frac{(n+1)g}{r}} \qquad (39)$$

By substituting (39) into (37) we obtain the expression of the average power of the *gravitational motor*, i.e.,

$$P = \tfrac{1}{2} m_i \sqrt{(n+1)^3 g^3 r} \qquad (40)$$

Now consider an electric generator coupling to the gravitational motor in order to produce electric energy.

Since $\omega = 2\pi f$ then for $f = 60Hz$ we have $\omega = 120\pi rad.s^{-1} = 3600\, rpm$ .

Therefore for $\omega = 120\pi rad.s^{-1}$ and $n = 788$ $(B \cong 0.22T)$ the Eq. (40) tell us that we must have

$$r = \frac{(n+1)g}{\omega^2} = 0.0545m$$

Since $r = R/3$ and $m_i = \rho\pi R^2 h$ where $\rho$, $R$ and $h$ are respectively the mass density, the radius and the height of the rotor then for $h = 0.5m$ and $\rho = 7800 Kg.m^{-3}$ (iron) we obtain

$$m_i = 327.05kg$$

Then Eq. (40) gives

$$P \cong 2.19\times10^5\, watts \cong 219\ KW \cong 294 HP \qquad (41)$$

This shows that the *gravitational motor* can be used to yield electric energy at large scale.

The possibility of gravity control leads to a new concept of spacecraft which is presented in Fig. 10. Due to the *Meissner effect*, the magnetic field *B* is expelled from the *superconducting shell*. The Eq. (35) shows that a magnetic field, $B$, through the *aluminum shell* of the spacecraft reduces its gravitational mass according to the following expression:

$$m_{g(Al)} = \left\{1 - 2\left[\sqrt{1+\left(\frac{B^2}{\mu c^2 \rho_{(Al)}} n_{r(Al)}\right)^2} - 1\right]\right\} m_{i(Al)} \qquad (42)$$

If the frequency of the magnetic field is $f = 10^{-4}\,Hz$ then we have that $\sigma_{(Al)} >> \omega\varepsilon$ since the electric conductivity of the aluminum is $\sigma_{(Al)} = 3.82\times10^7\,S.m^{-1}$. In this case, the Eq. (11) tell us that

$$n_{r(Al)} = \sqrt{\frac{\mu c^2 \sigma_{(Al)}}{4\pi f}} \qquad (43)$$

Substitution of (43) into (42) yields

$$m_{g(Al)} = \left\{1 - 2\left[\sqrt{1+\frac{\sigma_{(Al)} B^4}{4\pi f \mu \rho_{(Al)}^2 c^2}} - 1\right]\right\} m_{i(Al)} \qquad (44)$$

Since the mass density of the Aluminum is $\rho_{(Al)} = 2700\,kg.m^{-3}$ then the Eq. (44) can be rewritten in the following form:

$$\chi_{Al} = \frac{m_{g(Al)}}{m_{i(Al)}} = \left\{1 - 2\left[\sqrt{1+3.68\times10^{-8} B^4} - 1\right]\right\} \qquad (45)$$

In practice it is possible to adjust $B$ in order to become, for example, $\chi_{Al} \cong 10^{-9}$. This occurs to $B \cong 76.3T$. (Novel superconducting magnets are able to produce up to $14.7T$ [10, 11]).

Then the gravity acceleration in any direction *inside* the spacecraft, $g_l'$, will be reduced and given by

$$g_l' = \frac{m_{g(Al)}}{m_{i(Al)}} g_l = \chi_{Al} g_l \cong -10^{-9} g_l \qquad l = 1,2,..,n$$

Where $g_l$ is the *external* gravity in the direction $l$. We thus conclude that the gravity acceleration inside the spacecraft becomes negligible if $g_l << 10^9\,m.s^{-2}$. This means that the aluminum shell, under these conditions, works like a gravity shielding.

Consequently, the gravitational forces between anyone point inside the spacecraft with gravitational mass, $m_{gj}$, and another external to the spacecraft (gravitational mass $m_{gk}$) are given by

$$\vec{F}_j = -\vec{F}_k = -G\frac{m_{gj} m_{gk}}{r_{jk}^2}\hat{\mu}$$

where $m_{gk} \cong m_{ik}$ and $m_{gj} = \chi_{Al} m_{ij}$. Therefore we can rewrite equation above in the following form

$$\vec{F}_j = -\vec{F}_k = -\chi_{Al} G\frac{m_{ij} m_{ik}}{r_{jk}^2}\hat{\mu}$$

Note that when $B = 0$ the *initial gravitational forces* are

$$\vec{F}_j = -\vec{F}_k = -G\frac{m_{ij} m_{ik}}{r_{jk}^2}\hat{\mu}$$

Thus, if $\chi_{Al} \cong -10^{-9}$ then the initial gravitational forces are reduced from $10^9$ times and become repulsives.

According to the new expression for the *inertial forces* [1], $\vec{F} = m_g \vec{a}$, we see that these forces have origin in the *gravitational interaction* between a particle and the others of the Universe, just as *Mach's principle* predicts. Hence mentioned expression incorporates the Mach's principle into Gravitation Theory, and furthermore reveals that the inertial effects upon a body can be strongly reduced by means of the decreasing of its gravitational mass.

Consequently, we conclude that if the *gravitational forces* upon the spacecraft are reduced from $10^9$ times then also the *inertial forces* upon the spacecraft will be reduced from $10^9$ times when $\chi_{Al} \cong -10^{-9}$. Under these conditions, the inertial effects on the crew would be strongly decreased. Obviously this leads to a new concept of aerospace flight.

Inside the spacecraft the gravitational forces between the dielectric with gravitational mass, $M_g$ and the man (gravitational mass, $m_g$), when $B = 0$ are

$$\vec{F}_m = -\vec{F}_M = -G\frac{M_g m_g}{r^2}\hat{\mu} \qquad (46)$$

or

$$\vec{F}_m = -G\frac{M_g}{r^2}m_g\hat{\mu} = -m_g g_M\hat{\mu} \qquad (47)$$

$$\vec{F}_M = +G\frac{m_g}{r^2}M_g\hat{\mu} = +M_g g_m\hat{\mu} \qquad (48)$$

If the *superconducting box* under $M_g$ (Fig. 10) is filled with *air* at ultra-low pressure (3×10^-12 torr, 300K for example) then, when $B \neq 0$, the gravitational mass of the *air* will be reduced according to (35). Consequently, we have

$$g'_M = (\chi_{steel})^2\chi_{air}g_M \cong \chi_{air}g_M \qquad (49)$$

$$g'_m = (\chi_{steel})^2\chi_{air}g_m \cong \chi_{air}g_m \qquad (50)$$

Then the forces $\vec{F}_m$ and $\vec{F}_M$ become

$$\vec{F}_m = -m_g(\chi_{air}g_M)\hat{\mu} \qquad (51)$$

$$\vec{F}_M = +M_g(\chi_{air}g_m)\hat{\mu} \qquad (52)$$

Therefore if $\chi_{air} = -n$ we will have

$$\vec{F}_m = +nm_g g_M\hat{\mu} \qquad (53)$$

$$\vec{F}_M = -nM_g g_m\hat{\mu} \qquad (54)$$

Thus, $\vec{F}_m$ and $\vec{F}_M$ become *repulsive*. Consequently, the man inside the spacecraft is subjected to a gravity acceleration given by

$$\vec{a}_{man} = ng_M\hat{\mu} = -\chi_{air}G\frac{M_g}{r^2}\hat{\mu} \qquad (55)$$

Inside the GCC we have,

$$\chi_{air} = \frac{m_{g(air)}}{m_{i(air)}} = \left\{1 - 2\left[\sqrt{1 + \frac{\sigma_{(air)}B^4}{4\pi f\mu\rho^2_{(air)}c^2}} - 1\right]\right\} \qquad (56)$$

By ionizing the air inside the GCC (Fig. 10), for example, by means of a *radioactive* material, it is possible to increase the *air conductivity* inside the GCC up to $\sigma_{(air)} \cong 10^6 S.m^{-1}$. Then for $f = 10\ Hz$; $\rho_{(air)} = 4.94 \times 10^{-15}kg.m^{-3}$ (Air at 3 ×10^-12 torr, 300K) and we obtain

$$\chi_{air} = \left\{2\left[\sqrt{1 + 2.8\times10^{21}B^4} - 1\right] - 1\right\} \qquad (57)$$

For $B = B_{GCC} = 0.1T$ (note that, due to the *Meissner effect*, the magnetic field $B_{GCC}$ stay confined inside the *superconducting box*) the Eq. (57) yields

$$\chi_{air} \cong -10^9$$

Since there is no magnetic field through the *dielectric* presented in Fig.10 then, $M_g \cong M_i$. Therefore if $M_g \cong M_i = 100Kg$ and $r = r_0 \cong 1m$ the gravity acceleration upon the man, according to Eq. (55), is

$$a_{man} \cong 10m.s^{-1}$$

Consequently it is easy to see that this system is ideal to yield artificial gravity inside the spacecraft in the case of *inter-stellar travel*, when the gravity acceleration out of the spacecraft - due to the Universe - becomes negligible.

The *vertical* displacement of the spacecraft can be produced by means of *Gravitational Thrusters*. A schematic diagram of a Gravitational Thruster is shown in Fig.11. The Gravitational Thrusters can also provide the *horizontal* displacement of the spacecraft.

The concept of Gravitational Thruster results from the theory of the *Gravity Control Battery*, showed in Fig. 8 (b). Note that the number of GCC increases the thrust of the thruster. For example, if the thruster has *three* GCCs then the gravity acceleration upon the gas sprayed inside the thruster will be *repulsive* in respect to $M_g$ (See Fig. 11(a)) and given by

$$a_{gas} = (\chi_{air})^3(\chi_{steel})^4 g \cong -(\chi_{air})^3 G\frac{M_g}{r_0^2}$$

Thus, if inside the GCCs, $\chi_{air} \cong -10^9$

(See Eq. 56 and 57) then the equation above gives

$$a_{gas} \cong +10^{27} G \frac{M_i}{r_0^2}$$

For $M_i \cong 10kg$, $r_0 \cong 1m$ and $m_{gas} \cong 10^{-12} kg$ the thrust is

$$F = m_{gas} a_{gas} \cong 10^5 N$$

Thus, the Gravitational Thrusters are able to produce strong thrusts.

Note that in the case of very strong $\chi_{air}$, for example $\chi_{air} \cong -10^9$, the gravity accelerations upon the boxes of the second and third GCCs become very strong (Fig.11 (a)). Obviously, the walls of the mentioned boxes cannot to stand the enormous pressures. However, it is possible to build a similar system with 3 or more GCCs, *without material boxes*. Consider for example, a surface with several radioactive sources (Am-241, for example). The *alpha* particles emitted from the Am-241 cannot reach besides 10cm of air. Due to the trajectory of the alpha particles, three or more successive layers of air, with different electrical conductivities $\sigma_1$, $\sigma_2$ and $\sigma_3$, will be established in the ionized region (See Fig.11 (b)). It is easy to see that the gravitational shielding effect produced by these three layers is similar to the effect produced by the 3 GCCs shown in Fig. 11 (a).

It is important to note that if $F$ is force produced by a thruster then the spacecraft acquires acceleration $a_{spacecraft}$ given by [1]

$$a_{spacecraft} = \frac{F}{M_{g(spacecraft)}} = \frac{F}{\chi_{Al} M_{i(inside)} + m_{i(Al)}}$$

Therefore if $\chi_{Al} \cong 10^{-9}$; $M_{i(inside)} = 10^4 Kg$ and $m_{i(Al)} = 100 Kg$ (inertial mass of the aluminum shell) then it will be necessary $F = 10kN$ to produce

$$a_{spacecraft} = 100 m.s^{-2}$$

Note that the concept of Gravitational Thrusters leads directly to the *Gravitational Turbo Motor* concept (See Fig. 12).

Let us now calculate the gravitational forces between two very close *thin* layers of the *air* around the spacecraft. (See Fig. 13).

The gravitational force $dF_{12}$ that $dm_{g1}$ exerts upon $dm_{g2}$, and the gravitational force $dF_{21}$ that $dm_{g2}$ exerts upon $dm_{g1}$ are given by

$$d\vec{F}_{12} = d\vec{F}_{21} = -G \frac{dm_{g2} dm_{g1}}{r^2} \hat{\mu} \qquad (58)$$

Thus, the gravitational forces between the *air layer* 1, gravitational mass $m_{g1}$, and the *air layer* 2, gravitational mass $m_{g2}$, around the spacecraft are

$$\vec{F}_{12} = -\vec{F}_{21} = -\frac{G}{r^2} \int_0^{m_{g1}} \int_0^{m_{g2}} dm_{g1} dm_{g2} \hat{\mu} =$$

$$= -G \frac{m_{g1} m_{g2}}{r^2} \hat{\mu} = -\chi_{air} \chi_{air} G \frac{m_{i1} m_{i2}}{r^2} \hat{\mu} \qquad (59)$$

At 100km altitude the air pressure is $5.691 \times 10^{-3} torr$ and $\rho_{(air)} = 5.998 \times 10^{-6} kg.m^{-3}$ [12]. By ionizing the air surround the spacecraft, for example, by means of an oscillating electric field, $E_{osc}$, starting from the surface of the spacecraft ( See Fig. 13) it is possible to increase the *air conductivity* near the spacecraft up to $\sigma_{(air)} \cong 10^6 S.m^{-1}$ . Since $f = 1Hz$ and, in this case $\sigma_{(air)} >> \omega\varepsilon$, then, according to Eq. (11), $n_r = \sqrt{\mu \sigma_{(air)} c^2 / 4\pi f}$ . From Eq.(56) we thus obtain

$$\chi_{air} = \frac{m_{g(air)}}{m_{i(air)}} = \left\{ 1 - 2\left[ \sqrt{1 + \frac{\sigma_{(air)} B^4}{4\pi f \mu_0 \rho_{(air)}^2 c^2}} - 1 \right] \right\} \qquad (60)$$

Then for $B = 763T$ the Eq. (60) gives

$$\chi_{air} = \left\{ 1 - 2\left[ \sqrt{1 + \sim 10^4 B^4} - 1 \right] \right\} \cong -10^8 \qquad (61)$$

By substitution of $\chi_{air} \cong -10^8$ into Eq., (59) we get

$$\vec{F}_{12} = -\vec{F}_{21} = -10^{16} G \frac{m_{i1} m_{i2}}{r^2} \hat{\mu} \qquad (62)$$

If $m_{i1} \cong m_{i2} = \rho_{air} \mathbf{V}_1 \cong \rho_{air} \mathbf{V}_2 \cong 10^{-8} kg$, and $r = 10^{-3} m$ we obtain

$$\vec{F}_{12} = -\vec{F}_{21} \cong -10^{-4} N \qquad (63)$$

These forces are much more intense than the *inter-atomic forces* (the forces which maintain joined atoms, and molecules that make the solids and liquids) whose intensities, according to the Coulomb's law, is of the order of 1-1000×10$^{-8}$N.

Consequently, the air around the spacecraft will be strongly compressed upon their surface, making an "*air shell*" that will accompany the spacecraft during its displacement and will protect the *aluminum shell* of the direct attrition with the Earth's atmosphere.

In this way, during the flight, the attrition would occur just between the "air shell" and the atmospheric air around her. Thus, the spacecraft would stay free of the thermal effects that would be produced by the direct attrition of the aluminum shell with the Earth's atmosphere.

Another interesting effect produced by the magnetic field $B$ of the spacecraft is the possibility of to lift a body from the surface of the Earth to the spacecraft as shown in Fig. 14. By ionizing the air surround the spacecraft, by means of an oscillating electric field, $E_{osc}$, the *air conductivity* near the spacecraft can reach, for example, $\sigma_{(air)} \cong 10^6 S.m^{-1}$. Then for $f = 1Hz$; $B = 40.8T$ and $\rho_{(air)} \cong 1.2 kg.m^{-3}$ (300K and 1 atm) the Eq. (56) yields

$$\chi_{air} = \left\{1 - 2\left[\sqrt{1 + 4.9 \times 10^{-7} B^4} - 1\right]\right\} \cong -0.1$$

Thus, the weight of the body becomes

$$P_{body} = m_{g(body)} g = \chi_{air} m_{i(body)} g = m_{i(body)} g'$$

Consequently, the body will be lifted on the direction of the spacecraft with acceleration

$$g' = \chi_{air} g \cong +0.98 m.s^{-1}$$

Let us now consider an important aspect of the flight dynamics of a Gravitational Spacecraft.

Before starting the flight, the *gravitational mass of the spacecraft*, $M_g$, must be strongly reduced, by means of a gravity control system, in order to produce – with a weak thrust $\vec{F}$, a strong acceleration, $\vec{a}$, given by [1]

$$\vec{a} = \frac{\vec{F}}{M_g}$$

In this way, the spacecraft could be strongly accelerated and quickly to reach very high speeds near speed of light.

If the gravity control system of the spacecraft is *sudden*ly turned off, the *gravitational mass* of the spacecraft becomes immediately equal to its *inertial mass*, $M_i$, $(M'_g = M_i)$ and the velocity $\vec{V}$ becomes equal to $\vec{V}'$. According to the *Momentum* Conservation Principle, we have that

$$M_g V = M'_g V'$$

Supposing that the spacecraft was traveling in space with speed $V \approx c$, and that its gravitational mass it was $M_g = 1Kg$ and $M_i = 10^4 Kg$ then the velocity of the spacecraft is reduced to

$$V' = \frac{M_g}{M'_g} V = \frac{M_g}{M_i} V \approx 10^{-4} c$$

Initially, when the velocity of the spacecraft is $\vec{V}$, its kinetic energy is $E_k = (M_g - m_g)c^2$. Where $M_g = m_g \Big/ \sqrt{1 - V^2/c^2}$.
At the instant in which the gravity control system of the spacecraft is turned off, the kinetic energy becomes $E'_k = (M'_g - m'_g)c^2$. Where $M'_g = m'_g \Big/ \sqrt{1 - V'^2/c^2}$.

We can rewritten the expressions of $E_k$ and $E'_k$ in the following form

$$E_k = (M_g V - m_g V)\frac{c^2}{V}$$

$$E'_k = (M'_g V' - m'_g V')\frac{c^2}{V'}$$

Substitution of $M_g V = M'_g V' = p$,

$m_g V = p\sqrt{1-V^2/c^2}$ and $m'_g V' = p\sqrt{1-V'^2/c^2}$ into the equations of $E_k$ and $E'_k$ gives

$$E_k = \left(1-\sqrt{1-V^2/c^2}\right)\frac{pc^2}{V}$$

$$E'_k = \left(1-\sqrt{1-V'^2/c^2}\right)\frac{pc^2}{V'}$$

Since $V \approx c$ then follows that

$$E_k \approx pc$$

On the other hand, since $V' << c$ we get

$$E'_k = \left(1-\sqrt{1-V'^2/c^2}\right)\frac{pc^2}{V'} =$$

$$\cong \left(1-\frac{1}{1+\frac{V'^2}{2c^2}+\ldots}\right)\frac{pc^2}{V'} \cong \left(\frac{V'}{2c}\right)pc$$

Therefore we conclude that $E_k >> E'_k$. Consequently, when the gravity control system of the spacecraft is turned off, occurs an *abrupt* decrease in the kinetic energy of the spacecraft, $\Delta E_k$, given by

$$\Delta E_k = E_k - E'_k \approx pc \approx M_g c^2 \approx 10^{17} J$$

By comparing the energy $\Delta E_k$ with the *inertial energy* of the spacecraft, $E_i = M_i c^2$, we conclude that

$$\Delta E_k \approx \frac{M_g}{M_i} E_i \approx 10^{-4} M_i c^2$$

The energy $\Delta E_k$ (several *megatons*) must be released in very short time interval. It is approximately the same amount of energy that would be released in the case of collision of the spacecraft‡. However, the situation is very different of a collision ($M_g$ just becomes suddenly equal to $M_i$), and possibly the energy $\Delta E_k$ is converted into a *High Power Electromagnetic Pulse*.

Obviously this electromagnetic pulse (EMP) will induce heavy currents in all electronic equipment that mainly contains semiconducting and conducting materials. This produces immense heat that melts the circuitry inside. As such, *while not being directly responsible for the loss of lives*, these EMP are capable of disabling electric/electronic systems. Therefore, we possibly have a new type of *electromagnetic bomb*. An *electromagnetic bomb* or *E-bomb* is a well-known weapon designed to disable electric/electronic systems on a wide scale with an intense electromagnetic pulse.

Based on the theory of the GCC it is also possible to build a *Gravitational Press* of *ultra-high* pressure as shown in Fig.15.

The chamber 1 and 2 are GCCs with air at $1\times10^{-4}$torr, 300K $\left(\sigma_{(air)} \approx 10^6 S.m^{-1}; \rho_{(air)} = 5\times10^{-8} kg.m^{-3}\right)$. Thus, for $f = 10Hz$ and $B = 0.107T$ we have

$$\chi_{air} = \left\{1-2\left[\sqrt{1+\frac{\sigma_{(air)}B^4}{4\pi f\mu_0\rho^2_{(air)}c^2}}-1\right]\right\} \cong -118$$

The gravity acceleration above the air of the chamber 1 is

$$\vec{g}_1 = \chi_{stell}\chi_{air} g\hat{\mu} \cong +1.15\times10^3\hat{\mu} \qquad (64)$$

Since, in this case, $\chi_{steel} \cong 1$; $\hat{\mu}$ is an *unitary vector* in the opposite direction of $\vec{g}$.

Above the air of the chamber 2 the gravity acceleration becomes

$$\vec{g}_2 = (\chi_{stell})^2(\chi_{air})^2 g\hat{\mu} \cong -1.4\times10^5\hat{\mu} \qquad (65)$$

Therefore the *resultant* force $\vec{R}$ acting on $m_2$, $m_1$ and $m$ is

‡ In this case, the collision of the spacecraft would release $\approx 10^{17}$J (several megatons) and it would be similar to a powerful *kinetic weapon*.

$$\vec{R} = \vec{F}_2 + \vec{F}_1 + \vec{F} = m_2\vec{g}_2 + m_1\vec{g}_1 + m\vec{g} =$$
$$= -1.4 \times 10^5 m_2\hat{\mu} + 1.15 \times 10^3 m_1\hat{\mu} - 9.81m\hat{\mu} =$$
$$\cong -1.4 \times 10^5 m_2\hat{\mu} \qquad (66)$$

where

$$m_2 = \rho_{steel} V_{disk\,2} = \rho_{steel}\left(\frac{\pi}{4}\phi_{inn}^2 H\right) \qquad (67)$$

Thus, for $\rho_{steel} \cong 10^4 kg.m^{-3}$ we can write that

$$F_2 \cong 10^9 \phi_{inn}^2 H$$

For the steel $\tau \cong 10^5 kg.cm^{-2} = 10^9 kg.m^{-2}$ consequently we must have $F_2/S_\tau < 10^9 kg.m^{-2}$ ($S_\tau = \pi\phi_{inn} H$ see Fig.15). This means that

$$\frac{10^9 \phi_{inn}^2 H}{\pi\phi_{inn} H} < 10^9 kg.m^{-2}$$

Then we conclude that

$$\phi_{inn} < 3.1m$$

For $\phi_{inn} = 2m$ and $H = 1m$ the Eq. (67) gives

$$m_2 \cong 3\times10^4 kg$$

Therefore from the Eq. (66) we obtain

$$R \cong 10^{10} N$$

Consequently, in the area $S = 10^{-4} m^2$ of the Gravitational Press, the pressure is

$$p = \frac{R}{S} \cong 10^{14} N.m^{-2}$$

This enormous pressure is much greater than the pressure in the center of the Earth ($3.617\times10^{11} N.m^{-2}$) [13]. It is near of the gas pressure in the *center of the sun* ($2\times10^{16} N.m^{-2}$). Under the action of such intensities new states of matter are created and astrophysical phenomena may be simulated in the lab for the first time, e.g. supernova explosions. Controlled thermonuclear fusion by inertial confinement, fast nuclear ignition for energy gain, novel collective acceleration schemes of particles and the numerous variants of material processing constitute examples of progressive applications of such *Gravitational Press* of ultra-high pressure.

The GCCs can also be applied on generation and detection of *Gravitational Radiation*.

Consider a cylindrical GCC (GCC antenna) as shown in Fig.16 (a). The *gravitational mass* of the *air* inside the GCC is

$$m_{g(air)} = \left\{1 - 2\left[\sqrt{1 + \frac{\sigma_{(air)} B^4}{4\pi f \mu \rho_{(air)}^2 c^2}} - 1\right]\right\} m_{i(air)} \qquad (68)$$

By varying $B$ one can varies $m_{g(air)}$ and consequently to vary the gravitational field generated by $m_{g(air)}$, producing then gravitational radiation. Then a GCC can work like a *Gravitational Antenna*.

Apparently, Newton's theory of gravity had no gravitational waves because, if a gravitational field changed in some way, that change took place *instantaneously* everywhere in space, and one can think that there is not a wave in this case. However, we have already seen that the gravitational interaction can be repulsive, besides attractive. Thus, as with electromagnetic interaction, the gravitational interaction must be produced by the exchange of "virtual" *quanta of* spin 1 and mass null, i.e., the *gravitationa*l "virtual" *quanta* (*graviphoton*) must have spin 1 and not 2. Consequently, the fact of a change in a gravitational field reach *instantaneously* everywhere in space occurs simply due to the speed of the *graviphoton* to be *infinite*. It is known that there is no speed limit for "*virtual*" photons. On the contrary, the *electromagnetic quanta* ("virtual" photons) could not communicate the *electromagnetic interaction* an infinite distance.

Thus, there are *two types* of gravitational radiation: the *real* and *virtual*, which is constituted of graviphotons; the *real* gravitational waves are ripples in the space-time generated by *gravitational field* changes. According to Einstein's theory of gravity the velocity of propagation of these waves is equal to the speed of light ($c$).

Unlike the electromagnetic waves the *real* gravitational waves have low interaction with matter and consequently low scattering. Therefore *real* gravitational waves are suitable as a means of transmitting information. However, when the distance between transmitter and receiver is too large, for example of the order of magnitude of several light-years, the transmission of information by means of gravitational waves becomes impracticable due to the long time necessary to receive the information. On the other hand, there is no delay during the transmissions by means of *virtual* gravitational radiation. In addition the scattering of this radiation is null. Therefore the *virtual* gravitational radiation is very suitable as a means of transmitting information at any distances including astronomical distances.

As concerns detection of the *virtual* gravitational radiation from GCC antenna, there are many options. Due to *Resonance Principle* a similar GCC antenna (receiver) *tuned at the same frequency* can absorb energy from an incident *virtual* gravitational radiation (See Fig.16 (b)). Consequently, the gravitational mass of the air inside the GCC receiver will vary such as the gravitational mass of the air inside the GCC transmitter. This will induce a magnetic field similar to the magnetic field of the GCC transmitter and therefore the current through the coil inside the GCC receiver will have the same characteristics of the current through the coil inside the GCC transmitter. However, the *volume* and *pressure* of the air inside the two GCCs must be exactly the same; also the *type* and the *quantity of atoms* in the air inside the two GCCs must be exactly the same. Thus, the GCC antennas are simple but they are not easy to build.

Note that a GCC antenna radiates *graviphotons* and *gravitational waves* simultaneously (Fig. 16 (a)). Thus, it is not only a gravitational antenna: it is a *Quantum Gravitational Antenna* because it can also emit and detect gravitational "virtual" *quanta* (graviphotons), which, in turn, can transmit information *instantaneously* from any distance in the Universe *without* scattering.

Due to the difficulty to build two similar GCC antennas and, considering that the electric current in the receiver antenna can be detectable even if the gravitational mass of the nuclei of the antennas are not *strongly* reduced, then we propose to replace the gas at the nuclei of the antennas by a thin *dielectric lamina*. The dielectric lamina with exactly $10^8$ atoms ($10^3$atoms $\times$ $10^3$atoms $\times$ $10^2$atoms) is placed between the plates (electrodes) as shown in Fig. 17. When the *virtual* gravitational radiation strikes upon the dielectric lamina, its gravitational mass varies similarly to the gravitational mass of the dielectric lamina of the transmitter antenna, inducing an electromagnetic field ($E$, $B$) similar to the transmitter antenna. Thus, the electric current in the receiver antenna will have the same characteristics of the current in the transmitter antenna. In this way, it is then possible to build two similar antennas whose nuclei have the same volumes and the same types and quantities of atoms.

Note that the Quantum Gravitational Antennas can also be used to transmit *electric power*. It is easy to see that the Transmitter and Receiver (Fig. 17(a)) can work with strong voltages and electric currents. This means that strong electric power can be transmitted among Quantum Gravitational Antennas. This obviously solves the problem of *wireless* electric power transmission.

The existence of *imaginary masses* has been predicted in a previous work [1]. Here we will propose a method and a device using GCCs for obtaining *images* of *imaginary bodies*.

It was shown that the *inertial* imaginary mass associated to an *electron* is given by

$$m_{ie(ima)} = \frac{2}{\sqrt{3}}\left(\frac{hf}{c^2}\right)i = \frac{2}{\sqrt{3}} m_{ie(real)}\, i \qquad (69)$$

Assuming that the correlation between the gravitational mass and the inertial mass (Eq.6) is the same for both imaginary and real masses then follows that the *gravitational* imaginary mass associated to an *electron* can be written in the following form:

$$m_{ge(ima)} = \left\{1-2\left[\sqrt{1+\left(\frac{U}{m_i c^2} n_r\right)^2}-1\right]\right\} m_{ie(ima)} \qquad (70)$$

Thus, the gravitational *imaginary* mass *associated to matter* can be *reduced*, made *negative* and *increased,* just as the gravitational *real* mass.

It was shown that also *photons* have imaginary mass. Therefore, the imaginary mass can be associated or *not* to the matter.

In a general way, the gravitational forces between two gravitational imaginary masses are then given by

$$\vec{F} = -\vec{F} = -G\frac{(iM_g)(im_g)}{r^2}\hat{\mu} = +G\frac{M_g m_g}{r^2}\hat{\mu} \qquad (71)$$

Note that these forces are *real* and *repulsive*.

Now consider a gravitational imaginary mass, $m_{g(ima)} = im_g$, *not associated with matter* (like the gravitational imaginary mass associated to the photons) and another gravitational imaginary mass $M_{g(ima)} = iM_g$ *associated to* a *material* body.

*Any material body has an imaginary mass associated to it, d*ue to the existence of imaginary masses associated to the electrons. We will choose a *quartz crystal* (for the material body with gravitational imaginary mass $M_{g(ima)} = iM_g$) because quartz crystals are widely used to detect forces (piezoelectric effect).

By using GCCs as shown in Fig. 18(b) and Fig.18(c), we can increase the gravitational acceleration, $\vec{a}$, produced by the imaginary mass $im_g$ upon the crystals. Then it becomes

$$a = -\chi^3_{air} G\frac{m_g}{r^2} \qquad (72)$$

As we have seen, the value of $\chi_{air}$ can be increased up to $\chi_{air} \cong -10^9$ (See Eq.57). Note that in this case, the gravitational forces become *attractive*. In addition, if $m_g$ is not small, the gravitational forces between the imaginary body of mass $im_g$ and the crystals can become sufficiently intense to be easily detectable.

Due to the piezoelectric effect, the gravitational force acting on the crystal will produce a voltage proportional to its intensity. Then consider a board with hundreds micro-crystals behind a set of GCCs, as shown in Fig.18(c). By amplifying the voltages generated in each micro-crystal and sending to an appropriated data acquisition system, it will be thus possible to obtain an image of the imaginary body of mass $m_{g(ima)}$ placed in front of the board.

In order to decrease strongly the gravitational effects produced by bodies placed behind the imaginary body of mass $im_g$, one can put five GCCs making a *Gravitational Shielding* as shown in Fig.18(c). If the GCCs are filled with air at 300K and $3\times10^{-12} torr$. Then $\rho_{air} = 4.94\times10^{-15} kg.m^{-3}$ and $\sigma_{air} \cong 1\times10^{-14} S.m^{-1}$. Thus, for $f = 60Hz$ and $B \cong 0.7T$ the Eq. (56) gives

$$\chi_{air} = \frac{m_{g(air)}}{m_{i(air)}} = \left\{1 - 2\left[\sqrt{1+5B^4} - 1\right]\right\} \cong -10^{-2} \qquad (73)$$

For $\chi_{air} \cong 10^{-2}$ the gravitational shielding presented in Fig.18(c) will reduce any value of $g$ to $\chi^5_{air} g \cong 10^{-10} g$. This will be sufficiently to reduce strongly the gravitational effects proceeding from both sides of the gravitational shielding.

Another important consequence of the correlation between gravitational mass and inertial mass expressed by Eq. (1) is the possibility of building *Energy Shieldings* around objects in order to protect them from *high-energy particles* and *ultra-intense fluxes of radiation*.

In order to explain that possibility, we start from the new expression [1] for the *momentum* $q$ of a particle with gravitational mass $M_g$ and velocity $V$, which is given by

$$q = M_g V \qquad (74)$$

where $M_g = m_g \big/ \sqrt{1 - V^2/c^2}$ and $m_g = \chi\, m_i$ [1]. Thus, we can write

$$\frac{m_g}{\sqrt{1 - V^2/c^2}} = \frac{\chi\, m_i}{\sqrt{1 - V^2/c^2}} \qquad (75)$$

Therefore, we get

$$M_g = \chi M_i \qquad (76)$$

It is known from the Relativistic Mechanics that

$$q = \frac{UV}{c^2} \qquad (77)$$

where $U$ is the *total* energy of the particle. This expression is valid for *any* velocity $V$ of the particle, including $V = c$.

By comparing Eq. (77) with Eq. (74) we obtain

$$U = M_g c^2 \quad (78)$$

It is a well-known experimental fact that

$$M_i c^2 = hf \quad (79)$$

Therefore, by substituting Eq. (79) and Eq. (76) into Eq. (74), gives

$$q = \frac{V}{c}\chi\frac{h}{\lambda} \quad (80)$$

Note that this expression is valid for *any* velocity $V$ of the particle. In the particular case of $V = c$, it reduces to

$$q = \chi\frac{h}{\lambda} \quad (81)$$

By comparing Eq. (80) with Eq. (77), we obtain

$$U = \chi hf \quad (82)$$

Note that only for $\chi = 1$ the Eq. (81) and Eq. (82) are reduced to the well=known expressions of DeBroglie $(q = h/\lambda)$ and Einstein $(U = hf)$.

Equations (80) and (82) show for example, that *any* real particle (material particles, real photons, etc) that penetrates a region (with density $\rho$ and electrical conductivity $\sigma$), where there is an ELF electric field $E$, will have its *momentum* $q$ and its energy $U$ reduced by the factor $\chi$, given by

$$\chi = \frac{m_g}{m_i} = \left\{1 - 2\left[\sqrt{1 + \frac{\mu}{4c^2}\left(\frac{\sigma}{4\pi f}\right)^3\frac{E^4}{\rho^2}} - 1\right]\right\} \quad (83)$$

The remaining amount of *momentum* and *energy*, respectively given by $(1-\chi)\left(\frac{V}{c}\right)\frac{h}{\lambda}$ and $(1-\chi)\,hf$, are *transferred to* the *imaginary* particle associated to the *real* particle§ (material particles or real photons) that penetrated the mentioned region.

It was previously shown that, when the *gravitational mass* of a particle is reduced to ranging between $+0.159M_i$ to $-0.159M_i$, i.e., when $\chi < 0.159$, it becomes *imaginary* [1], i.e., the gravitational and the inertial masses of the particle become *imaginary*. Consequently, the particle disappears from our ordinary space-time. It goes to the Imaginary Universe. On the other hand, when the gravitational mass of the particle becomes greater than $+0.159M_i$, or less than $-0.159M_i$, i.e., when $\chi > 0.159$, the particle return to our Universe.

Figure 19 (a) clarifies the phenomenon of reduction of the *momentum* for $\chi > 0.159$, and Figure 19 (b) shows the effect in the case of $\chi < 0.159$. In this case, the particles become imaginary and consequently, they go to the *imaginary space-time* when they penetrate the electric field $E$. However, the electric field $E$ stays at the *real* space-time. Consequently, the particles return immediately to the real space-time in order to return soon after to the *imaginary* space-time, due to the action of the electric field $E$. Since the particles are moving at a direction, they *appear* and *disappear* while they are crossing the region, up to collide with the plate (See Fig.19) with a *momentum,* $q_m = \chi\left(\frac{V}{c}\right)\frac{h}{\lambda}$, in the case of the *material particle*, and $q_r = \chi\frac{h}{\lambda}$ in the case of the *photon*. Note that by making $\chi \cong 0$, it is possible to block high-energy particles and ultra-intense fluxes of radiation. These *Energy Shieldings* can be built around objects in order to protect them from such particles and radiation.

It is also important to note that the gravity control process described here points to the possibility of obtaining *Controlled Nuclear Fusion* by means of increasing of the intensity of the gravitational interaction between the nuclei. When the gravitational forces $F_G = Gm_g m'_g / r^2$ become greater than the electrical forces $F_E = qq'/4\pi\varepsilon_0 r^2$ between the nuclei, then nuclear fusion reactions can occur.

Note that, according to Eq. (83), the gravitational mass can be strongly increased. Thus, if $E = E_m \sin\omega t$, then the average value for $E^2$ is equal to $\frac{1}{2}E_m^2$, because $E$ varies sinusoidaly ($E_m$ is the maximum value for $E$). On the other hand, $E_{rms} = E_m/\sqrt{2}$. Consequently, we can replace

§ As previously shown, there are *imaginary particles* associated to each *real particle* [1].

$E^4$ for $E^4_{rms}$. In addition, as $j = \sigma E$ (*Ohm's vectorial Law*), then Eq. (83) can be rewritten as follows

$$\chi = \frac{m_g}{m_{i0}} = \left\{1 - 2\left[\sqrt{1 + K\frac{\mu_r j^4_{rms}}{\sigma\rho^2 f^3}} - 1\right]\right\} \qquad (84)$$

where $K = 1.758\times10^{-27}$ and $j_{rms} = j/\sqrt{2}$.

Thus, the gravitational force equation can be expressed by

$$F_G = Gm_g m'_g/r^2 = \chi^2 Gm_{i0}m'_{i0}/r^2 =$$
$$= \left\{1 - 2\left[\sqrt{1 + K\frac{\mu_r j^4_{rms}}{\sigma\rho^2 f^3}} - 1\right]\right\}^2 Gm_{i0}m'_{i0}/r^2 \qquad (85)$$

In order to obtain $F_G > F_E$ we must have

$$\left\{1 - 2\left[\sqrt{1 + K\frac{\mu_r j^4_{rms}}{\sigma\rho^2 f^3}} - 1\right]\right\} > \sqrt{\frac{qq'/4\pi\varepsilon_0}{Gm_{i0}m'_{i0}}} \qquad (86)$$

The *carbon fusion* is a set of nuclear fusion reactions that take place in massive stars (at least $8M_{sun}$ at birth). It requires high temperatures ($>5\times10^8 K$) and densities ($>3\times10^9\,kg.m^{-3}$). The principal reactions are:

$$^{12}\text{C} + {}^{12}\text{C} \rightarrow \begin{cases} ^{23}\text{Na} + \text{p} + 2.24\ \text{MeV} \\ ^{20}\text{Ne} + \alpha + 4.62\ \text{MeV} \\ ^{24}\text{Mg} + \gamma + 13.93\ \text{MeV} \end{cases}$$

In the case of Carbon nuclei ($^{12}$C) of a *thin carbon wire* ($\sigma \cong 4\times10^4 S.m^{-1}$; $\rho = 2.2\times10^3 S.m^{-1}$) Eq. (86) becomes

$$\left\{1 - 2\left[\sqrt{1 + 9.08\times10^{-39}\frac{j^4_{rms}}{f^3}} - 1\right]\right\} > \sqrt{\frac{e^2}{16\pi\varepsilon_0 Gm_p^2}}$$

whence we conclude that the condition for the $^{12}$C + $^{12}$C fusion reactions occur is

$$j_{rms} > 1.7\times10^{18} f^{\frac{3}{4}} \qquad (87)$$

If the electric current through the carbon wire has Extremely-Low Frequency (ELF), for example, if $f = 1\mu Hz$, then the current density, $j_{rms}$, must have the following value:

$$j_{rms} > 5.4\times10^{13} A.m^{-2} \qquad (88)$$

Since $j_{rms} = i_{rms}/S$ where $S = \pi\phi^2/4$ is the area of the cross section of the wire, we can conclude that, for an *ultra-thin carbon* wire with $10\mu m$-diameter, it is necessary that the current through the wire, $i_{rms}$, have the following intensity

$$i_{rms} > 4.24\ kA$$

Obviously, this current will *explode* the carbon wire. However, this explosion becomes negligible in comparison with the very strong *gravitational implosion*, which occurs simultaneously due to the enormous increase in intensities of the gravitational forces among the carbon nuclei produced by means of the ELF current through the carbon wire as predicted by Eq. (85). Since, in this case, the gravitational forces among the carbon nuclei become greater than the repulsive electric forces among them the result is the production of $^{12}$C + $^{12}$C fusion reactions.

Similar reactions can occur by using a *lithium* wire. In addition, it is important to note that $j_{rms}$ is directly proportional to $f^{\frac{3}{4}}$ (Eq. 87). Thus, for example, if $f = 10^{-8} Hz$, the current necessary to produce the nuclear reactions will be $i_{rms} = 130A$.

## IV.CONCLUSION

The process described here is clearly the better way in order to control the gravity. This is because the *Gravity Control Cell* in this case is very easy to be built, the cost is low and it works at ambient temperature. The Gravity Control is the starting point for the generation of and detection of *Virtual Gravitational Radiation* (Quantum Gravitational *Transceiver*) also for the construction of the *Gravitational Motor* and the *Gravitational Spacecraft* which includes the system for generation of *artificial gravity* presented in Fig.10 and the *Gravitational Thruster* (Fig.11). While the *Gravitational Transceiver* leads to a new concept in *Telecommunication*, the Gravitational Motor changes the paradigm of *energy conversion* and the Gravitational Spacecraft points to a new concept in *aerospace flight*.

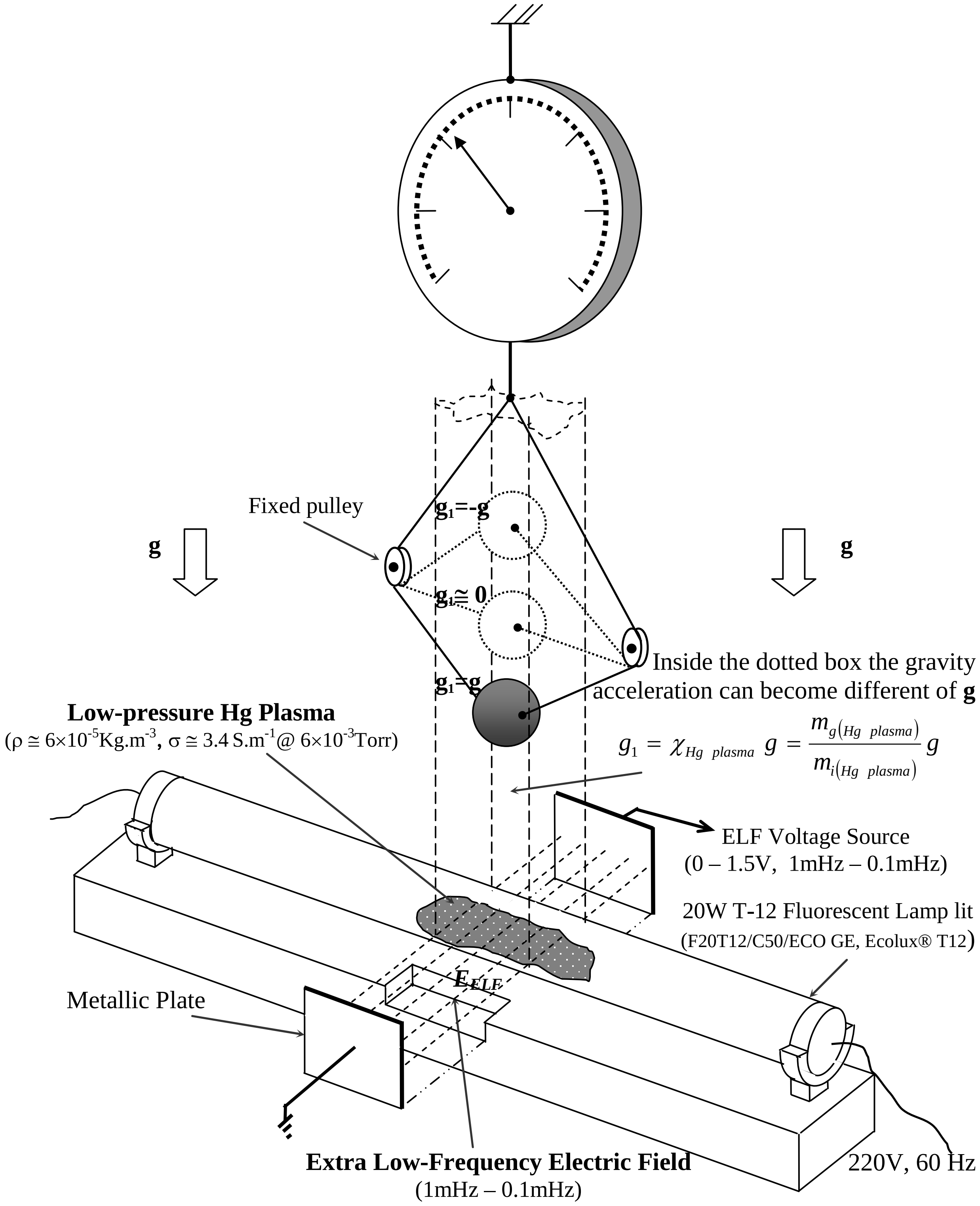

Fig. 1 – Gravitational Shielding Effect by means of an ELF electric field through low- pressure Hg Plasma.

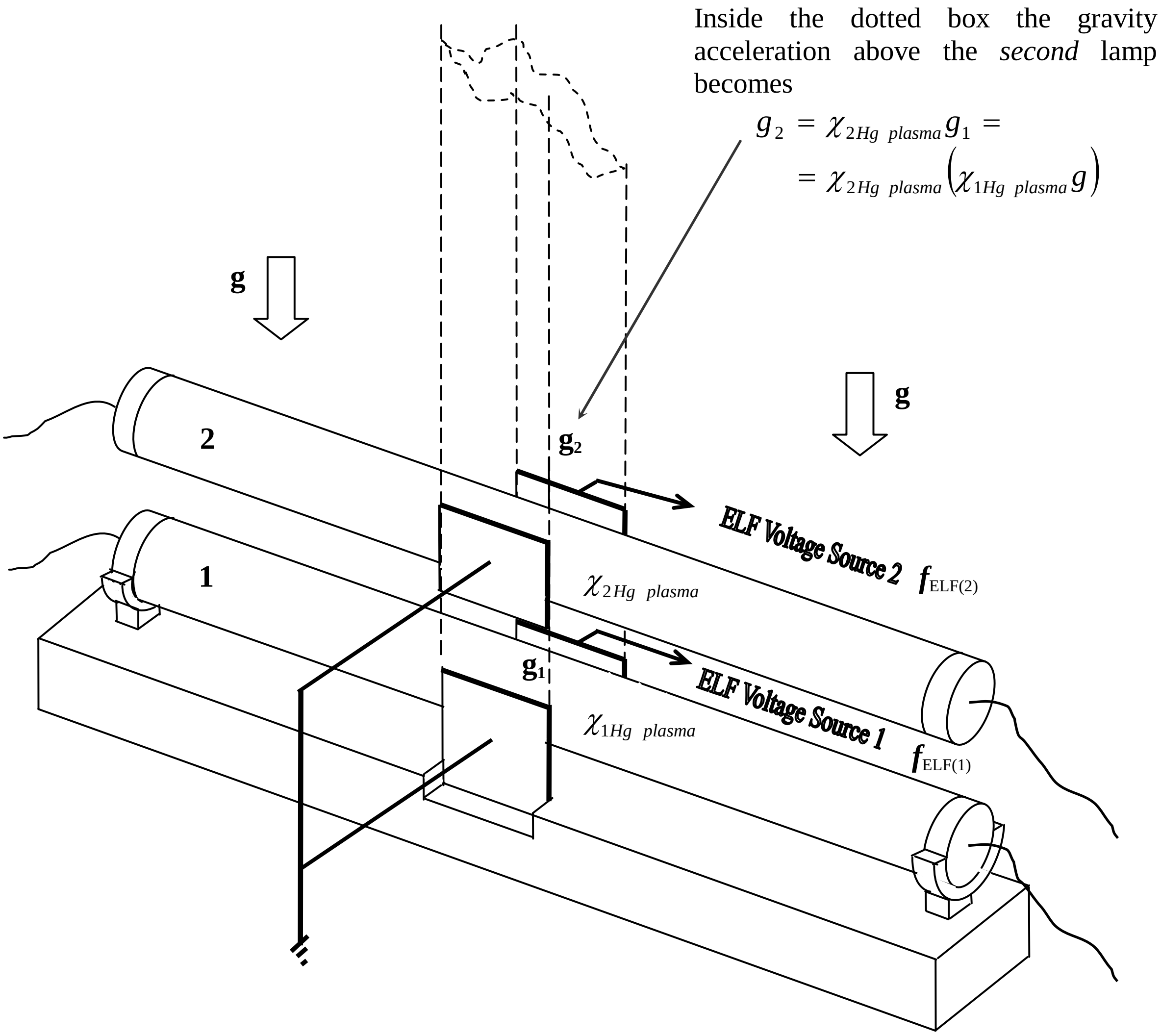

Fig. 2 – Gravity acceleration above a *second* fluorescent lamp.

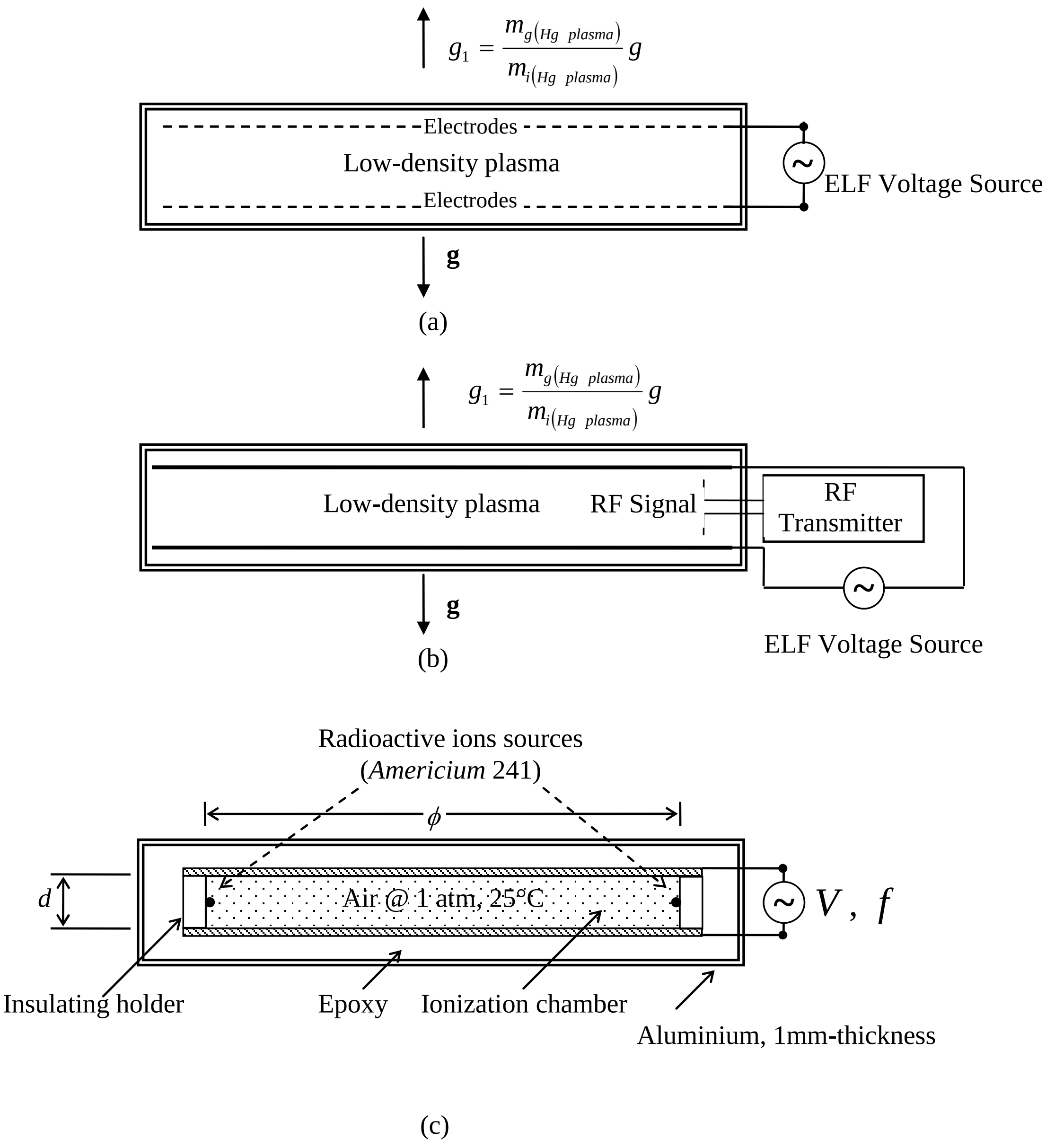

Fig. 3 – *Schematic diagram of Gravity Control Cells (GCCs).*
(a) GCC where the ELF electric field and the ionizing electric field can be the same. (b) GCC where the plasma is ionized by means of a RF signal. (c) GCC filled with *air* (at ambient temperature and 1 atm) strongly ionized by means of alpha particles emitted from radioactive ions sources (Am 241, *half-life* 432 years). Since the electrical conductivity of the ionized air depends on the amount of ions then it can be strongly increased by increasing the amount of Am 241 in the GCC. This GCC has 36 radioactive ions sources each one with 1/5000$^{th}$ of gram of Am 241, conveniently positioned around the ionization chamber, in order to obtain $\sigma_{air} \cong 10^3 S.m^{-1}$.

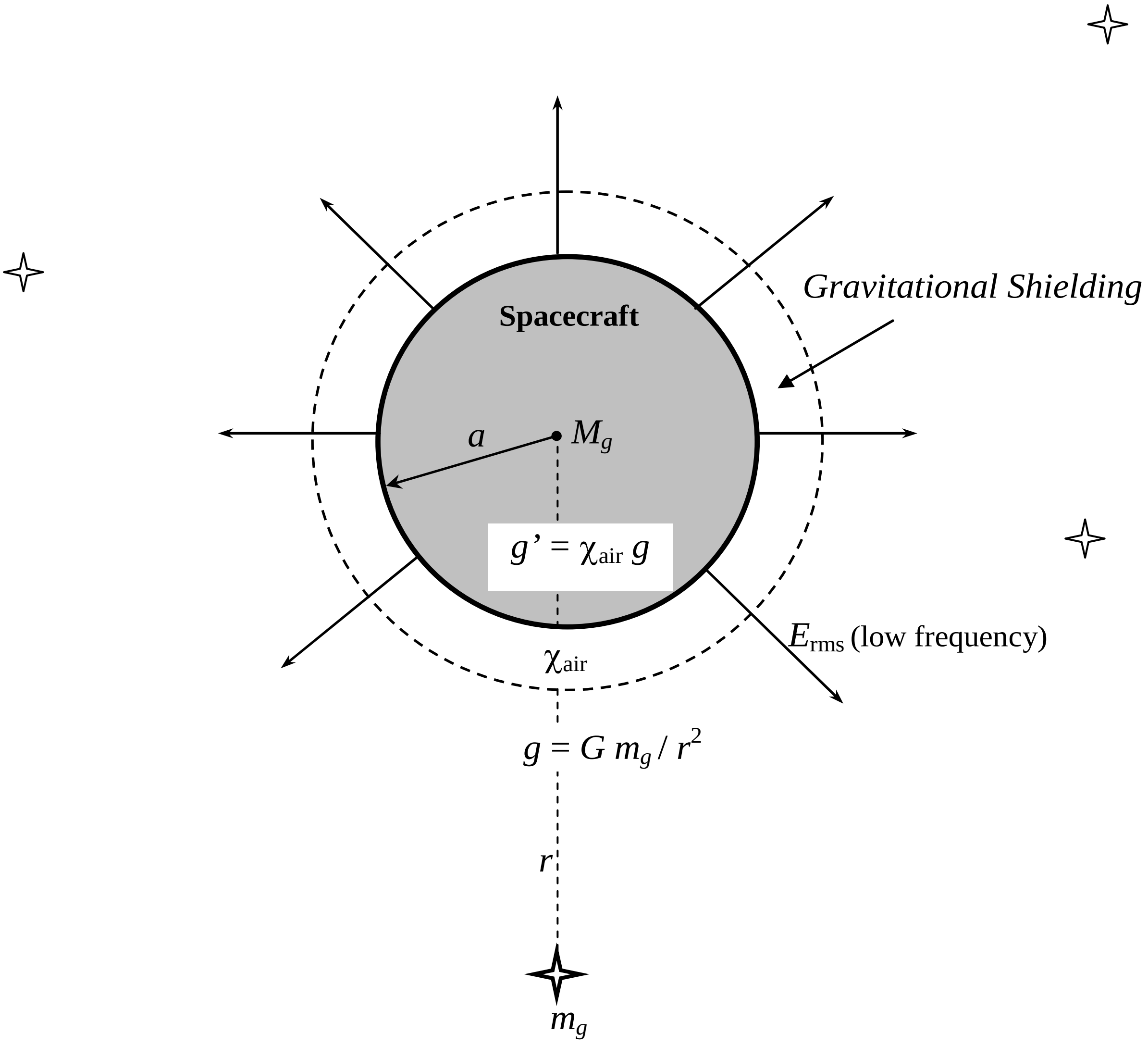

The *gravity accelerations* on the spacecraft (due to the rest of the Universe) can be controlled by means of the *gravitational shielding,* i.e.,

$$g'_i = \chi_{air}\, g_i \qquad i = 1, 2, 3 \ldots n$$

Thus,

$$F_{is} = F_{si} = M_g\, g'_i = M_g\, (\chi_{air}\, g_i)$$

Then the inertial forces acting on the spacecraft (s) can be strongly reduced. According to the *Mach's principle* this effect can reduce *the inertial properties of the spacecraft* and consequently, leads to a new concept of spacecraft and aerospace flight.

Fig. 4 – Gravitational Shielding surround a Spherical Spacecraft.

| V = $V_0$ (Volts) | t = T /4 (s) | ( min) | $\boldsymbol{E}_{ELF\,(1)}$ (V/m) | $\boldsymbol{f}_{ELF\,(1)}$ (mHz) | $\boldsymbol{g_1 / g}$ Exp. | Teo. | $\boldsymbol{E}_{ELF\,(2)}$ (V/m) | $\boldsymbol{f}_{ELF\,(2)}$ (mHz) | $\boldsymbol{g_2 / g}$ Exp. | Teo. |
|---|---|---|---|---|---|---|---|---|---|---|
| **1.0** V | 250 | **4.17** | 24.81 | 1 | - | **0.993** | 24.81 | 1 | - | **0.986** |
| | 312.5 | **5.21** | 24.81 | 0.8 | - | **0.986** | 24.81 | 0.8 | - | **0.972** |
| | 416.6 | **6.94** | 24.81 | 0.6 | - | **0.967** | 24.81 | 0.6 | - | **0.935** |
| | 625 | **10.42** | 24.81 | 0.4 | - | **0.890** | 24.81 | 0.4 | - | **0.792** |
| | 1250 | **20.83** | 24.81 | 0.2 | - | **0.240** | 24.81 | 0.2 | - | **0.058** |
| **1.5**V | 250 | **4.17** | 37.22 | 1 | - | **0.964** | 37.22 | 1 | - | **0.929** |
| | 312.5 | **5.21** | 37.22 | 0.8 | - | **0.930** | 37.22 | 0.8 | - | **0.865** |
| | 416.6 | **6.94** | 37.22 | 0.6 | - | **0.837** | 37.22 | 0.6 | - | **0.700** |
| | 625 | **10.42** | 37.22 | 0.4 | - | **0,492** | 37.22 | 0.4 | - | **0.242** |
| | 1250 | **20.83** | 37.22 | 0.2 | - | **-1,724** | 37.22 | 0.2 | - | **2.972** |

Table 1 – Theoretical Results.

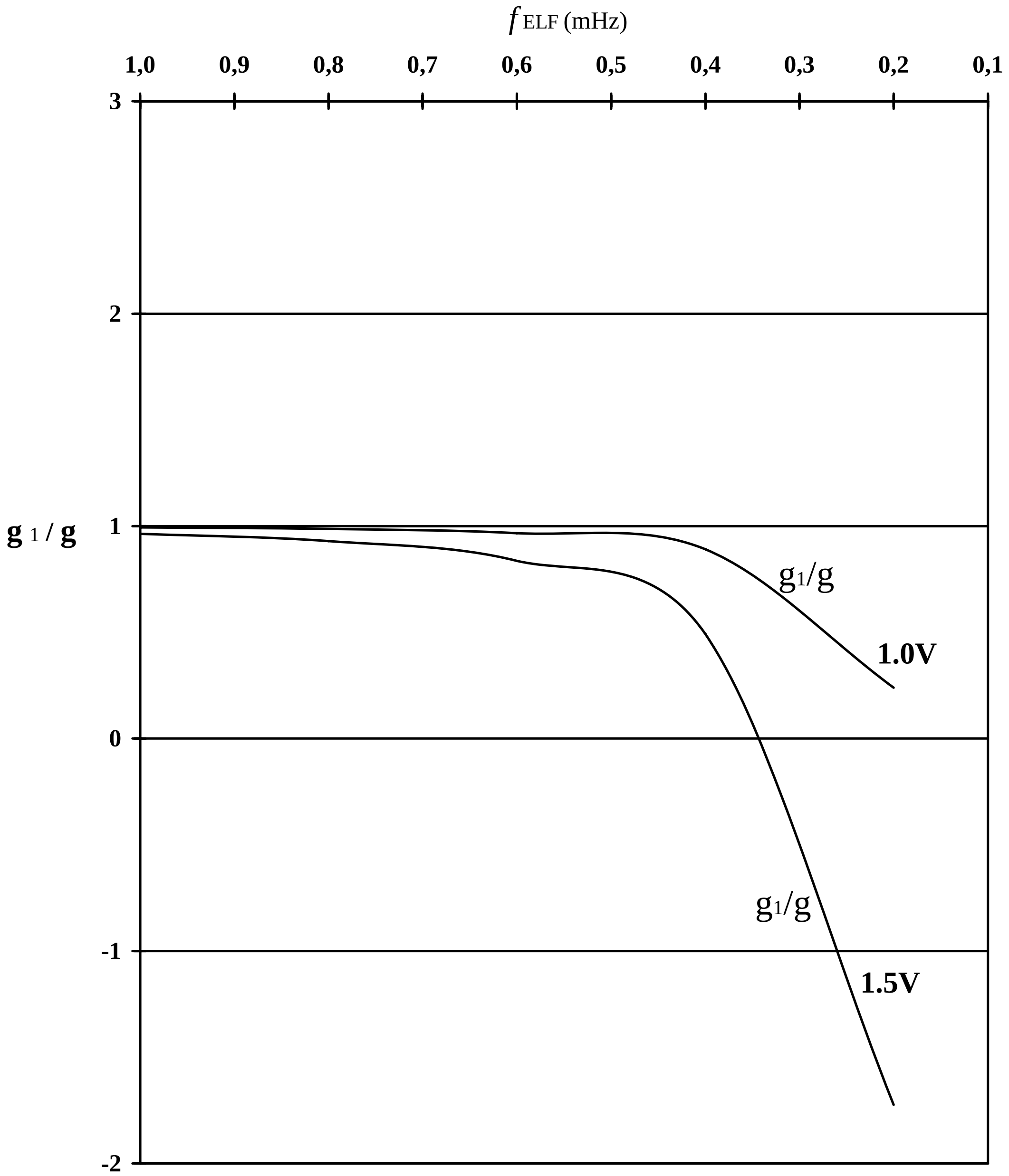

Fig. 5- Distribution of the correlation **$g_1/g$** as a function of $f_{ELF}$

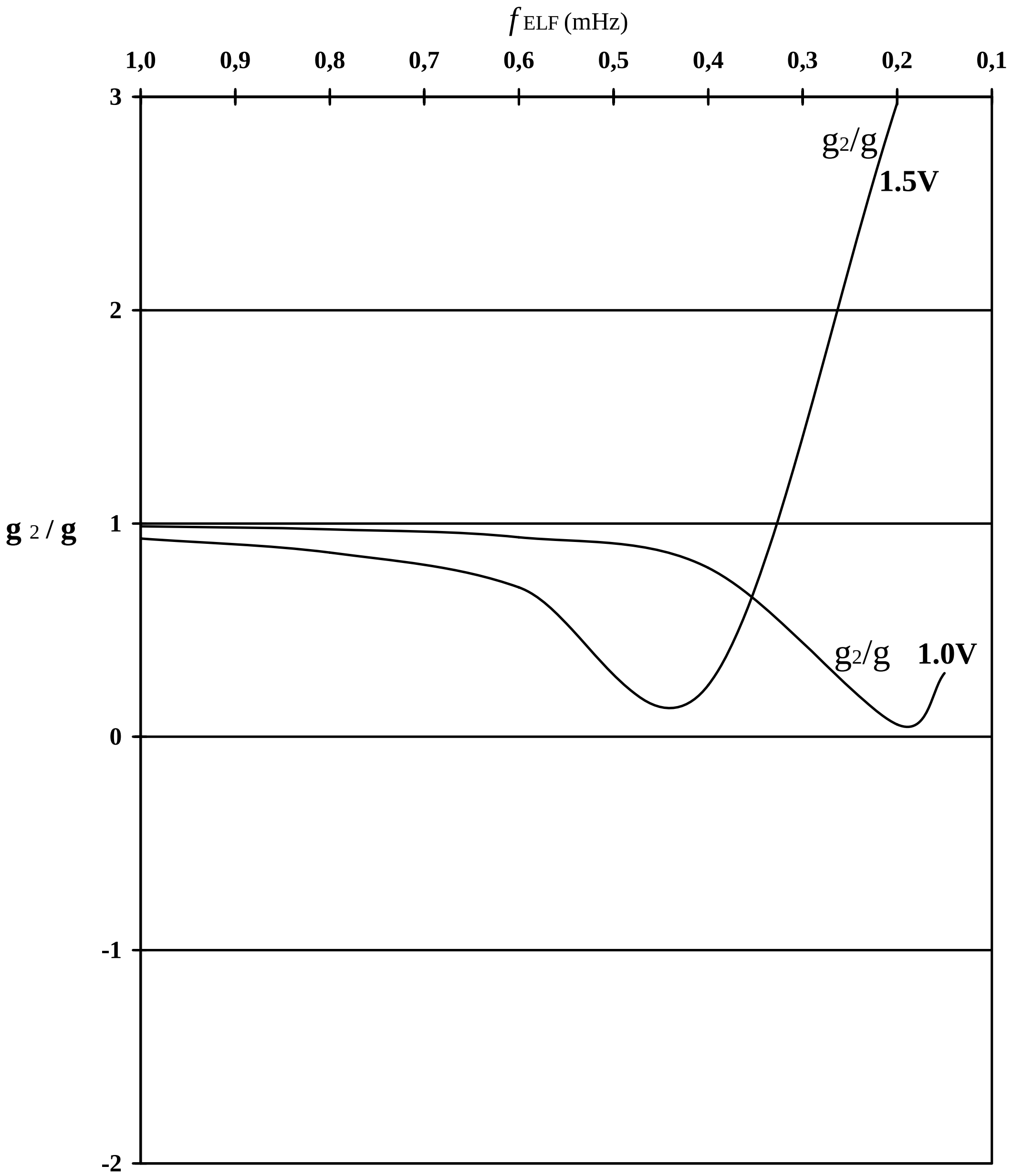

Fig. 6- Distribution of the correlation $\mathbf{g}_2 / \mathbf{g}$ as a function of $f_{ELF}$

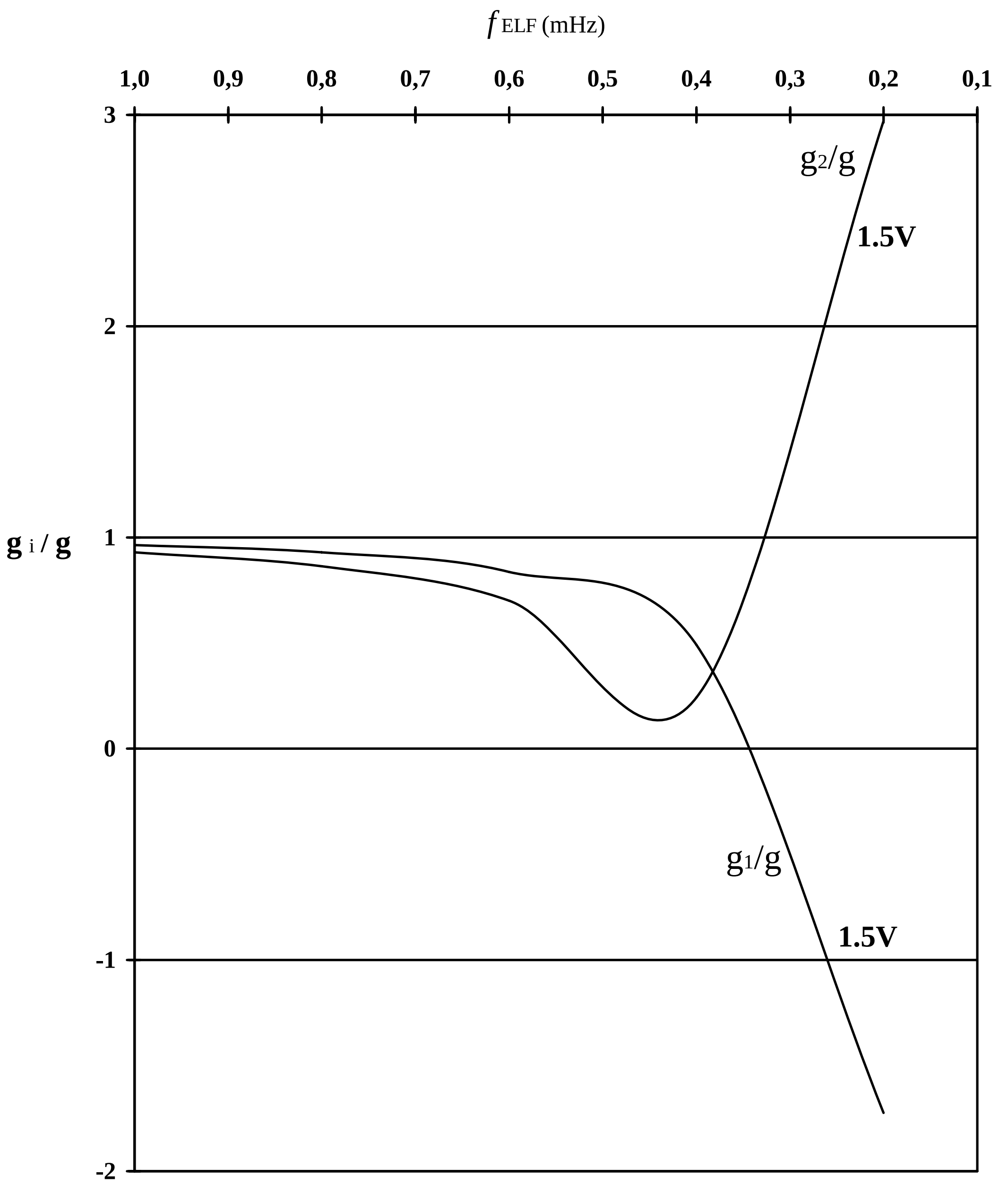

Fig. 7- Distribution of the correlations $\mathbf{g}_i / \mathbf{g}$ as a function of $f_{ELF}$

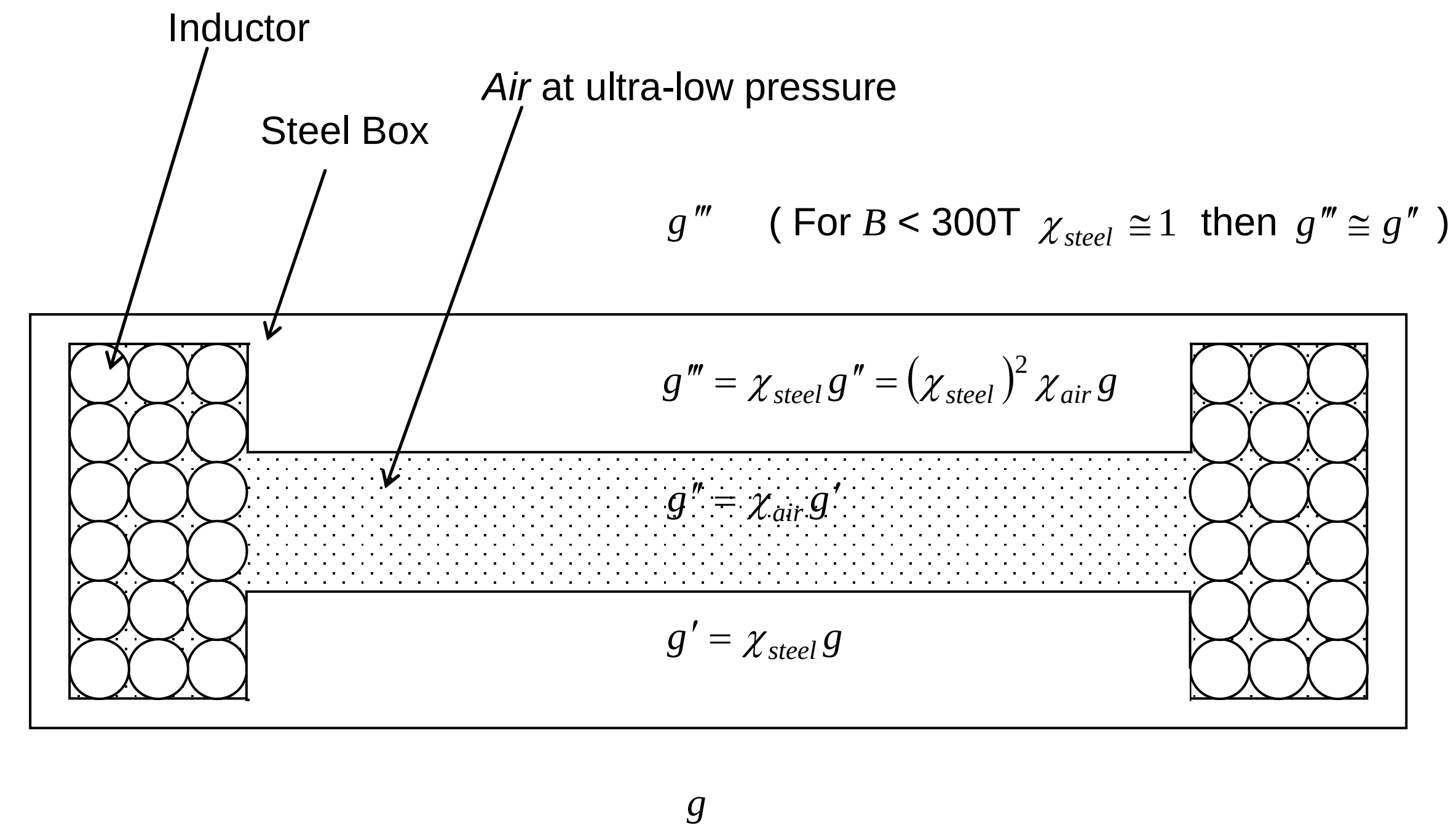

(a)

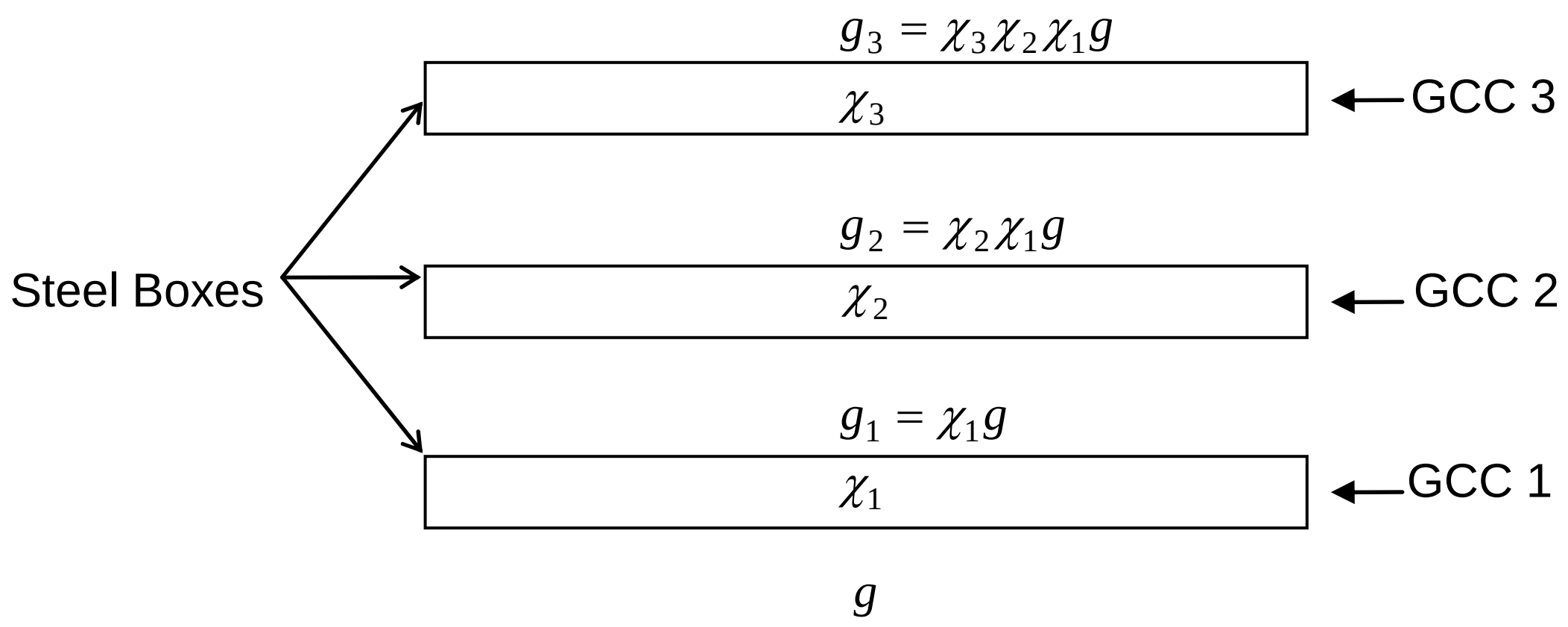

(b)

Fig. 8 – (a) Gravity Control Cell (GCC) filled with *air* at ultra-low pressure.
(b) Gravity Control Battery (Note that if $\chi_1 = \chi_2^{-1} = -1$ then $g'' = g$ )

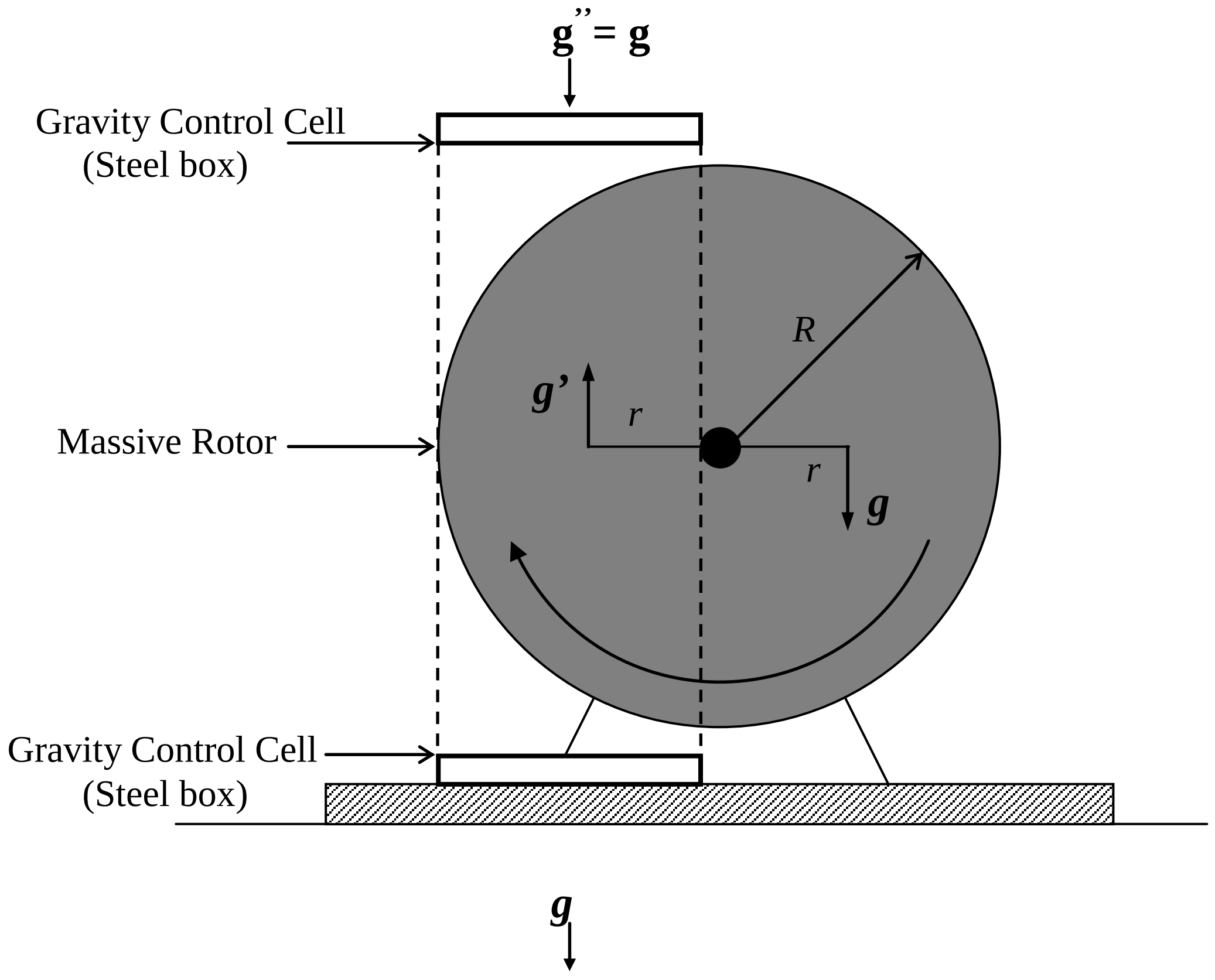

Note that $g' = (\chi_{steel})^2 \chi_{air} g$ and $g'' = (\chi_{steel})^4 (\chi_{air})^2 g$ therefore for $\chi_{steel} \cong 1$ and $\chi_{air(1)} = \chi^{-1}_{air(2)} = -n$ we get $g' \cong -ng$ and $g'' = g$

Fig. 9 – The Gravitational Motor

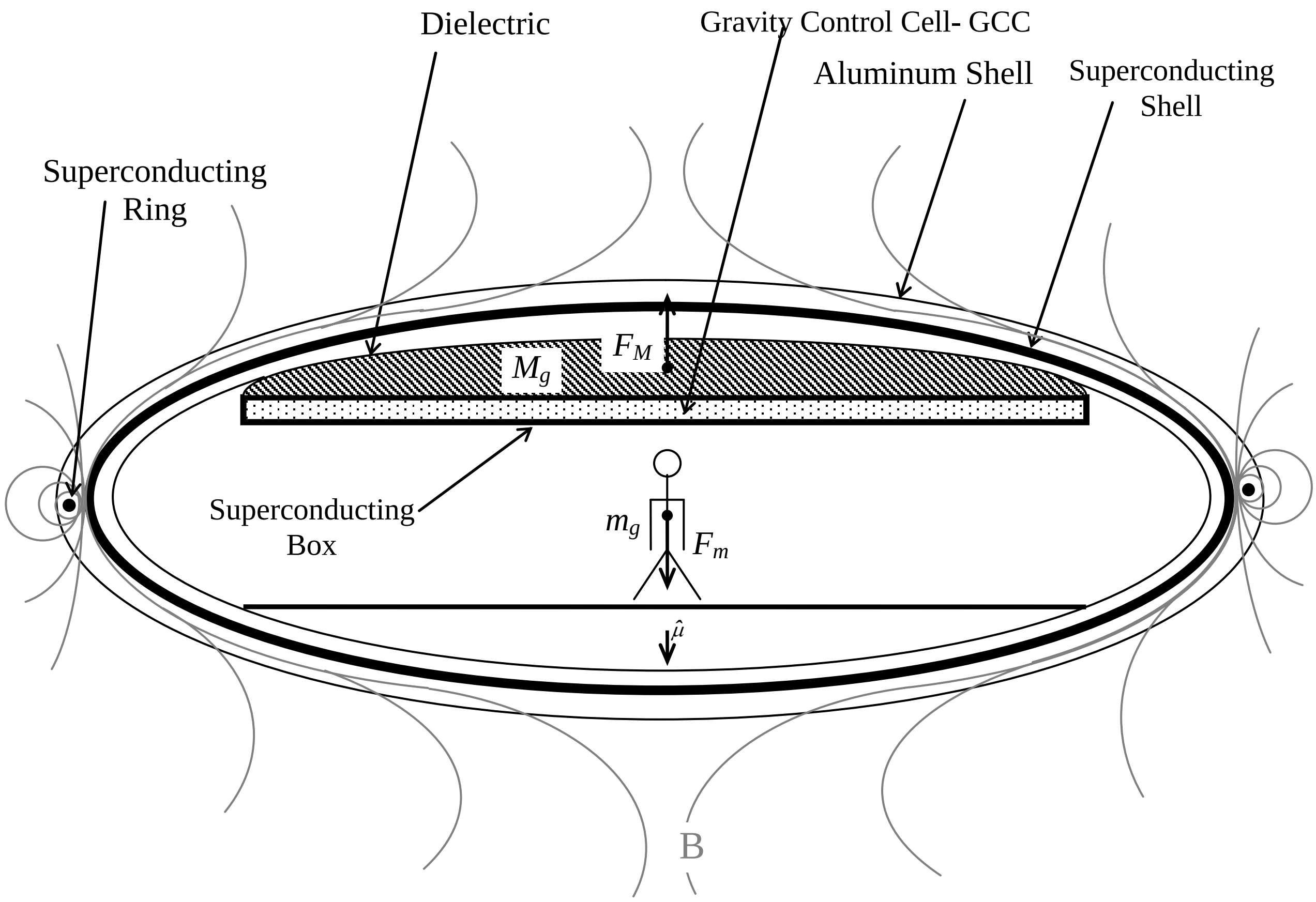

Fig. 10 – The Gravitational Spacecraft – Due to the *Meissner effect*, the magnetic field *B* is expelled from the *superconducting shell*. Similarly, the magnetic field $B_{GCC}$, of the GCC stay confined inside the *superconducting box*.

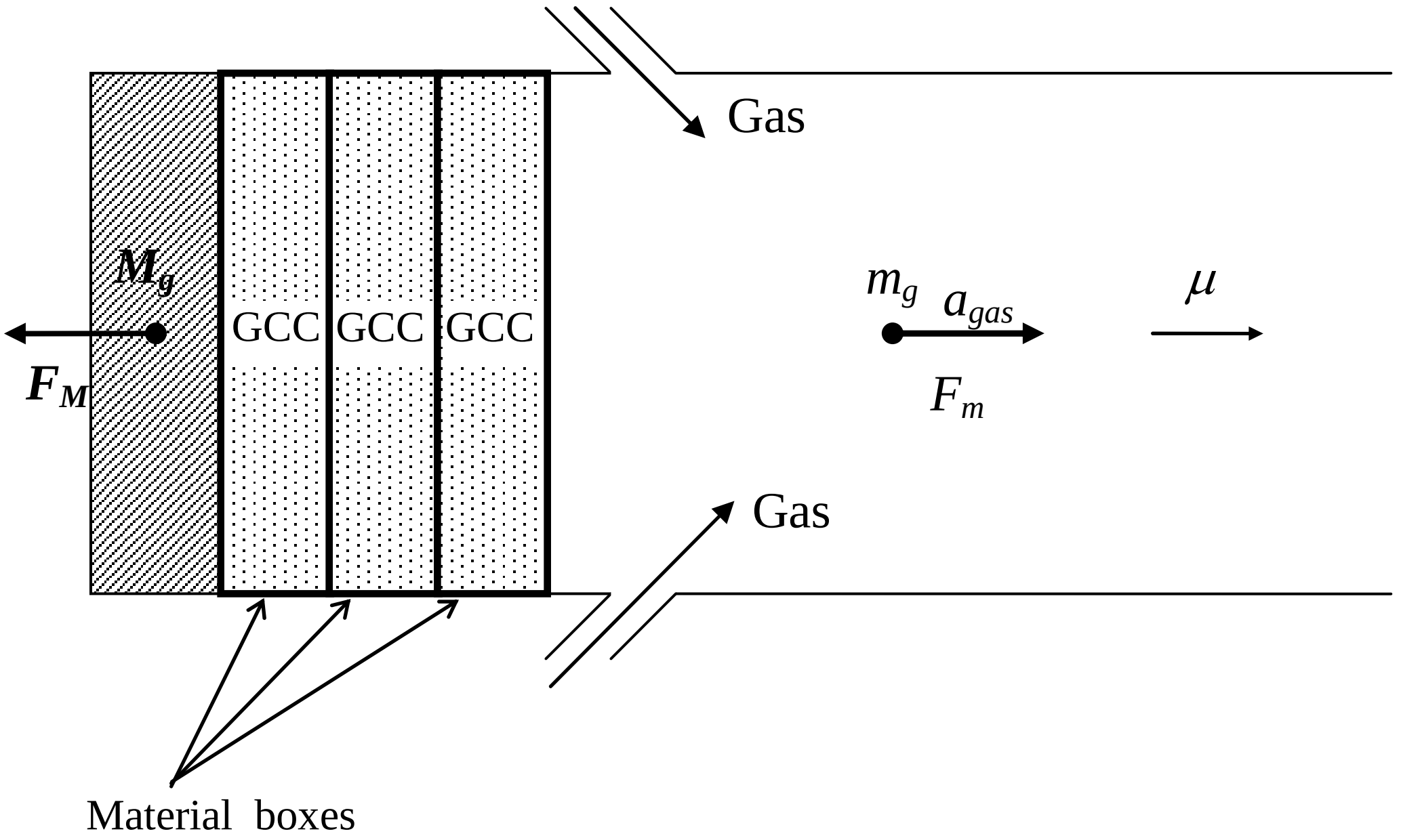

(a)

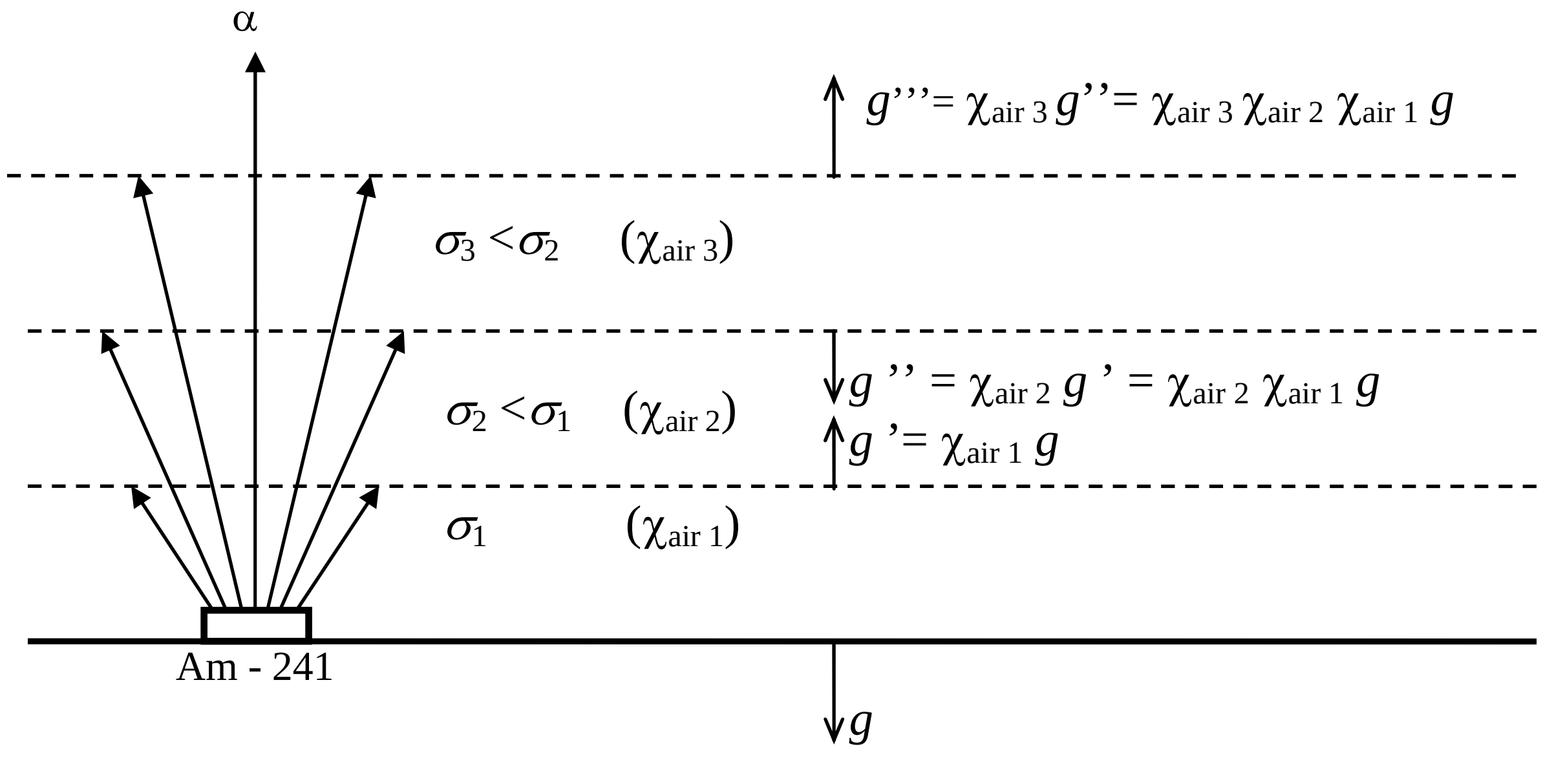

(b)

Fig. 11 – The Gravitational Thruster .
(a) Using material boxes. (b) Without material boxes

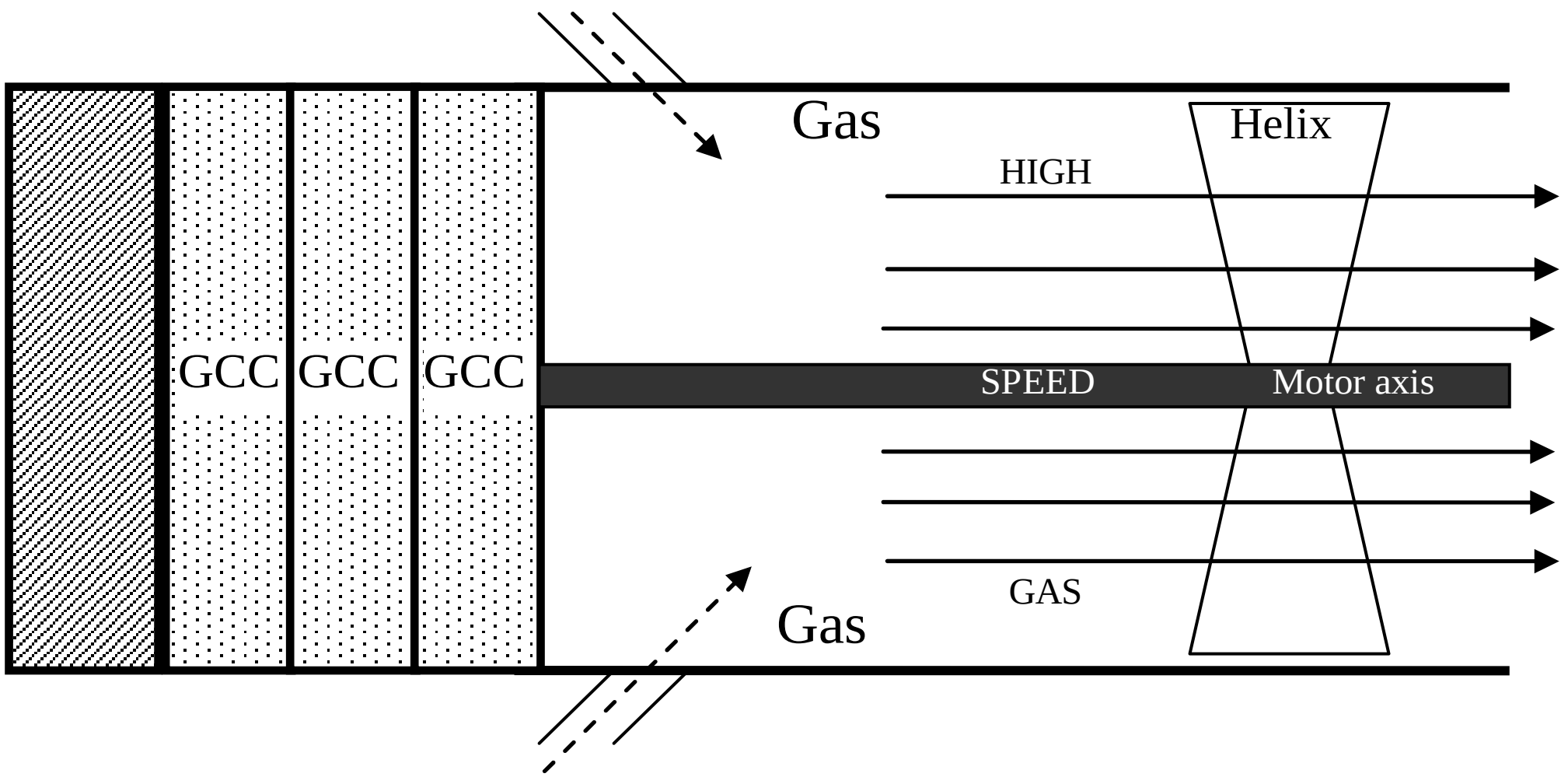

Fig. 12 - The Gravitational Turbo Motor – The gravitationally accelerated gas, by means of the GCCs, propels the helix which movies the motor axis.

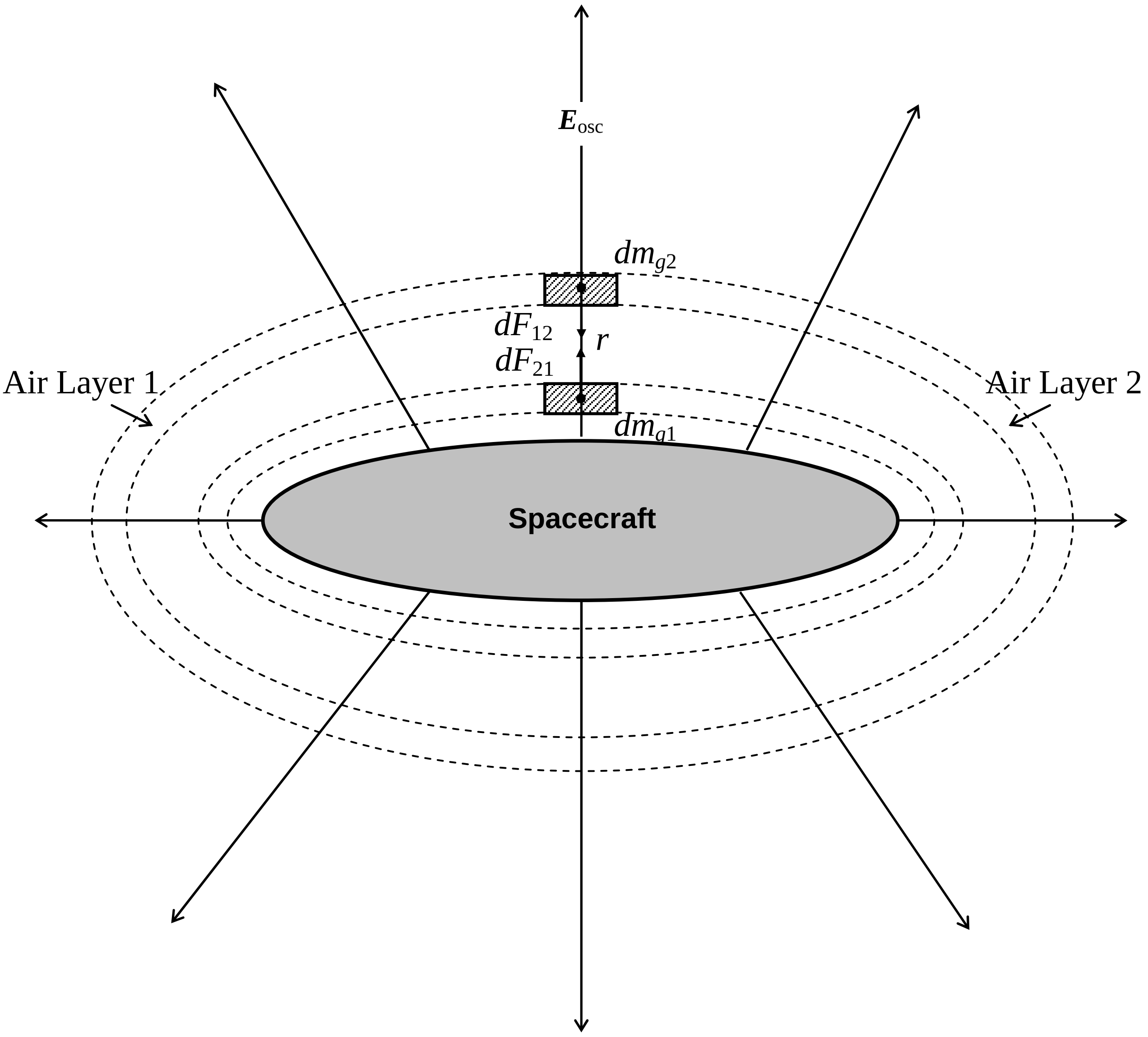

Fig. 13 – Gravitational forces between two layers of the "*air shell*". The electric field $\boldsymbol{E}_{osc}$ provides the *ionization* of the *air*.

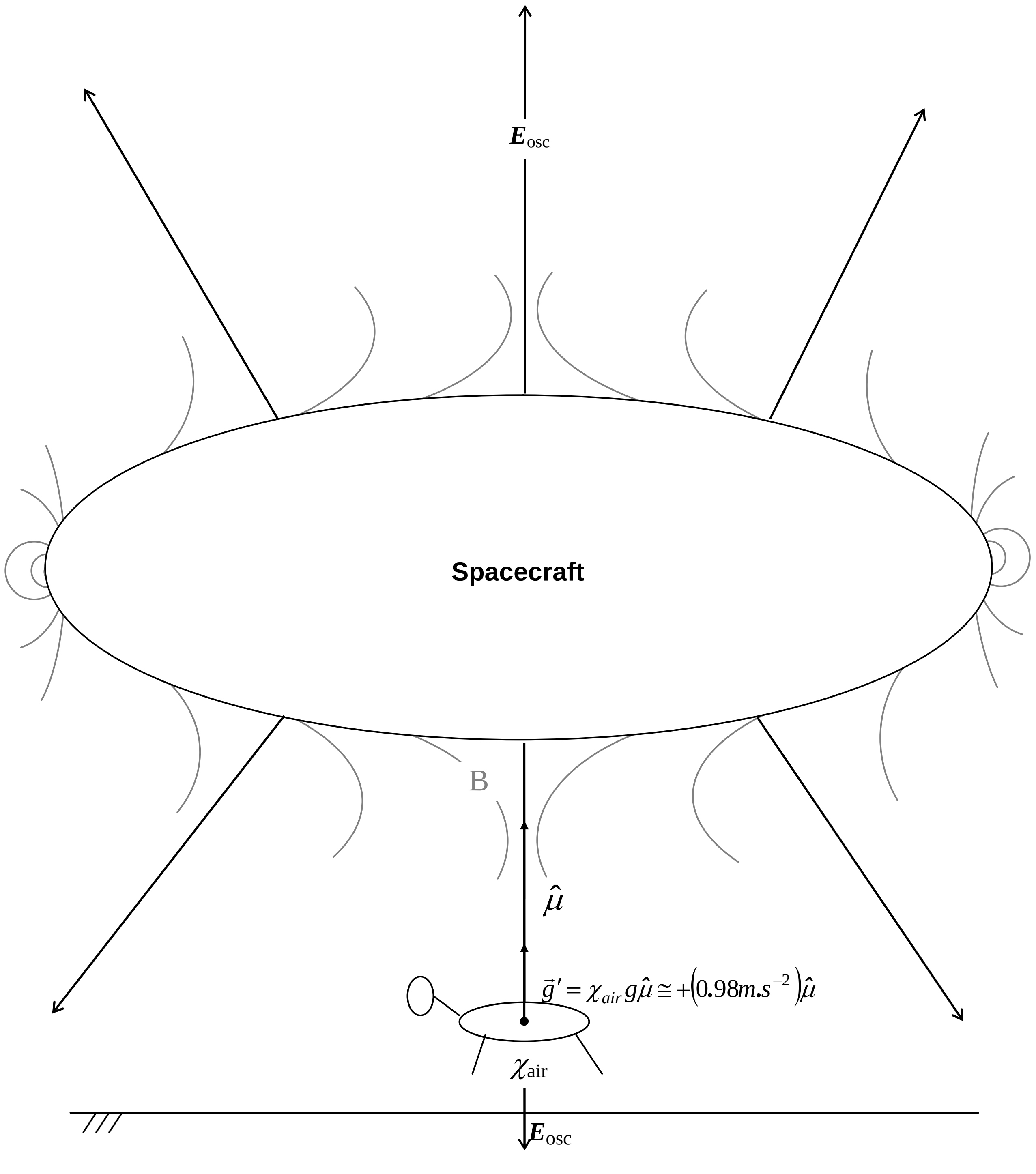

Fig. 14 – The Gravitational Lifter

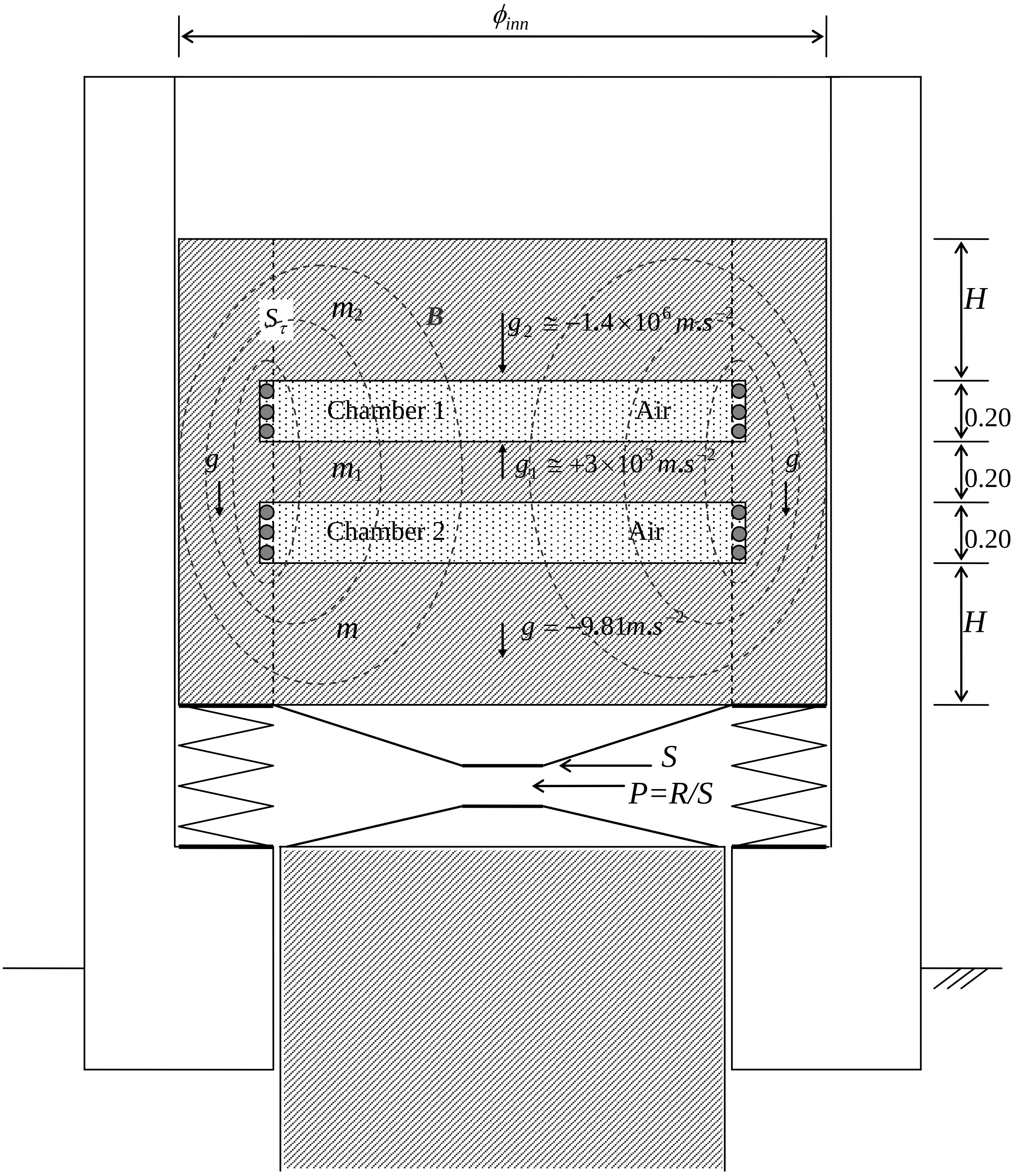

Fig. 15 – *Gravitational Press*

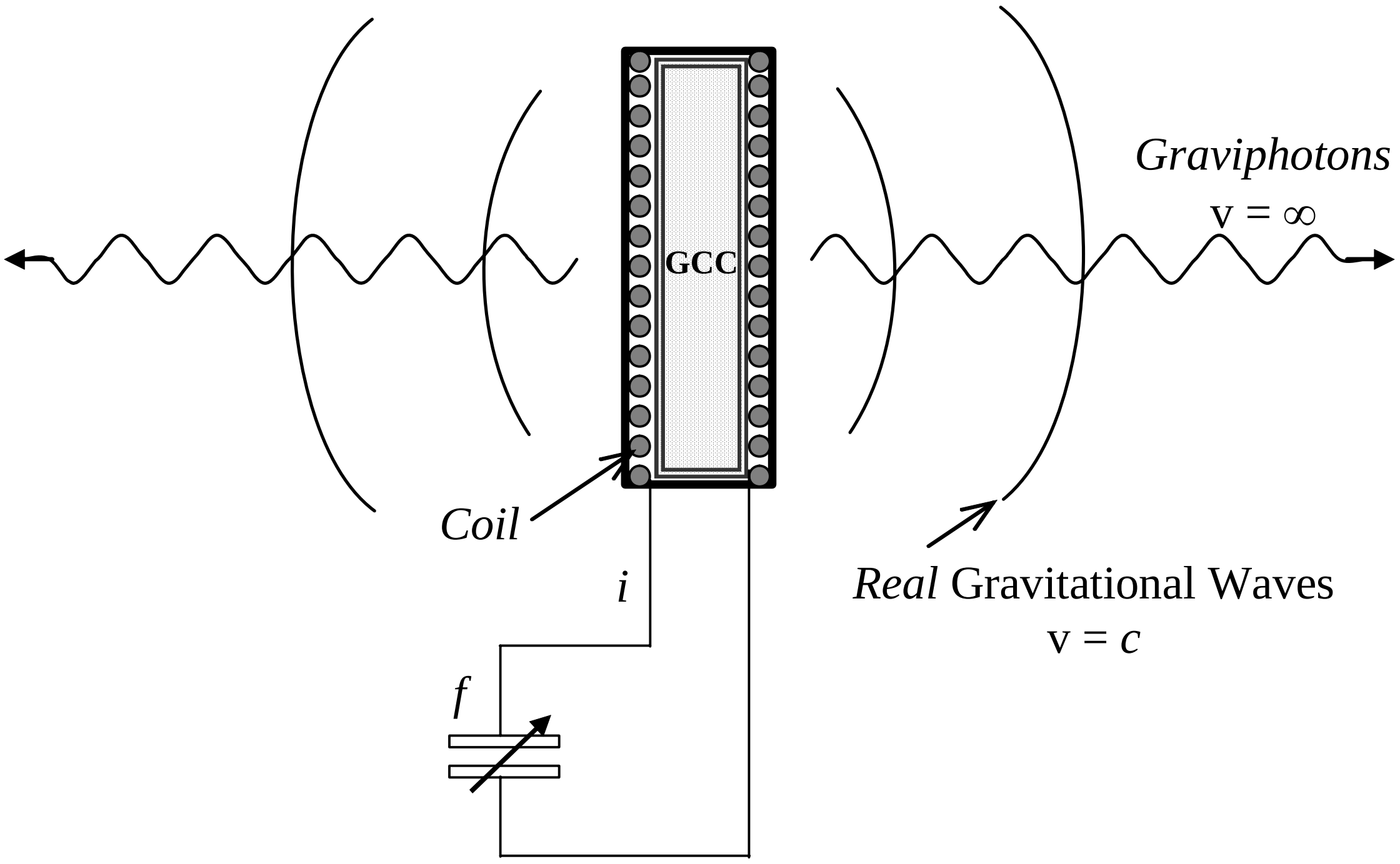

(a) GCC Antenna

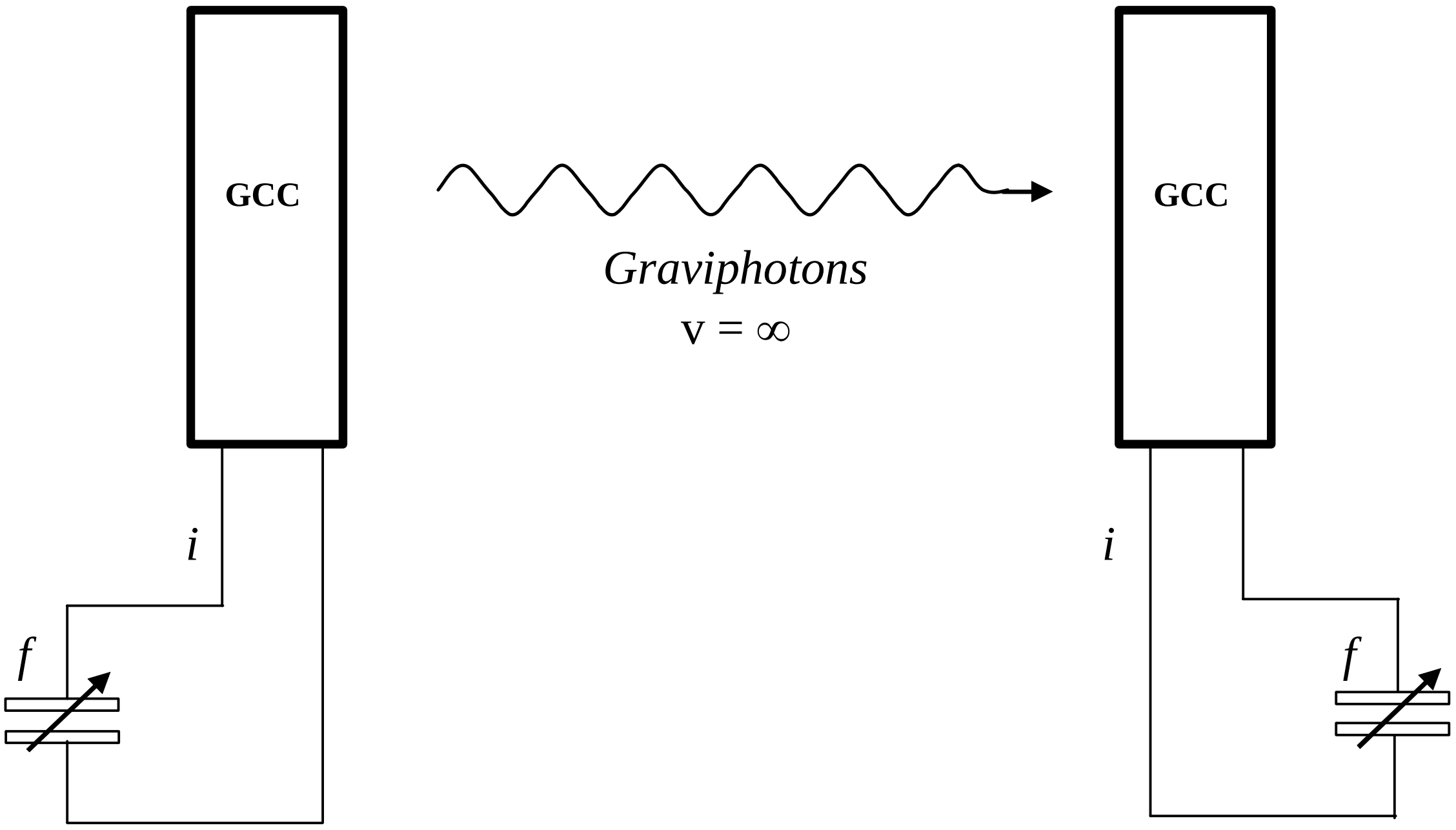

Transmitter (b) Receiver

Fig. 16 - Transmitter and Receiver of *Virtual* Gravitational Radiation.

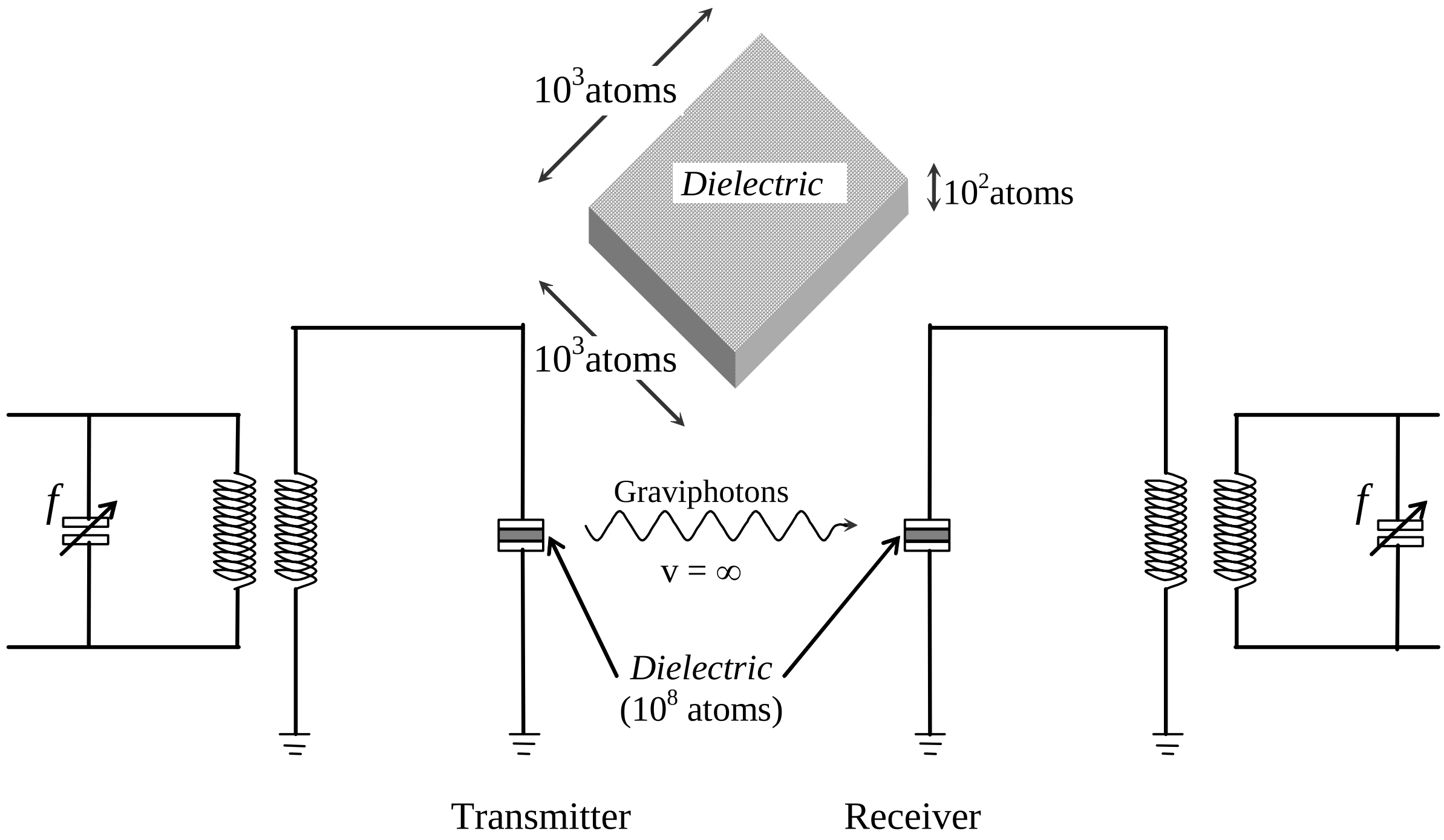

(a)

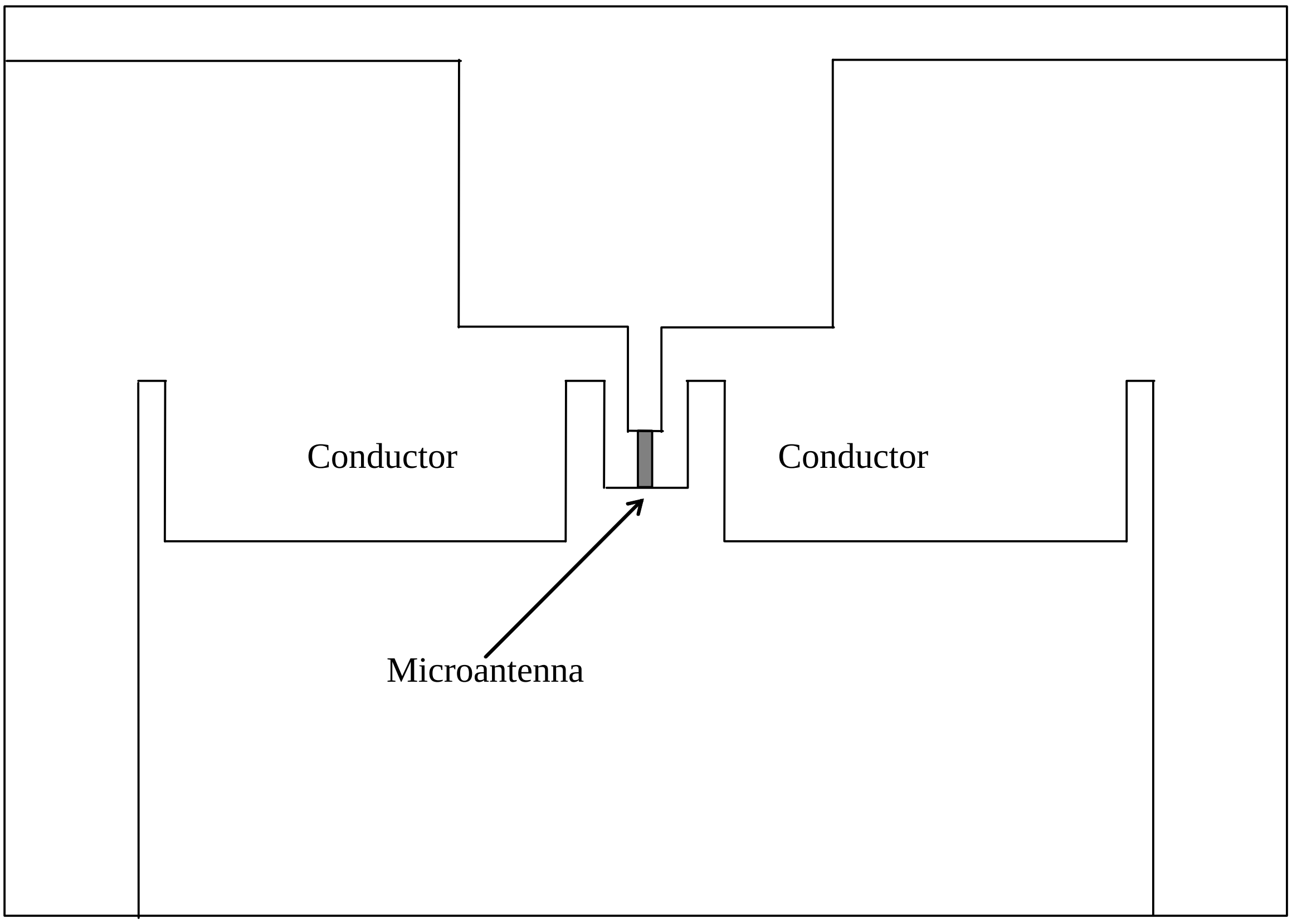

(b)

Fig. 17 – Quantum Gravitational Microantenna

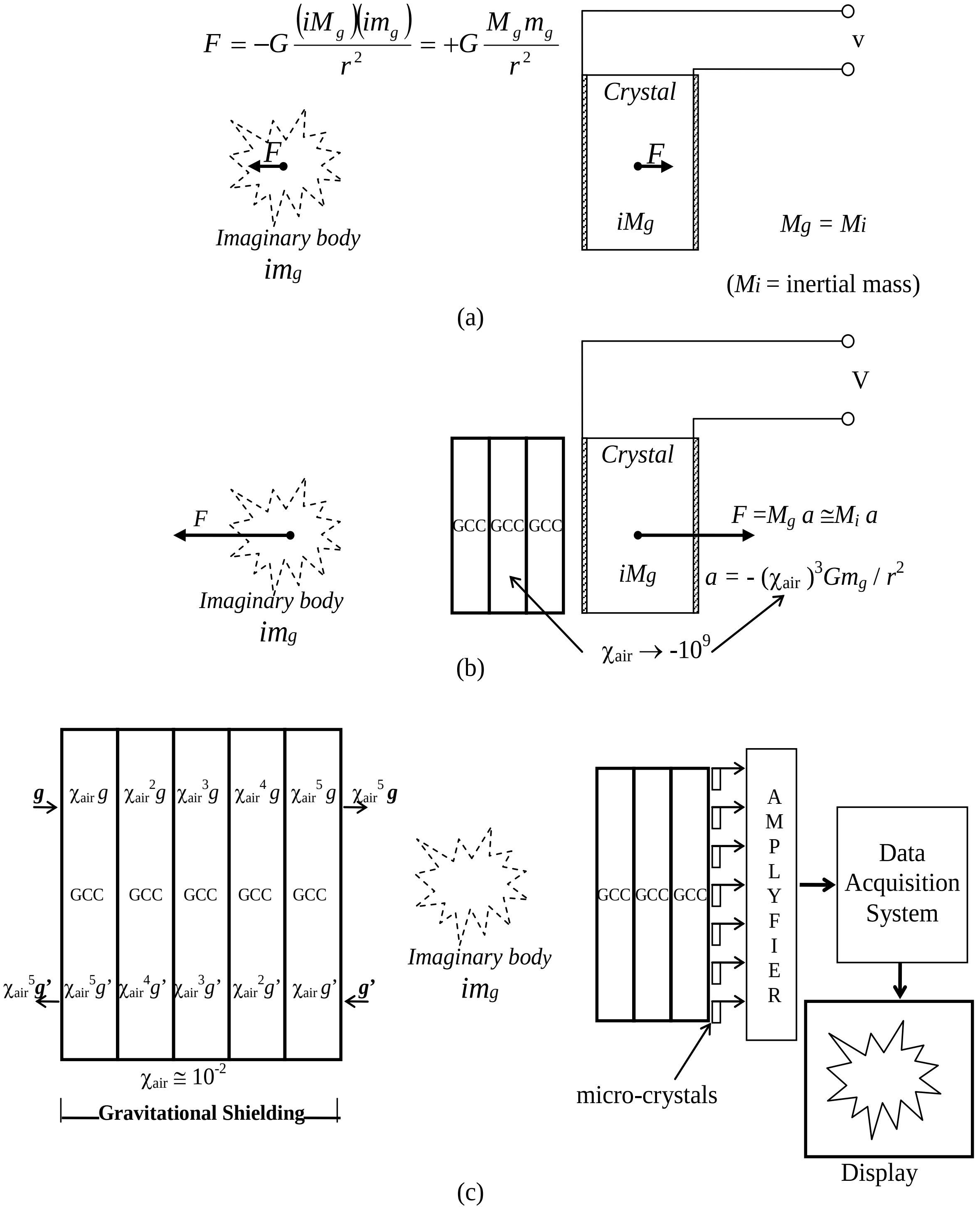

Fig.18 – Method and device using GCCs for obtaining *images* of *imaginary bodies*.

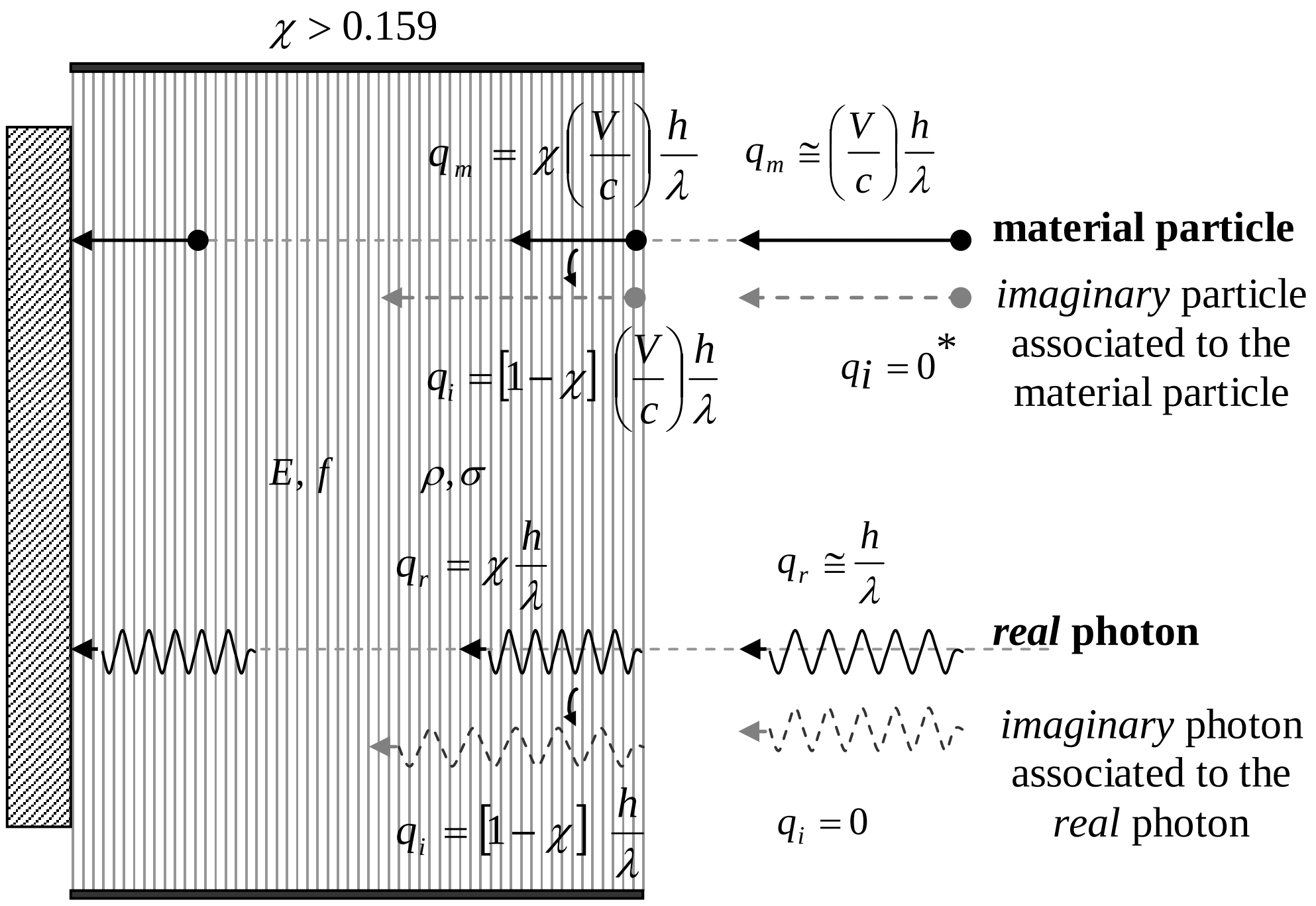

(a)

* There are a type of neutrino, called "ghost" neutrino, predicted by General Relativity, with *zero mass* and *zero momentum*. In spite its *momentum be zero*, it is known that there are wave functions that describe these neutrinos and that prove that really they exist.

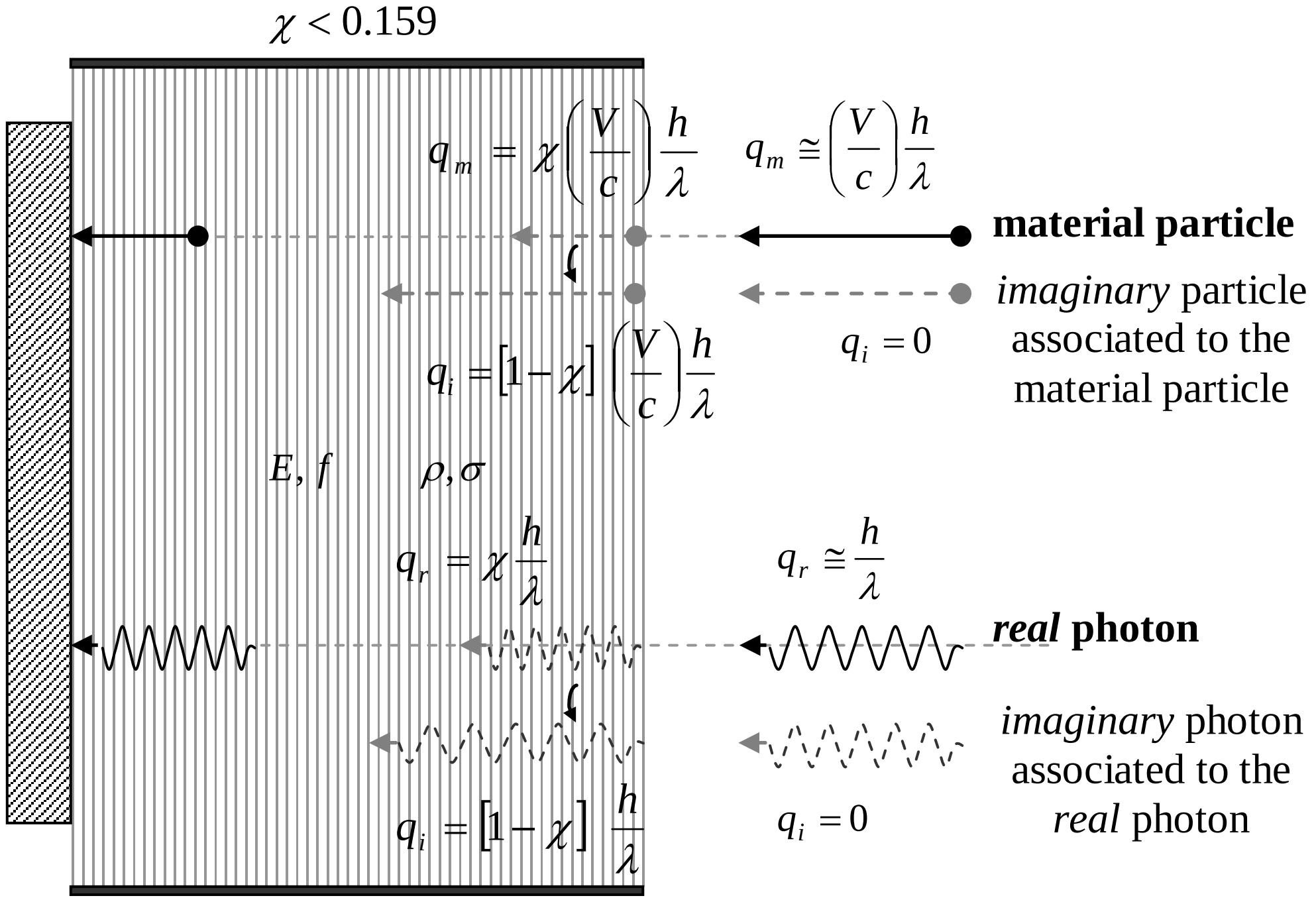

(b)

Fig. 19 – *The phenomenon of reduction of the momentum*. (a) Shows the reduction of *momentum* for $\chi > 0.159$. (b) Shows the effect when $\chi < 0.159$. Note that in both cases, the *material* particles collide with the cowl with the *momentum* $q_m = \chi(V/c)(h/\lambda)$, and the photons with $q_r = \chi\frac{h}{\lambda}$. Therefore, that by making $\chi \cong 0$, it is possible to block high-energy particles and ultra-intense fluxes of radiation.

**APPENDIX A: THE SIMPLEST METHOD TO CONTROL THE GRAVITY**

In this Appendix we show the simplest method to control the gravity.

Consider a body with mass density $\rho$ and the following electric characteristics: $\mu_r, \varepsilon_r, \sigma$ (relative permeability, relative permittivity and electric conductivity, respectively). Through this body, passes an electric current $I$, which is the sum of a sinusoidal current $i_{osc} = i_0 \sin \omega t$ and the DC current $I_{DC}$, i.e., $I = I_{DC} + i_0 \sin \omega t$ ; $\omega = 2\pi f$. If $i_0 << I_{DC}$ then $I \cong I_{DC}$. Thus, the current $I$ varies with the frequency $f$, but the variation of its intensity is quite small in comparison with $I_{DC}$, i.e., $I$ will be practically constant (Fig. 1A). This is of fundamental importance for maintaining the value of the gravitational mass of the body, $m_g$, sufficiently stable during all the time.

The *gravitational mass* of the *body* is given by [1]

$$m_g = \left\{1 - 2\left[\sqrt{1 + \left(\frac{n_r U}{m_{i0} c^2}\right)^2} - 1\right]\right\} m_{i0} \qquad (A1)$$

where $U$, is the electromagnetic energy absorbed by the body and $n_r$ is the index of refraction of the body.

Equation (A1) can also be rewritten in the following form

$$\frac{m_g}{m_{i0}} = \left\{1 - 2\left[\sqrt{1 + \left(\frac{n_r W}{\rho c^2}\right)^2} - 1\right]\right\} \qquad (A2)$$

where, $W = U/V$ is the *density of electromagnetic energy* and $\rho = m_{i0}/V$ is the density of inertial mass.

The *instantaneous values* of the density of electromagnetic energy in an *electromagnetic* field can be deduced from Maxwell's equations and has the following expression

$$W = \tfrac{1}{2}\varepsilon E^2 + \tfrac{1}{2}\mu H^2 \qquad (A3)$$

where $E = E_m \sin \omega t$ and $H = H \sin \omega t$ are the *instantaneous values* of the electric field and the magnetic field respectively.

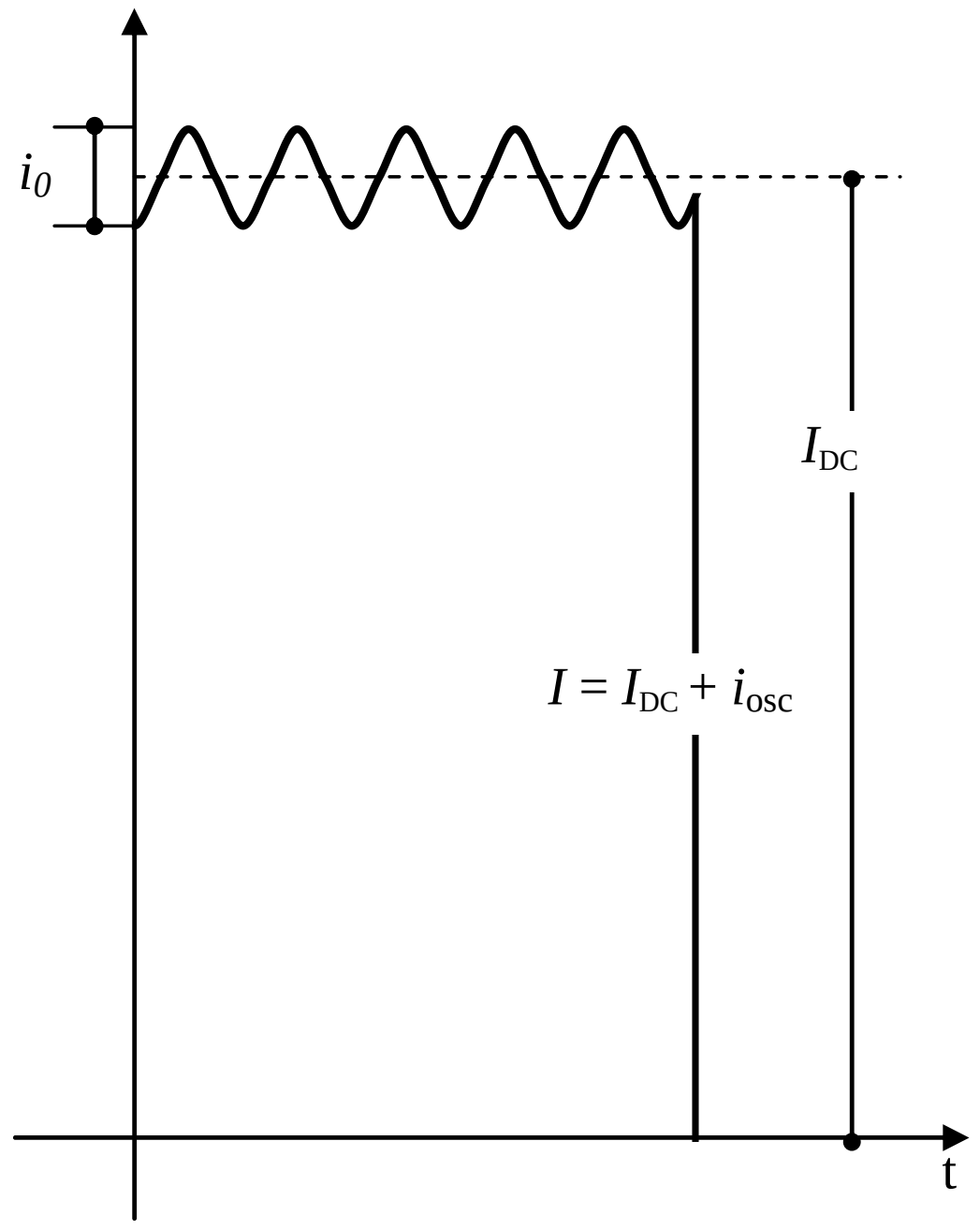

Fig. A1 - The electric current $I$ varies with frequency $f$. But the variation of $I$ is quite small in comparison with $I_{DC}$ due to $i_o << I_{DC}$. In this way, we can consider $I \cong I_{DC}$.

It is known that $B = \mu H$, $E/B = \omega/k_r$ [11] and

$$v = \frac{dz}{dt} = \frac{\omega}{\kappa_r} = \frac{c}{\sqrt{\frac{\varepsilon_r \mu_r}{2}\left(\sqrt{1 + (\sigma/\omega\varepsilon)^2} + 1\right)}} \qquad (A4)$$

where $k_r$ is the real part of the *propagation vector* $\vec{k}$ (also called *phase constant*); $k = \left|\vec{k}\right| = k_r + ik_i$ ; $\varepsilon$, $\mu$ and $\sigma$, are the electromagnetic characteristics of the medium in which the incident (or emitted) radiation is propagating($\varepsilon = \varepsilon_r \varepsilon_0$; $\varepsilon_0 = 8.854 \times 10^{-12} F/m$ ; $\mu = \mu_r \mu_0$ where $\mu_0 = 4\pi \times 10^{-7} H/m$). It is known that for *free-space* $\sigma = 0$ and $\varepsilon_r = \mu_r = 1$. Then Eq. (A4) gives

$$v = c$$

From (A4), we see that the *index of refraction* $n_r = c/v$ is given by

$$n_r = \frac{c}{v} = \sqrt{\frac{\varepsilon_r \mu_r}{2}\left(\sqrt{1 + (\sigma/\omega\varepsilon)^2} + 1\right)} \qquad (A5)$$

Equation (A4) shows that $\omega/\kappa_r = v$. Thus, $E/B = \omega/k_r = v$, i.e.,

$$E = vB = v\mu H \qquad (A6)$$

Then, Eq. (A3) can be rewritten in the following form:

$$W = \tfrac{1}{2}\left(\varepsilon v^2 \mu\right)\mu H^2 + \tfrac{1}{2}\mu H^2 \qquad (A7)$$

For $\sigma << \omega\varepsilon$, Eq. (A4) reduces to

$$v = \frac{c}{\sqrt{\varepsilon_r \mu_r}}$$

Then, Eq. (A7) gives

$$W = \tfrac{1}{2}\left[\varepsilon\left(\frac{c^2}{\varepsilon_r \mu_r}\right)\mu\right]\mu H^2 + \tfrac{1}{2}\mu H^2 = \mu H^2$$

This equation can be rewritten in the following forms:

$$W = \frac{B^2}{\mu} \qquad (A8)$$

or

$$W = \varepsilon\ E^2 \qquad (A9)$$

For $\sigma >> \omega\varepsilon$, Eq. (A4) gives

$$v = \sqrt{\frac{2\omega}{\mu\sigma}} \qquad (A10)$$

Then, from Eq. (A7) we get

$$W = \tfrac{1}{2}\left[\varepsilon\left(\frac{2\omega}{\mu\sigma}\right)\mu\right]\mu H^2 + \tfrac{1}{2}\mu H^2 = \left(\frac{\omega\varepsilon}{\sigma}\right)\mu H^2 + \tfrac{1}{2}\mu H^2 \cong$$
$$\cong \tfrac{1}{2}\mu H^2 \qquad (A11)$$

Since $E = vB = v\mu H$, we can rewrite (A11) in the following forms:

$$W \cong \frac{B^2}{2\mu} \qquad (A12)$$

or

$$W \cong \left(\frac{\sigma}{4\omega}\right)E^2 \qquad (A13)$$

By comparing equations (A8) (A9) (A12) and (A13), we can see that Eq. (A13) shows that the best way to obtain a strong value of $W$ *in practice* is by applying an *Extra Low-Frequency* (ELF) *electric field* $(w = 2\pi f << 1Hz)$ through a *medium with high electrical conductivity*.

Substitution of Eq. (A13) into Eq. (A2), gives

$$m_g = \left\{1 - 2\left[\sqrt{1 + \frac{\mu}{4c^2}\left(\frac{\sigma}{4\pi f}\right)^3 \frac{E^4}{\rho^2}} - 1\right]\right\} m_{i0} =$$
$$= \left\{1 - 2\left[\sqrt{1 + \left(\frac{\mu_0}{256\pi^3 c^2}\right)\left(\frac{\mu_r \sigma^3}{\rho^2 f^3}\right)E^4} - 1\right]\right\} m_{i0} =$$
$$= \left\{1 - 2\left[\sqrt{1 + 1.758\times 10^{-27}\left(\frac{\mu_r \sigma^3}{\rho^2 f^3}\right)E^4} - 1\right]\right\} m_{i0} \qquad (A14)$$

Note that $E = E_m \sin \omega t$. The average value for $E^2$ is equal to $\frac{1}{2}E_m^2$ because $E$ varies sinusoidaly ($E_m$ is the maximum value for $E$). On the other hand, $E_{rms} = E_m/\sqrt{2}$. Consequently, we can change $E^4$ by $E_{rms}^4$, and the equation above can be rewritten as follows

$$m_g = \left\{1 - 2\left[\sqrt{1 + 1.758\times 10^{-27}\left(\frac{\mu_r \sigma^3}{\rho^2 f^3}\right)E_{rms}^4} - 1\right]\right\} m_{i0}$$

Substitution of the well-known equation of the *Ohm's vectorial Law*: $j = \sigma E$ into (A14), we get

$$m_g = \left\{1 - 2\left[\sqrt{1 + 1.758\times 10^{-27}\frac{\mu_r j_{rms}^4}{\sigma \rho^2 f^3}} - 1\right]\right\} m_{i0} \qquad (A15)$$

where $j_{rms} = j/\sqrt{2}$.

Consider a 15 cm square *Aluminum thin foil* of *10.5 microns thickness* with the following characteristics: $\mu_r = 1$ ; $\sigma = 3.82\times 10^7 S.m^{-1}$; $\rho = 2700\,Kg.m^{-3}$. Then, (A15) gives

$$m_g = \left\{1 - 2\left[\sqrt{1 + 6.313\times 10^{-42}\frac{j_{rms}^4}{f^3}} - 1\right]\right\} m_{i0} \qquad (A16)$$

Now, consider that the ELF electric current $I = I_{DC} + i_0 \sin \omega t$, $(i_0 << I_{DC})$ passes through that Aluminum foil. Then, the current density is

$$j_{rms} = \frac{I_{rms}}{S} \cong \frac{I_{DC}}{S} \qquad (A17)$$

where

$$S = 0.15m\left(10.5\times 10^{-6} m\right) = 1.57\times 10^{-6} m^2$$

If the ELF electric current has frequency $f = 2\mu Hz = 2\times 10^{-6} Hz$, then, the gravitational mass of the aluminum foil, given by (A16), is expressed by

$$m_g = \left\{1 - 2\left[\sqrt{1 + 7.89\times10^{-25}\frac{I_{DC}^4}{S^4}} - 1\right]\right\}m_{i0} =$$

$$= \left\{1 - 2\left[\sqrt{1 + 0.13 I_{DC}^4} - 1\right]\right\}m_{i0} \qquad (A18)$$

Then,

$$\chi = \frac{m_g}{m_{i0}} \cong \left\{1 - 2\left[\sqrt{1 + 0.13 I_{DC}^4} - 1\right]\right\} \qquad (A19)$$

For $I_{DC} = 2.2A$, the equation above gives

$$\chi = \left(\frac{m_g}{m_{i0}}\right) \cong -1 \qquad (A20)$$

This means that *the gravitational shielding* produced by the aluminum foil can change the gravity acceleration *above* the foil down to

$$g' = \chi\, g \cong -1g \qquad (A21)$$

Under these conditions, the Aluminum foil works basically as a Gravity Control Cell (GCC).

In order to check these theoretical predictions, we suggest an experimental set-up shown in Fig.A2.

A 15cm square Aluminum foil of *10.5 microns thickness* with the following composition: Al 98.02%; Fe 0.80%; Si 0.70%; Mn 0.10%; Cu 0.10%; Zn 0.10%; Ti 0.08%; Mg 0.05%; Cr 0.05%, and with the following characteristics: $\mu_r = 1$; $\sigma = 3.82\times10^7 S.m^{-1}; \rho = 2700 Kg.m^{-3}$, is fixed on a 17 cm square *Foam Board* ** plate of 6mm thickness as shown in Fig.A3. This device (the simplest Gravity Control Cell GCC) is placed on a pan balance shown in Fig.A2.

Above the Aluminum foil, a *sample* (any type of material, any mass) connected to a dynamometer will check the decrease of the *local gravity acceleration* upon the sample $(g' = \chi\, g)$, due to the gravitational shielding produced by the decreasing of gravitational mass of the Aluminum foil $(\chi = m_g/m_{i0})$. Initially, the sample lies 5 cm above the Aluminum foil. As shown in Fig.A2, the board with the dynamometer can be displaced up to few meters in height. Thus, the initial distance between the Aluminum foil and the sample can be increased in order to check the reach of the gravitational shielding produced by the Aluminum foil.

In order to generate the ELF electric current of $f = 2\mu Hz$, we can use the widely-known Function Generator HP3325A (Op.002 High Voltage Output) that can generate sinusoidal voltages with *extremely-low* frequencies down to $f = 1\times10^{-6}Hz$ and amplitude up to 20V (40$V_{pp}$ into $500\Omega$ load). The maximum output current is $0.08A_{pp}$; output impedance <2Ω at ELF.

Figure A4 shows the equivalent electric circuit for the experimental set-up. The electromotive forces are: $\varepsilon_1$ (HP3325A) and $\varepsilon_2$ (*12V* DC Battery).The values of the *resistors* are : $R_1 = 500\Omega - 2W$ ; $r_{i1} < 2\Omega$ ; $R_2 = 4\Omega - 40W$ ; $r_{i2} < 0.1\Omega$ ; $R_p = 2.5\times10^{-3}\Omega$ ; *Rheostat* (0≤ *R* ≤10Ω - 90W). The *coupling transformer* has the following characteristics: air core with diameter $\phi = 10mm$ ; area $S = \pi\phi^2/4 = 7.8\times10^{-5}m^2$ ; wire#12AWG; $N_1 = N_2 = N = 20$; $l = 42mm$; $L_1 = L_2 = L = \mu_0 N^2 (S/l) = 9.3\times10^{-7}H$ .Thus, we get

$$Z_1 = \sqrt{(R_1 + r_{i1})^2 + (\omega L)^2} \cong 501\Omega$$

and

$$Z_2 = \sqrt{(R_2 + r_{i2} + R_p + R)^2 + (\omega L)^2}$$

For $R = 0$ we get $Z_2 = Z_2^{min} \cong 4\Omega$; for $R = 10\Omega$ the result is $Z_2 = Z_2^{max} \cong 14\Omega$ . Thus,

$$Z_{1,total}^{min} = Z_1 + Z_{1,reflected}^{min} = Z_1 + Z_2^{min}\left(\frac{N_1}{N_2}\right)^2 \cong 505\Omega$$

$$Z_{1,total}^{max} = Z_1 + Z_{1,reflected}^{max} = Z_1 + Z_2^{max}\left(\frac{N_1}{N_2}\right)^2 \cong 515\Omega$$

The maxima *rms* currents have the following values:

$$I_1^{max} = \tfrac{1}{\sqrt{2}}40V_{pp}\big/Z_{1,total}^{min} = 56mA$$

(The maximum output current of the Function Generator HP3325A (Op.002 High Voltage Output) is $80mA_{pp} \cong 56.5mA_{rms}$);

$$I_2^{max} = \frac{\varepsilon_2}{Z_2^{min}} = 3A$$

and

$$I_3^{max} = I_2^{max} + I_1^{max} \cong 3A$$

The new expression for the *inertial forces*, (Eq.5) $\vec{F}_i = M_g \vec{a}$ , shows that the inertial forces are proportional to *gravitational mass*. Only in the particular case of $m_g = m_{i0}$ , the expression above reduces to the well-known Newtonian expression $\vec{F}_i = m_{i0}\vec{a}$ . The equivalence

** *Foam board* is a very strong, *lightweight* (density: 24.03 kg.m$^{-3}$) and easily cut material used for the mounting of photographic prints, as backing in picture framing, in 3D design, and in painting. It consists of three layers — an inner layer of polystyrene clad with outer facing of either white clay coated paper or brown Kraft paper.

between gravitational and inertial forces $\left(\vec{F}_i \equiv \vec{F}_g\right)$ [1] shows then that a balance measures the *gravitational mass* subjected to acceleration $a = g$. Here, the decrease in the *gravitational mass* of the Aluminum foil will be measured by a pan balance with the following characteristics: range 0-200g; readability 0.01g.

The mass of the Foam Board plate is: $\cong 4.17g$, the mass of the Aluminum foil is: $\cong 0.64g$, the total mass of the ends and the electric wires of connection is $\cong 5g$. Thus, *initially* the balance will show $\cong 9.81g$. According to (A18), when the electric current through the Aluminum foil (resistance $r_p^* = l/\sigma S = 2.5\times10^{-3}\Omega$) reaches the value: $I_3 \cong 2.2A$, we will get $m_{g(Al)} \cong -m_{i0(Al)}$. Under these circumstances, the balance will show:

$$9.81g - 0.64g - 0.64g \cong 8.53g$$

and the gravity acceleration $g'$ *above* the Aluminum foil, becomes $g' = \chi\, g \cong -1g$.

It was shown [1] that, when the gravitational mass of a particle is reduced to the gravitational mass ranging between $+0.159M_i$ to $-0.159M_i$, it becomes *imaginary*, i.e., the gravitational and the inertial masses of the particle become *imaginary*. Consequently, the particle *disappears* from our ordinary space-time. This phenomenon can be observed in the proposed experiment, i.e., *the Aluminum foil will disappear* when its gravitational mass becomes smaller than $+0.159M_i$. It will become visible again, only when its gravitational mass becomes smaller than $-0.159M_i$, or when it becomes greater than $+0.159M_i$.

Equation (A18) shows that the gravitational mass of the Aluminum foil, $m_{g(Al)}$, goes *close to zero* when $I_3 \cong 1.76A$. Consequently, the gravity acceleration *above* the Aluminum foil also goes close to zero since $g' = \chi\, g = m_{g(Al)}/m_{i0(Al)}$. Under these circumstances, the Aluminum foil remains *invisible*.

Now consider a rigid Aluminum wire # 14 AWG. The area of its cross section is

$$S = \pi\left(1.628\times10^{-3}m\right)^2/4 = 2.08\times10^{-6}m^2$$

If an ELF electric current with frequency $f = 2\mu Hz = 2\times10^{-6}Hz$ passes through this wire, its gravitational mass, given by (A16), will be expressed by

$$m_g = \left\{1-2\left[\sqrt{1+6.313\times10^{-42}\frac{j_{rms}^4}{f^3}}-1\right]\right\}m_{i0} =$$

$$= \left\{1-2\left[\sqrt{1+7.89\times10^{-25}\frac{I_{DC}^4}{S^4}}-1\right]\right\}m_{i0} =$$

$$= \left\{1-2\left[\sqrt{1+0.13I_{DC}^4}-1\right]\right\}m_{i0} \qquad (A22)$$

For $I_{DC} \cong 3A$ the equation above gives

$$m_g \cong -3.8m_{i0}$$

Note that we can replace the Aluminum foil for this wire in the experimental set-up shown in Fig.A2. It is important also to note that an ELF electric current that passes through a wire - which makes a spherical form, as shown in Fig A5 - reduces the gravitational mass of the wire (Eq. A22), and the gravity *inside sphere* at the same proportion, $\chi = m_g/m_{i0}$, (Gravitational Shielding Effect). In this case, that effect can be checked by means of the Experimental set-up 2 (Fig.A6). Note that the spherical form can be transformed into an ellipsoidal form or a disc in order to coat, for example, a Gravitational Spacecraft. It is also possible to coat with a wire several forms, such as cylinders, cones, cubes, etc.

The circuit shown in Fig.A4 (a) can be modified in order to produce a new type of Gravitational Shielding, as shown in Fig.A4 (b). In this case, the Gravitational Shielding will be produced in the Aluminum plate, with thickness $h$, of the parallel plate capacitor connected in the point $P$ of the circuit (See Fig.A4 (b)). Note that, in this circuit, the Aluminum foil (resistance $R_p$) (Fig.A4(a)) has been replaced by a Copper wire # 14 AWG with *1cm* length ($l = 1cm$) in order to produce a resistance $R_\phi = 5.21\times10^{-5}\Omega$. Thus, the voltage in the point $P$ of the circuit will have the maximum value $V_p^{max} = 1.1\times10^{-4}V$ when the resistance of the rheostat is null $(R = 0)$ and the minimum value $V_p^{min} = 4.03\times10^{-5}V$ when $R = 10\Omega$. In this way, the voltage $V_p$ (with frequency $f = 2\mu Hz$) applied on the capacitor will produce an electric field $E_p$ with intensity $E_p = V_p/h$ through the Aluminum plate of thickness $h = 3mm$. It is important to note that *this plate cannot be connected to ground* (earth), in other words, cannot be grounded, because, in this case, the electric field through it will be *null* ††.

According to Eq. A14, when $E_p^{max} = V_p^{max}/h = 0.036\, V/m$, $f = 2\mu Hz$ and $\sigma_{Al} = 3.82 \times 10^7\, S/m$, $\rho_{Al} = 2700 kg/m^3$ (Aluminum), we get

$$\chi = \frac{m_{(Al)}}{m_{i(Al)}} \cong -0.9$$

Under these conditions, the maximum *current density* through the plate with thickness $h$ will be given by $j^{max} = \sigma_{Al} E_p^{max} = 1.4 \times 10^6 A/m^2$ (It is well-known that the maximum current density supported by the Aluminum is $\approx 10^8 A/m^2$ ).

Since the area of the plate is $A = (0.2)^2 = 4 \times 10^{-2} m^2$, then the maximum current is $i^{max} = j^{max} A = 56kA$. Despite this enormous current, the maximum dissipated power will be just $P^{max} = \left(i^{max}\right)^2 R_{plate} = 6.2W$, because the resistance of the plate is very small, i.e., $R_{plate} = h/\sigma_{Al} A \cong 2 \times 10^{-9}\Omega$.

Note that the area $A$ of the plate (where the Gravitational Shielding takes place) can have several geometrical configurations. For example, it can be the area of the external surface of an ellipsoid, sphere, etc. Thus, it can be the area of the external surface of a Gravitational Spacecraft. In this case, if $A \cong 100m^2$, for example, the maximum dissipated power will be $P^{max} \cong 15.4kW$, i.e., approximately $154W/m^2$.

All of these systems work with Extra-Low Frequencies $\left(f << 10^{-3} Hz\right)$. Now, we show that, by simply changing *the geometry of the surface of the Aluminum foil*, it is possible to increase the working frequency $f$ up to more than *1Hz*.

Consider the Aluminum foil, now with several semi-spheres stamped on its surface, as shown in Fig. A7 . The semi-spheres have radius $r_0 = 0.9\ mm$, and are joined one to another. The Aluminum foil is now coated by an insulation layer with relative permittivity $\varepsilon_r$ and dielectric strength $k$. A voltage source is connected to the Aluminum foil in order to provide a voltage $V_0$ (rms) with frequency $f$. Thus, the electric potential $V$ at a distance $r$, in the interval from $r_0$ to $a$, is given by

$$V = \frac{1}{4\pi\varepsilon_r\varepsilon_0}\frac{q}{r} \qquad (A23)$$

In the interval $a < r \le b$ the electric potential is

$$V = \frac{1}{4\pi\varepsilon_0}\frac{q}{r} \qquad (A24)$$

since for the air we have $\varepsilon_r \cong 1$.

Thus, on the surface of the metallic spheres $(r = r_0)$ we get

$$V_0 = \frac{1}{4\pi\varepsilon_r\varepsilon_0}\frac{q}{r_0} \qquad (A25)$$

Consequently, the electric field is

$$E_0 = \frac{1}{4\pi\varepsilon_r\varepsilon_0}\frac{q}{r_0^2} \qquad (A26)$$

By comparing (A26) with (A25), we obtain

$$E_0 = \frac{V_0}{r_0} \qquad (A27)$$

The electric potential $V_b$ at $r = b$ is

$$V_b = \frac{1}{4\pi\varepsilon_0}\frac{q}{b} = \frac{\varepsilon_r V_0 r_0}{b} \qquad (A28)$$

Consequently, the electric field $E_b$ is given by

$$E_b = \frac{1}{4\pi\varepsilon_0}\frac{q}{b^2} = \frac{\varepsilon_r V_0 r_0}{b^2} \qquad (A29)$$

From $r = r_0$ up to $r = b = a + d$ *the electric field is approximately constant* (See Fig. A7). Along the distance $d$ it will be called $E_{air}$. For $r > a + d$, the electric field stops being constant. Thus, the intensity of the electric field at $r = b = a + d$ is approximately equal to $E_0$, i.e., $E_b \cong E_0$. Then, we can write that

$$\frac{\varepsilon_r V_0 r_0}{b^2} \cong \frac{V_0}{r_0} \qquad (A30)$$

whence we get

$$b \cong r_0\sqrt{\varepsilon_r} \qquad (A31)$$

Since the intensity of the electric field through the air, $E_{air}$, is $E_{air} \cong E_b \cong E_0$, then, we can write that

$$E_{air} = \frac{1}{4\pi\varepsilon_0}\frac{q}{b^2} = \frac{\varepsilon_r V_0 r_0}{b^2} \qquad (A32)$$

Note that $\varepsilon_r$ refers to the *relative permittivity of*

†† When the voltage $V_p$ is applied on the capacitor, the charge distribution in the dielectric induces positive and negative charges, respectively on opposite sides of the Aluminum plate with thickness *h*. If the plate is not connected to the ground (Earth) this charge distribution produces an electric field $E_p = V_p/h$ through the plate. However, if the plate is connected to the ground, the negative charges (electrons) escapes for the ground and the positive charges are redistributed along the entire surface of the Aluminum plate making *null* the electric field through it.

*the insulation layer, which is covering the Aluminum foil*.

If the intensity of this field is greater than the dielectric strength of the air $\left(3\times10^{6}V/m\right)$ there will occur the well-known *Corona effect*. Here, this effect is necessary in order to increase the electric conductivity of the air at this region (layer with thickness *d*). Thus, we will assume

$$E_{air}^{min}=\frac{\varepsilon_r V_0^{min} r_0}{b^2}=\frac{V_0^{min}}{r_0}=3\times10^{6}V/m$$

and

$$E_{air}^{max}=\frac{\varepsilon_r V_0^{max} r_0}{b^2}=\frac{V_0^{max}}{r_0}=1\times10^{7}V/m \qquad (A33)$$

The electric field $E_{air}^{min}\le E_{air}\le E_{air}^{max}$ will produce an *electrons flux* in a direction and an *ions flux* in an opposite direction. From the viewpoint of electric current, the ions flux can be considered as an "electrons" flux at the same direction of the real electrons flux. Thus, the current density through the air, $j_{air}$, will be the *double* of the current density expressed by the well-known equation of Langmuir-Child

$$j=\frac{4}{9}\varepsilon_r\varepsilon_0\sqrt{\frac{2e}{m_e}}\frac{V^{\frac{3}{2}}}{d^2}=\alpha\frac{V^{\frac{3}{2}}}{d^2}=2.33\times10^{-6}\frac{V^{\frac{3}{2}}}{d^2} \qquad (A34)$$

where $\varepsilon_r\cong1$ for the *air*; $\alpha=2.33\times10^{-6}$ is the called *Child's constant*.

Thus, we have

$$j_{air}=2\alpha\frac{V^{\frac{3}{2}}}{d^2} \qquad (A35)$$

where $d$, in this case, is the thickness of the air layer where the electric field is approximately constant and $V$ is the voltage drop given by

$$V=V_a-V_b=\frac{1}{4\pi\varepsilon_0}\frac{q}{a}-\frac{1}{4\pi\varepsilon_0}\frac{q}{b}=$$

$$=V_0 r_0\varepsilon_r\left(\frac{b-a}{ab}\right)=\left(\frac{\varepsilon_r r_0 d}{ab}\right)V_0 \qquad (A36)$$

By substituting (A36) into (A35), we get

$$j_{air}=\frac{2\alpha}{d^2}\left(\frac{\varepsilon_r r_0 d V_0}{ab}\right)^{\frac{3}{2}}=\frac{2\alpha}{d^{\frac{1}{2}}}\left(\frac{\varepsilon_r r_0 V_0}{b^2}\right)^{\frac{3}{2}}\left(\frac{b}{a}\right)^{\frac{3}{2}}=$$

$$=\frac{2\alpha}{d^{\frac{1}{2}}}E_{air}^{\frac{3}{2}}\left(\frac{b}{a}\right)^{\frac{3}{2}} \qquad (A37)$$

According to the equation of the *Ohm's vectorial Law*: $j=\sigma E$, we can write that

$$\sigma_{air}=\frac{j_{air}}{E_{air}} \qquad (A38)$$

Substitution of (A37) into (A38) yields

$$\sigma_{air}=2\alpha\left(\frac{E_{air}}{d}\right)^{\frac{1}{2}}\left(\frac{b}{a}\right)^{\frac{3}{2}} \qquad (A39)$$

If the insulation layer has thickness $\Delta=0.6\ mm$, $\varepsilon_r\cong3.5$ (1- 60Hz), $k=17kV/mm$ (Acrylic sheet 1.5mm thickness), and the semi-spheres stamped on the metallic surface have $r_0=0.9\ mm$ (See Fig.A7) then $a=r_0+\Delta=1.5\ mm$. Thus, we obtain from Eq. (A33) that

$$V_0^{min}=2.7kV$$

$$V_0^{max}=9kV \qquad (A40)$$

From equation (A31), we obtain the following value for $b$:

$$b=r_0\sqrt{\varepsilon_r}=1.68\times10^{-3}m \qquad (A41)$$

Since $b=a+d$ we get

$$d=1.8\times10^{-4}m$$

Substitution of $a$, $b$, $d$ and A(32) into (A39) produces

$$\sigma_{air}=4.117\times10^{-4}E_{air}^{\frac{1}{2}}=1.375\times10^{-2}V_0^{\frac{1}{2}}$$

Substitution of $\sigma_{air}$, $E_{air}(rms)$ and $\rho_{air}=1.2\ kg.m^{-3}$ into (A14) gives

$$\frac{m_{g(air)}}{m_{i0(air)}}=\left\{1-2\left[\sqrt{1+1.758\times10^{-27}\frac{\sigma_{air}^3E_{air}^4}{\rho_{air}^2f^3}}-1\right]\right\}=$$

$$=\left\{1-2\left[\sqrt{1+4.923\times10^{-21}\frac{V_0^{5.5}}{f^3}}-1\right]\right\} \qquad (A42)$$

For $V_0=V_0^{max}=9kV$ and $f=2Hz$, the result is

$$\frac{m_{g(air)}}{m_{i0(air)}}\cong-1.2$$

Note that, by increasing $V_0$, the values of $E_{air}$ and $\sigma_{air}$ are increased. Thus, as show (A42), there are two ways for decrease the value of $m_{g(air)}$: increasing the value of $V_0$ or decreasing the value of $f$.

Since $E_0^{max}=10^{7}V/m=10kV/mm$ and $\Delta=0.6\ mm$ then the dielectric strength of the insulation must be $\ge16.7kV/mm$. As mentioned above, the dielectric strength of the acrylic is $17kV/mm$.

It is important to note that, due to the strong value of $E_{air}$ (Eq. A37) the *drift velocity* $v_d$, $\left(v_d=j_{air}/ne=\sigma_{air}E_{air}/ne\right)$ of the free charges inside the ionized air put them at a distance $x = v_d/t = 2fv_d \cong 0.4m$, which is much greater than the distance $d = 1.8\times10^{-4}m$. Consequently, the number $n$ of free charges decreases strongly inside the air layer of thickness $d$ [‡‡], except, obviously, in a thin layer, very close to the dielectric, where the number of free charges remains sufficiently increased, to maintain the air conductivity with $\sigma_{air} \cong 1.1S/m$ (Eq. A39).

The thickness $h$ of this thin air layer close to the dielectric can be easily evaluated starting from the charge distribution in the neighborhood of the dielectric, and of the repulsion forces established among them. The result is $h = \sqrt{0.06e/4\pi\varepsilon_0 E} \cong 4\times10^{-9}m$. This is, therefore, the thickness of the *Air* Gravitational Shielding. If the area of this Gravitational Shielding is equal to the area of a format A4 sheet of paper, i.e., $A = 0.20\times0.291 = 0.0582m^2$, we obtain the following value for the resistance $R_{air}$ of the Gravitational Shielding: $R_{air} = h/\sigma_{air}A \cong 6\times10^{-8}\Omega$. Since the maximum electrical current through this air layer is $i^{max} = j^{max}A \cong 400kA$, then the maximum power radiated from the Gravitational Shielding is $P_{air}^{max} = R_{air}\left(i_{air}^{max}\right)^2 \cong 10kW$. This means that a very strong light will be radiated from this type of Gravitational Shielding. Note that this device can also be used as a lamp, which will be much more efficient than conventional lamps.

Coating a ceiling with this lighting system enables the entire area of ceiling to produce light. This is a form of lighting very different from those usually known.

Note that the value $P_{air}^{max} \cong 10kW$, defines the power of the transformer shown in Fig.A10. Thus, the maximum current in the secondary is $i_s^{max} = 9kV/10kW = 0.9A$.

Above the Gravitational Shielding, $\sigma_{air}$ is reduced to the normal value of conductivity of the atmospheric air $\left(\approx 10^{-14}S/m\right)$. Thus, the power radiated from this region is

$$P_{air}^{max} = (d-h)\left(i_{air}^{max}\right)^2/\sigma_{air}A =$$
$$= (d-h)A\sigma_{air}\left(E_{air}^{max}\right)^2 \cong 10^{-4}W$$

Now, we will describe a method to coat the Aluminum semi-spheres with acrylic in the necessary dimensions $(\Delta = a - r_0)$, we propose the following method. First, take an Aluminum plate with $21cm\times29.1cm$ (A4 format). By means of a convenient process, several semi-spheres can be stamped on its surface. The semi-spheres have radius $r_0 = 0.9\ mm$, and are joined one to another. Next, take an acrylic sheet (A4 format) with 1.5mm thickness (See Fig.A8 (a)). Put a heater below the Aluminum plate in order to heat the Aluminum (Fig.A8 (b)). When the Aluminum is sufficiently heated up, the acrylic sheet and the Aluminum plate are pressed, one against the other, as shown in Fig. A8 (c). The two D devices shown in this figure are used in order to impede that the press compresses the acrylic and the aluminum to a distance shorter than $y + a$. After some seconds, remove the press and the heater. The device is ready to be subjected to a voltage $V_0$ with frequency $f$, as shown in Fig.A9. Note that, in this case, the balance is not necessary, because *the substance that produces the gravitational shielding* is an *air layer* with thickness $d$ *above* the acrylic sheet. This is, therefore, more a type of Gravity Control Cell (GCC) with *external gravitational shielding*.

It is important to note that this GCC can be made very thin and as flexible as a fabric. Thus, it can be used to produce *anti- gravity clothes*. These clothes can be extremely useful, for example, to walk on the surface of high gravity planets.

Figure A11 shows some geometrical forms that can be stamped on a metallic surface in order to produce a Gravitational Shielding effect, similar to the produced by the *semi-spherical form*.

An obvious evolution from the semi-spherical form is the *semi-cylindrical* form shown in Fig. A11 (b); Fig.A11(c) shows *concentric metallic rings* stamped on the metallic surface, an evolution from Fig.A11 (b). These geometrical forms produce the same effect as the semi-spherical form, shown in Fig.A11 (a). By using concentric metallic rings, it is possible to build *Gravitational Shieldings* around bodies or spacecrafts with several formats (spheres, ellipsoids, etc); Fig. A11 (d) shows a Gravitational Shielding around a Spacecraft with *ellipsoidal form*.

The previously mentioned Gravitational Shielding, produced on a thin layer of ionized air, has a *behavior different from* the Gravitational Shielding produced on a *rigid substance*. When the gravitational masses of the air molecules, inside the shielding, are reduced to within the range $+0.159m_i < m_g < -0.159m_i$, they go to the *imaginary space-time*, as previously shown in this article. However, the electric field $E_{air}$ stays at the real space-time. Consequently, the molecules return immediately to the real space-time in order to return soon after to the *imaginary* space-time, due to the action of the electric field $E_{air}$.

[‡‡] Reducing therefore, the conductivity $\sigma_{air}$, to the normal value of conductivity of the atmospheric air.

In the case of the Gravitational Shielding produced on a *solid substance*, when the molecules of the substance go to the *imaginary* space-time, *the electric field that produces the effect, also goes to the imaginary space-time together with them*, since in this case, the substance of the Gravitational Shielding is rigidly connected to the metal that produces the electric field. (See Fig. A12 (b)). This is the fundamental difference between the *non-solid* and *solid* Gravitational Shieldings.

Now, consider a Gravitational Spacecraft that is able to produce an *Air* Gravitational Shielding and also a *Solid* Gravitational Shielding, as shown in Fig. A13 (a) [§§]. Assuming that the intensity of the electric field, $E_{air}$, necessary to reduce the gravitational mass of the *air molecules* to within the range $+0.159m_i < m_g < -0.159m_i$, *is much smaller* than the intensity of the electric field, $E_{rs}$, necessary to reduce the gravitational mass of the *solid substance* to within the range $+0.159m_i < m_g < -0.159m_i$, then we conclude that the Gravitational Shielding made of ionized air goes to the imaginary space-time *before* the Gravitational Shielding made of *solid substance*. When this occurs the spacecraft does not go to the imaginary space-time together with the Gravitational Shielding of air, because the air molecules are not rigidly connected to the spacecraft. Thus, while the air molecules go into the imaginary space-time, the spacecraft stays in the *real space-time*, and remains subjected to the effects of the Gravitational Shielding around it, since the shielding does not stop to work, due to its extremely short permanence at the imaginary space-time. Under these circumstances, the gravitational mass of the Gravitational Shielding can be reduced to $m_g \cong 0$. For example, $m_g \cong 10^{-4}kg$. Thus, if the *inertial mass* of the Gravitational Shielding is $m_{i0} \cong 1kg$, then $\chi = m_g / m_{i0} \cong 10^{-4}$. As we have seen, this means that *the inertial effects on the spacecraft* will be reduced by $\chi \cong 10^{-4}$. Then, in spite of the effective acceleration of the spacecraft be, for example, $a = 10^5 m.s^{-2}$, the effects on the crew of the spacecraft will be equivalent to an acceleration of only

$$a' = \frac{m_g}{m_{i0}} a = \chi \ a \approx 10 m.s^{-1}$$

This is the magnitude of the acceleration upon the passengers in a contemporary commercial jet.

Then, it is noticed that Gravitational Spacecrafts can be subjected to enormous *accelerations* (or *decelerations*) without imposing any harmful impacts whatsoever on the spacecrafts or its crew.

Now, imagine that the intensity of the electric field that produces the Gravitational Shielding around the spacecraft is *increased* up to reaching the value $E_{rs}$ that reduces the gravitational mass of the *solid* Gravitational Shielding to within the range $+0.159m_i < m_g < -0.159m_i$. Under these circumstances, the *solid* Gravitational Shielding goes to the imaginary space-time and, since it is rigidly connected to the spacecraft, also the spacecraft goes to the imaginary space-time together with the Gravitational Shielding. Thus, the spacecraft can travel within the imaginary space-time and make use of the Gravitational Shielding around it.

As we have already seen, the maximum velocity of propagation of the interactions in the imaginary space-time is *infinite* (in the real space-time this limit is equal to the light velocity $c$). This means that *there are no limits for the velocity of the spacecraft in the imaginary space-time*. Thus, the acceleration of the spacecraft can reach, for example, $a = 10^9 m.s^{-2}$, which leads the spacecraft to attain velocities $V \approx 10^{14} m.s^{-1}$ (about 1 million times the speed of light) after one day of trip. With this velocity, after 1 month of trip the spacecraft would have traveled about $10^{21} m$. In order to have idea of this distance, it is enough to remind that the diameter of our Universe (visible Universe) is of the order of $10^{26} m$.

---

[§§] The *solid* Gravitational Shielding can also be obtained by means of *an ELF electric current through a metallic lamina* placed *between the semi-spheres and the Gravitational Shielding of Air* (See Fig.A13 (a)). The gravitational mass of the solid Gravitational Shielding will be controlled just by means of the intensity of the ELF electric current. Recently, it was discovered that Carbon nanotubes (CNTs) can be added to *Alumina* ($Al_2O_3$) to convert it into a good electrical conductor. It was found that the electrical conductivity increased up to 3375 S/m at 77°C in samples that were 15% nanotubes by volume [12]. It is known that the density of α-Alumina is 3.98kg.m$^{-3}$ and that it can withstand 10-20 KV/mm. Thus, these values show that the Alumina-CNT can be used to make a *solid* Gravitational Shielding. In this case, the electric field produced by means of the semi-spheres will be used to control the gravitational mass of the Alumina-CNT.

Due to the extremely low density of the *imaginary* bodies, the collision between them cannot have the same consequences of the collision between the real bodies.

Thus, *for a Gravitational Spacecraft in imaginary state, the problem of the collision in high-speed doesn't exist.* Consequently, the Gravitational Spacecraft can transit freely in the imaginary Universe and, in this way, reach easily any point of our real Universe once they can make the transition back to our Universe by only increasing the gravitational mass of the Gravitational Shielding of the spacecraft in such way that it leaves the range of $+0.159M_i$ to $-0.159M_i$.

The return trip would be done in similar way. That is to say, the spacecraft would transit in the imaginary Universe back to the departure place where would reappear in our Universe. Thus, trips through our Universe that would delay millions of years, at speeds close to the speed of light, could be done in just a few *months* in the imaginary Universe.

In order to produce the acceleration of $a \approx 10^9 m.s^{-2}$ upon the spacecraft we propose a Gravitational Thruster with 10 GCCs (10 Gravitational Shieldings) of the type with several semi-spheres stamped on the metallic surface, as previously shown, or with the *semi-cylindrical* form shown in Figs. A11 (b) and (c). The 10 GCCs are filled with air at 1 atm and 300K. If the insulation layer is made with *Mica* $(\varepsilon_r \cong 5.4)$ and has thickness $\Delta = 0.1\ mm$, and the semi-spheres stamped on the metallic surface have $r_0 = 0.4\ mm$ (See Fig.A7) then $a = r_0 + \Delta = 0.5\ mm$. Thus, we get

$$b = r_0\sqrt{\varepsilon_r} = 9.295\times10^{-4} m$$

and

$$d = b - a = 4.295\times10^{-4} m$$

Then, from Eq. A42 we obtain

$$\chi_{air} = \frac{m_{g(air)}}{m_{i0(air)}} = \left\{1 - 2\left[\sqrt{1 + 1.758\times10^{-27}\frac{\sigma_{air}^3 E_{air}^4}{\rho_{air}^2 f^3}} - 1\right]\right\} =$$

$$= \left\{1 - 2\left[\sqrt{1 + 1.0\times10^{-18}\frac{V_0^{5.5}}{f^3}} - 1\right]\right\}$$

For $V_0 = V_0^{max} = 15.6kV$ and $f = 0.12Hz$, the result is

$$\chi_{air} = \frac{m_{g(air)}}{m_{i0(air)}} \cong -1.6\times10^4$$

Since $E_0^{max} = V_0^{max}/r_0$ is now given by $E_0^{max} = 15.6kV/0.9mm = 17.3kV/mm$ and $\Delta = 0.1\ mm$ then the dielectric strength of the insulation must be $\geq 173kV/mm$. As shown in the table below***, *0.1mm - thickness of* Mica *can withstand 17.6 kV* (that is greater than $V_0^{max} = 15.6kV$), in such way that the dielectric strength is *176 kV/mm*.

The Gravitational Thrusters are positioned at the spacecraft, as shown in Fig. A13 (b). Then, when the spacecraft is in the *intergalactic space*, the gravity acceleration upon the gravitational mass $m_{gt}$ of the bottom of the thruster (See Fig.A13 (c)), is given by [2]

$$\vec{a} \cong (\chi_{air})^{10}\vec{g}_M \cong -(\chi_{air})^{10} G\frac{M_g}{r^2}\hat{\mu}$$

where $M_g$ is the gravitational mass in front of the spacecraft.

For simplicity, let us consider just the effect of a hypothetical volume $V = 10\times10^3\times10^3 = 10^7 m^3$ of intergalactic matter in front of the spacecraft $(r \cong 30m)$. The average density of matter in the *intergalactic medium (IGM)* is $\rho_{ig} \approx 10^{-26}kg.m^{-3}$)†††. Thus, for $\chi_{air} \cong -1.6\times10^4$ we get

$$a = -(-1.6\times10^4)^{10}(6.67\times10^{-11})\left(\frac{10^{-19}}{30^2}\right) =$$

$$= -10^9 m.s^{-2}$$

In spite of this gigantic acceleration, the inertial effects for the crew of the spacecraft can be strongly reduced if, for example, the gravitational mass of the Gravitational Shielding is reduced

---

*** The *dielectric strength* of some dielectrics can have different values in lower thicknesses. This is, for example, the case of the *Mica*.

| *Dielectric* | *Thickness* (mm) | *Dielectric Strength* (kV/mm) |
|---|---|---|
| Mica | 0.01 mm | 200 |
| **Mica** | **0.1 mm** | **176** |
| Mica | 1 mm | 61 |

††† Some theories put the average density of the Universe as the equivalent of *one hydrogen atom per cubic meter* [13,14]. The density of the universe, however, is clearly not uniform. Surrounding and stretching between galaxies, there is a rarefied plasma [15] that is thought to possess a cosmic filamentary structure [16] and that is slightly denser than the average density in the universe. This material is called the *intergalactic medium (IGM)* and is mostly ionized hydrogen; i.e. a plasma consisting of equal numbers of electrons and protons. The IGM is thought to exist at a density of 10 to 100 times the average density of the Universe (10 to 100 hydrogen atoms per cubic meter, i.e., $\approx 10^{-26} kg.m^{-3}$).

down to $m_g \cong 10^{-6} kg$ and its inertial mass is $m_{i0} \cong 100 kg$. Then, we get $\chi = m_g / m_{i0} \cong 10^{-8}$. Therefore, *the inertial effects on the spacecraft* will be reduced by $\chi \cong 10^{-8}$, and consequently, the inertial effects on the crew of the spacecraft would be *equivalent to* an acceleration $a'$ of only

$$a' = \frac{m_g}{m_{i0}} a = \left(10^{-8}\right)\left(10^{9}\right) \approx 10 m.s^{-2}$$

Note that the Gravitational Thrusters in the spacecraft must have a very small diameter (of the order of *millimeters*) since, obviously, the hole through the Gravitational Shielding cannot be large. Thus, these thrusters are in fact, *Micro-Gravitational Thrusters*. As shown in Fig. A13 (b), it is possible to place several micro-gravitational thrusters in the spacecraft. This gives to the Gravitational Spacecraft, several degrees of freedom and shows the enormous superiority of this spacecraft in relation to the contemporaries spacecrafts.

The density of matter in the *intergalactic medium (IGM)* is about *10* $^{-26}$ *kg.m*$^{-3}$, which is very less than the density of matter in the *interstellar medium* (*~10*$^{-21}$ *kg.m*$^{-3}$) that is less than the density of matter in the *interplanetary medium* (*~10*$^{-20}$ *kg.m*$^{-3}$). The density of matter is enormously increased inside the Earth's atmosphere (1.2*kg.m*$^{-3}$ near to Earth's surface). Figure A14 shows the gravitational acceleration acquired by a Gravitational Spacecraft, in these media, using Micro-Gravitational thrusters.

In relation to the *Interstellar* and *Interplanetary medium*, the *Intergalactic medium* requires the greatest value of $\chi_{air}$ ($\chi$ inside the *Micro-Gravitational Thrusters*), i.e., $\chi_{air} \cong -1.6 \times 10^4$. This value strongly decreases when the spacecraft is within the Earth's atmosphere. In this case, it is sufficient only[‡‡‡] $\chi_{air} \cong -10$ in order to obtain:

$$a = -\left(\chi_{air}\right)^{10} G \frac{\rho_{atm} V}{r^2} \cong$$

$$\cong -\left(-10\right)^{10}\left(6.67 \times 10^{-11}\right)\frac{1.2\left(10^7\right)}{\left(20\right)^2} \cong 10^4 m.s^{-2}$$

With this acceleration the Gravitational Spacecraft can reach about *50000 km/h* in a few seconds. Obviously, the Gravitational Shielding of the spacecraft will reduce strongly *the inertial effects upon the crew* of the spacecraft, in such way that the inertial effects of this strong acceleration will not be felt. In addition, the *artificial atmosphere*, which is possible to build around the spacecraft, by means of gravity control technologies shown in this article (See Fig.6) and [2], will protect it from the *heating* produced by the friction with the Earth's atmosphere. Also, the gravity can be controlled inside of the Gravitational Spacecraft in order to maintain a value close to the Earth's gravity as shown in Fig.3.

Finally, it is important to note that a Micro-Gravitational Thruster does not work *outside* a Gravitational Shielding, because, in this case, *the resultant upon the thruster is null* due to the symmetry (See Fig. A15 (a)). Figure A15 (b) shows a micro-gravitational thruster inside a Gravitational Shielding. This thruster has 10 Gravitational Shieldings, in such way that the gravitational acceleration upon the *bottom* of the thruster, due to a gravitational mass $M_g$ *in front* of the thruster, is $a_{10} = \chi_{air}^{10} a_0$ where $a_0 = -G M_g / r^2$ is the gravitational acceleration acting on the front of the micro-gravitational thruster. *In the opposite direction*, the gravitational acceleration upon the bottom of the thruster, produced by a gravitational mass $M_g$, is

$$a_0' = \chi_s \left(-G M_g / r'^2\right) \cong 0$$

since $\chi_s \cong 0$ due to the Gravitational Shielding around the micro-thruster (See Fig. A15 (b)). Similarly, the acceleration in front of the thruster is

$$a_{10}' = \chi_{air}^{10} a_0' = \left[\chi_{air}^{10}\left(-G M_g / r'^2\right)\right] \chi_s$$

where $\left[\chi_{air}^{10}\left(-G M_g / r'^2\right)\right] < a_{10}$, since $r' > r$. Thus, for $a_{10} \cong 10^9 m.s^{-2}$ and $\chi_s \approx 10^{-8}$ we conclude that $a_{10}' < 10 m.s^{-2}$. This means that $a_{10}' << a_{10}$. Therefore, we can write that the resultant on the micro-thruster can be expressed by means of the following relation

$$R \cong F_{10} = \chi_{air}^{10} F_0$$

Figure A15 (c) shows a Micro-Gravitational Thruster with *10 Air Gravitational Shieldings* (10 GCCs). Thin Metallic laminas are placed after each *Air* Gravitational Shielding in order to retain the electric field $E_b = V_0/x$, produced by metallic *surface behind* the semi-spheres. The laminas with semi-spheres stamped on its surfaces are connected to the ELF voltage source $V_0$ and the thin laminas in front of the Air Gravitational Shieldings are grounded. The air inside this Micro-Gravitational Thruster is at 300K, 1atm.

‡‡‡ This value is within the range of values of $\chi$ ($\chi < -10^3$. *See Eq.A15*), which can be produced by means of *ELF electric currents* through metals as *Aluminum*, etc. This means that, in this case, if convenient, we can replace *air* inside the GCCs of the Gravitational Micro-thrusters by metal laminas with *ELF electric currents* through them.

We have seen that the insulation layer of a GCC can be made up of Acrylic, Mica, etc. Now, we will design a GCC using *Water* (*distilled water*, $\varepsilon_{r(H_2O)} = 80$) and Aluminum *semi-cylinders* with radius $r_0 = 1.3mm$. Thus, for $\Delta = 0.6mm$, the new value of $a$ is $a = 1.9mm$. Then, we get

$$b = r_0\sqrt{\varepsilon_{r(H_2O)}} = 11.63\times10^{-3}m \qquad (A43)$$

$$d = b - a = 9.73\times10^{-3}m \qquad (A44)$$

and

$$E_{air} = \frac{1}{4\pi\varepsilon_{r(air)}\varepsilon_0}\frac{q}{b^2} =$$

$$= \varepsilon_{r(H_2O)}\frac{V_0 r_0}{\varepsilon_{r(air)}b^2} =$$

$$= \frac{V_0/r_0}{\varepsilon_{r(air)}} \cong \frac{V_0}{r_0} = 1111.1\ V_0 \qquad (A45)$$

Note that

$$E_{(H_2O)} = \frac{V_0/r_0}{\varepsilon_{r(H_2O)}}$$

and

$$E_{(acrylic)} = \frac{V_0/r_0}{\varepsilon_{r(acrylic)}}$$

Therefore, $E_{(H_2O)}$ and $E_{(acrylic)}$ are much smaller than $E_{air}$. Note that for $V_0 \leq 9kV$ the intensities of $E_{(H_2O)}$ and $E_{(acrylic)}$ are not sufficient to produce the ionization effect, which increases the electrical conductivity. Consequently, the conductivities of the water and the acrylic remain $<< 1\ S.m^{-1}$. In this way, with $E_{(H_2O)}$ and $E_{(acrylic)}$ much smaller than $E_{air}$, and $\sigma_{(H2O)} << 1$, $\sigma_{(acrylic)} << 1$, the decrease in both the gravitational mass of the acrylic and the gravitational mass of water, according to Eq.A14, is negligible. This means that only in the air layer the decrease in the gravitational mass will be relevant.

Equation A39 gives the electrical conductivity of the air layer, i.e.,

$$\sigma_{air} = 2\alpha\left(\frac{E_{air}}{d}\right)^{\frac{1}{2}}\left(\frac{b}{a}\right)^{\frac{3}{2}} = 0.029V_0^{\frac{1}{2}} \qquad (A46)$$

Note that $b = r_0\sqrt{\varepsilon_{r(H_2O)}}$. Therefore, here the value of $b$ is larger than in the case of the acrylic. Consequently, *the electrical conductivity of the air layer will be larger here than in the case of acrylic.*

Substitution of $\sigma_{(air)}$, $E_{air}$ (*rms*) and $\rho_{air} = 1.2kg.m^{-3}$ into Eq. A14, gives

$$\frac{m_{g(air)}}{m_{i0(air)}} = \left\{1 - 2\left[\sqrt{1 + 4.54\times10^{-20}\frac{V_0^{5.5}}{f^3}} - 1\right]\right\} \qquad (A47)$$

For $V_0 = V_0^{max} = 9kV$ and $f = 2Hz$, the result is

$$\frac{m_{g(air)}}{m_{i0(air)}} \cong -8.4$$

This shows that, by using *water* instead of acrylic, the result is much better.

In order to build the GCC based on the calculations above (See Fig. A16), take an Acrylic plate with *885mm* X *885m* and *2mm* thickness, then paste on it an Aluminum sheet with *895.2mm* X *885mm* and *0.5mm* thickness(note that two edges of the Aluminum sheet are bent as shown in Figure A16 (b)). Next, take *342* Aluminum yarns with *884mm* length and *2.588mm* diameter (wire # 10 AWG) and insert them side by side on the Aluminum sheet. See in Fig. A16 (b) the detail of fixing of the yarns on the Aluminum sheet. Now, paste acrylic strips (with *13.43mm* height and *2mm* thickness) around the Aluminum/Acrylic, making a box. Put *distilled water* (approximately *1 litter*) inside this box, up to a height of exactly *3.7mm* from the edge of the acrylic base. Afterwards, paste an Acrylic lid (*889mm* X *889mm* and *2mm* thickness) on the box. Note that above the water there is an *air* layer with *885mm* X *885mm* and *7.73mm* thickness (See Fig. A16). This thickness plus the acrylic lid thickness (*2mm*) is equal to $d = b - a = 9.73mm$ where $b = r_0\sqrt{\varepsilon_{r(H_2O)}} = 11.63mm$ and $a = r_0 + \Delta = 1.99mm$, since $r_0 = 1.3mm$, $\varepsilon_{r(H_2O)} = 80$ and $\Delta = 0.6mm$.

Note that the gravitational action of the electric field $E_{air}$, extends itself only up to the distance $d$, which, in this GCC, is given by the sum of the Air layer thickness (*7.73mm*) plus the thickness of the Acrylic lid (*2mm*).

Thus, it is ensured the gravitational effect on the air layer while it is practically nullified in the acrylic sheet above the air layer, since $E_{(acrylic)} << E_{air}$ and $\sigma_{(acrylic)} << 1$.

With this GCC, we can carry out an experiment where the *gravitational mass of the air layer* is progressively reduced when the voltage applied to the GCC is increased (or when the frequency is decreased). A precision balance is placed below the GCC in order to measure the mentioned mass decrease for comparison with the values predicted by Eq. A(47). In total, this GCC weighs about *6kg*; the *air layer 7.3grams*. The balance has the following characteristics: *range 0-6kg; readability 0.1g*. Also, in order to prove the *Gravitational Shielding Effect*, we can put a *sample* (connected to a dynamometer) above the GCC in order to check the gravity acceleration in this region.

In order to prove *the exponential effect* produced by the superposition of the Gravitational Shieldings, we can take three similar GCCs and put them one above the other, in such way that above the GCC 1 the gravity acceleration will be $g' = \chi\, g$; above the GCC2 $g'' = \chi^2 g$, and above the GCC3 $g''' = \chi^3 g$. Where $\chi$ is given by Eq. (A47).

It is important to note that the intensity of the electric field through the air *below* the GCC is *much smaller* than the intensity of the electric field through the air layer inside the GCC. In addition, the electrical conductivity of the air below the GCC is much smaller than the conductivity of the air layer inside the GCC. Consequently, the decrease of the gravitational mass of the air below the GCC, according to Eq.A14, is negligible. This means that the GCC1, GCC2 and GCC3 can be simply overlaid, on the experiment proposed above. However, since it is necessary to put samples among them in order to measure the gravity above each GCC, we suggest a spacing of 30cm or more among them.

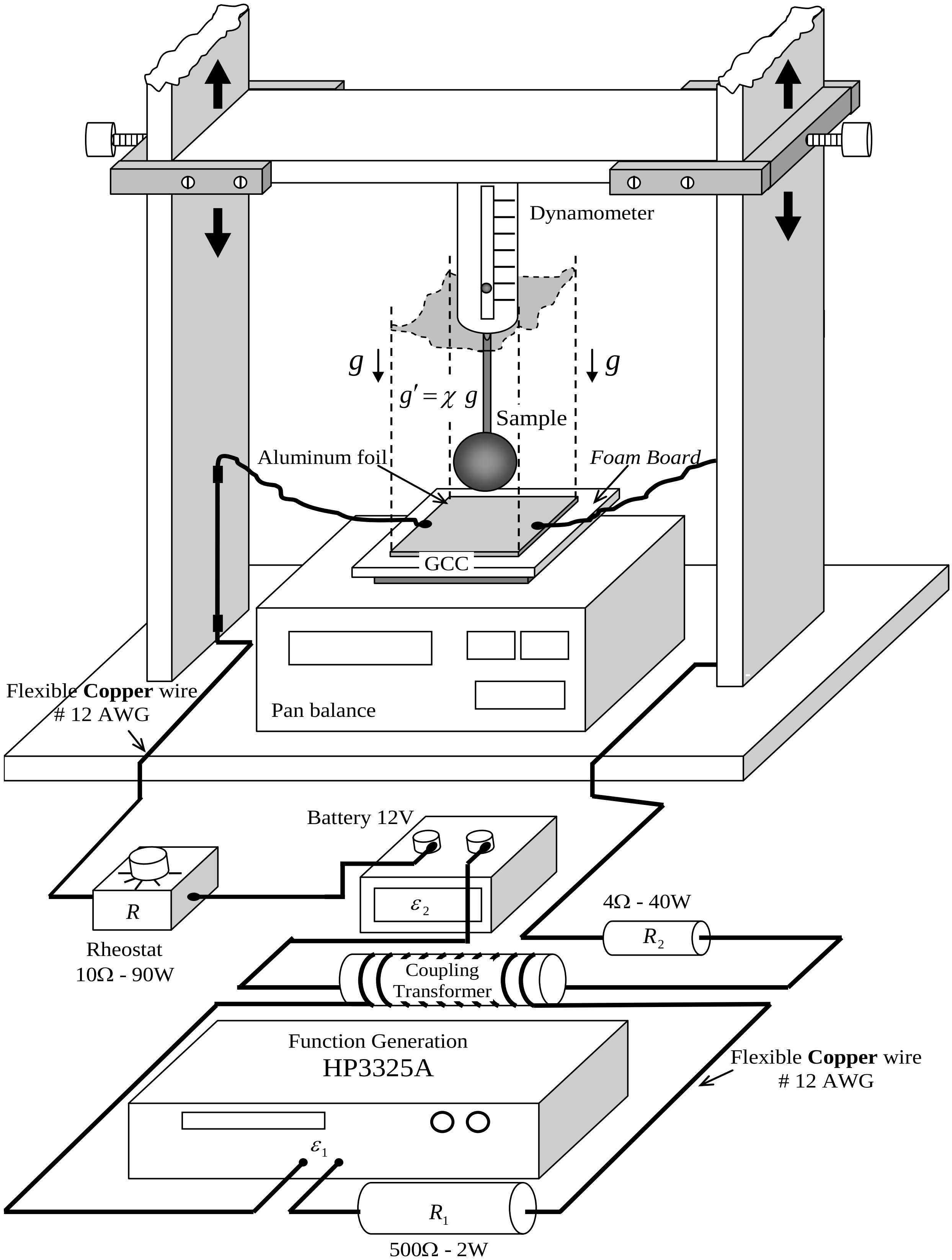

Figure A2 – Experimental Set-up 1.

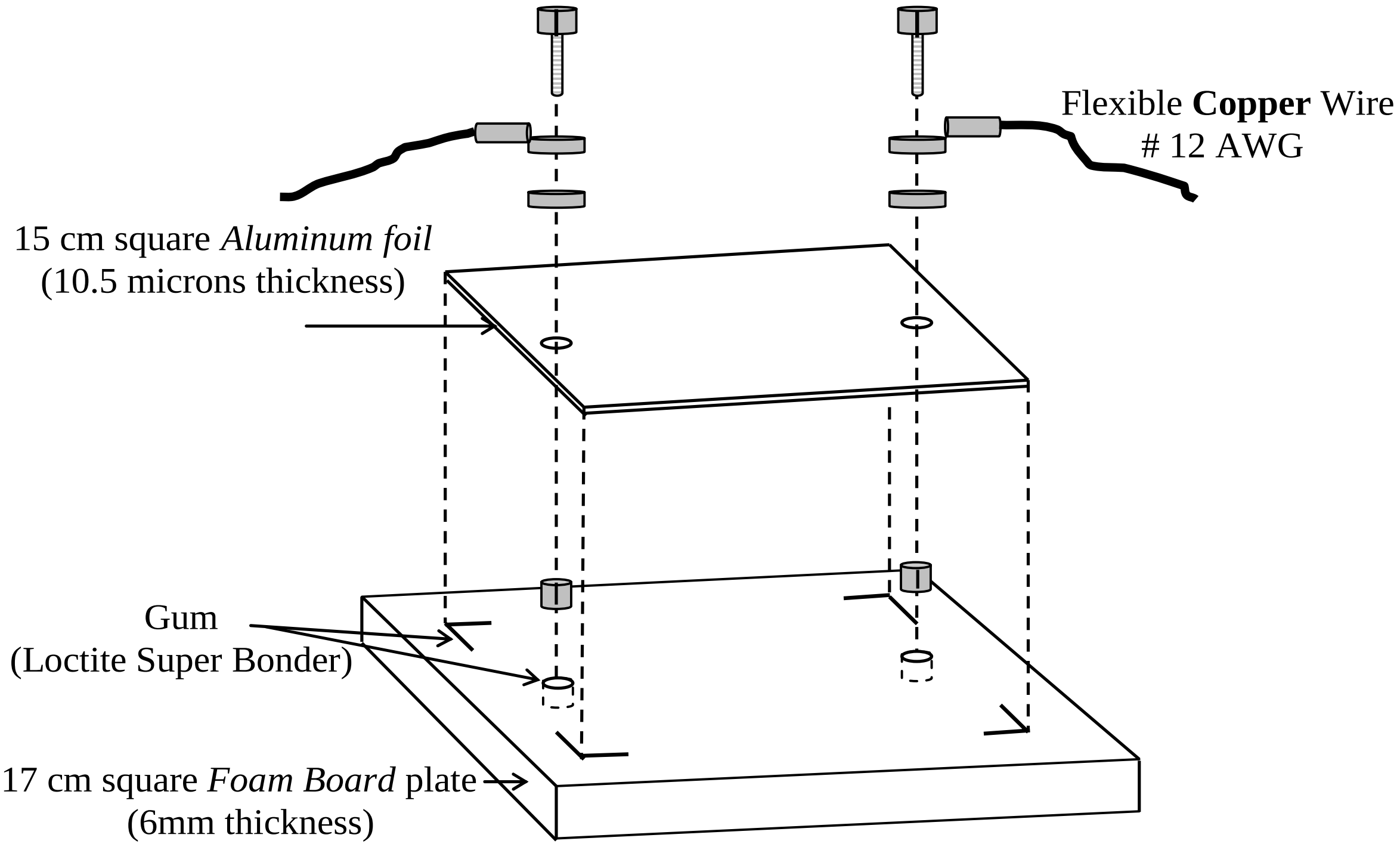

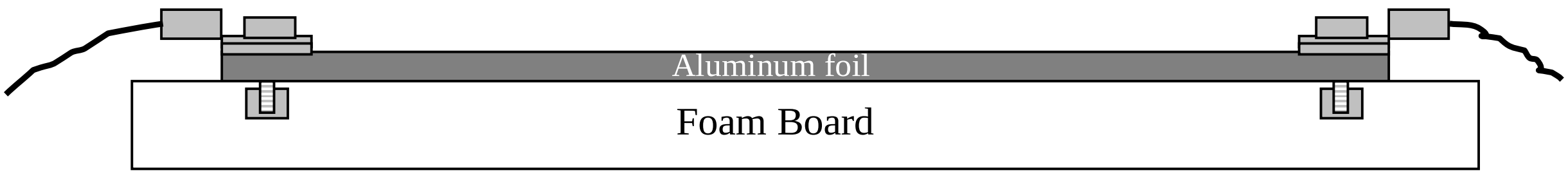

Figure A3 – The Simplest *Gravity Control Cell* (GCC).

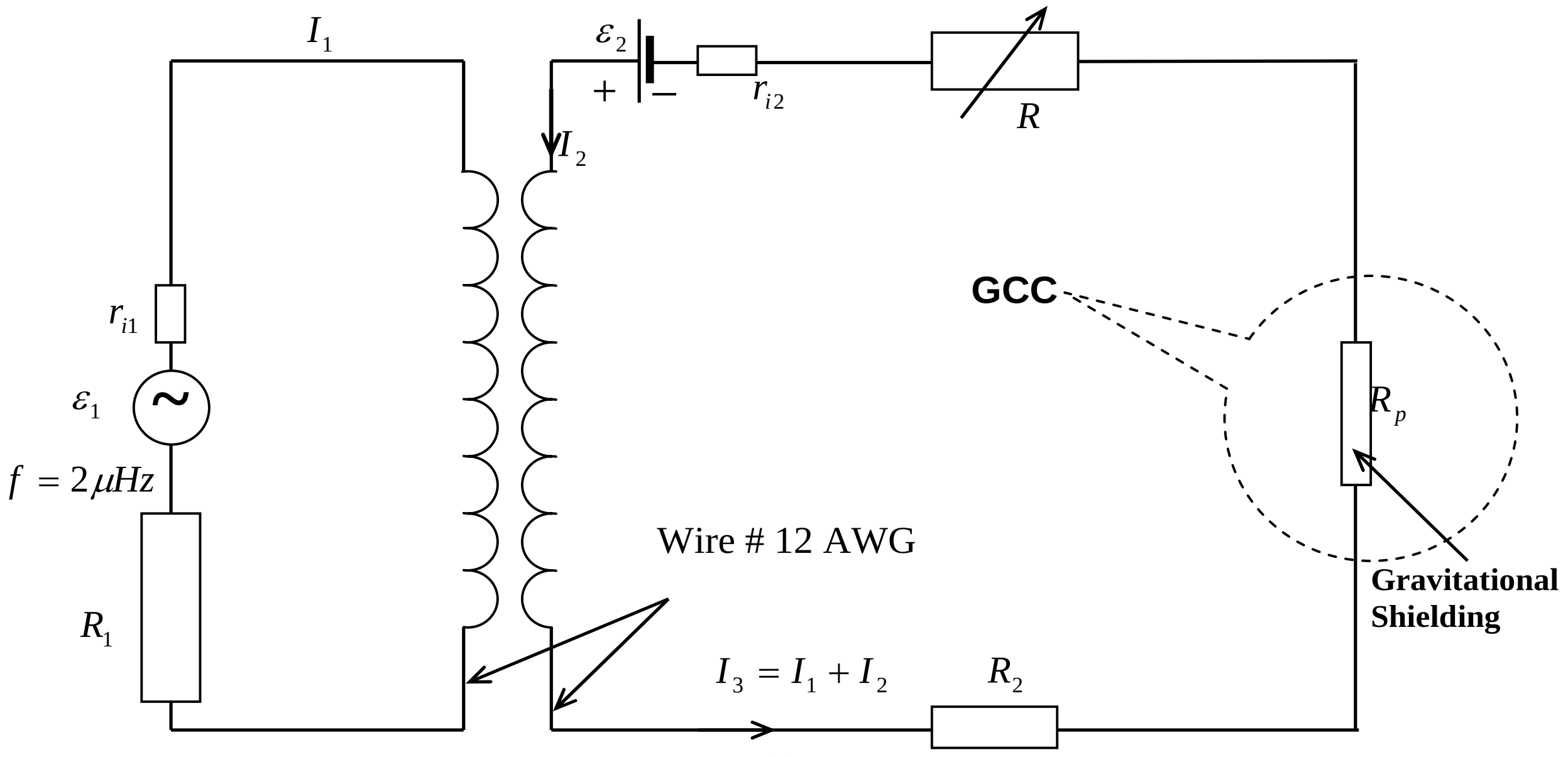

(a)

$\varepsilon_1$ = Function Generator HP3325A(*Option* 002 High Voltage Output)

$r_{i1} < 2\Omega$; $R_1 = 500\Omega - 2\ W$; $\varepsilon_2 = 12V\ DC$; $r_{i2} < 0.1\Omega$ (*Battery*);

$R_2 = 4\Omega - 40W$; $R_p = 2.5\times10^{-3}\Omega$; $Reostat = 0 \le R \le 10\Omega - 90W$

$I_1^{max} = 56mA$ $(rms)$; $I_2^{max} = 3A$ ; $I_3^{max} \cong 3A$ $(rms)$

*Coupling Transformer* to isolate the *Function Generator* from the Battery

• Air core 10-mm diameter; wire #12 AWG; $N_1 = N_2 = 20; l = 42mm$

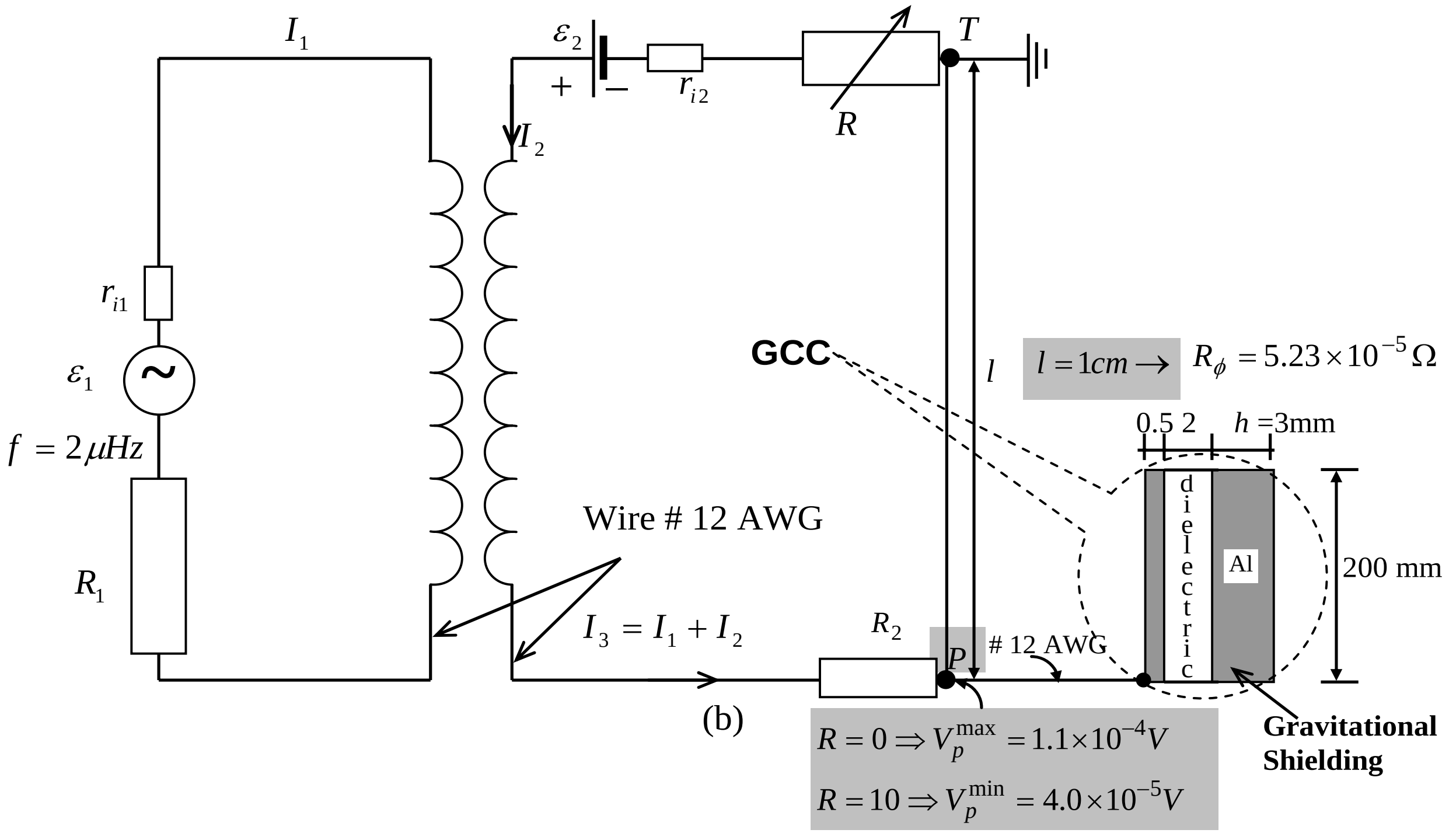

Fig. A4 – Equivalent Electric Circuits

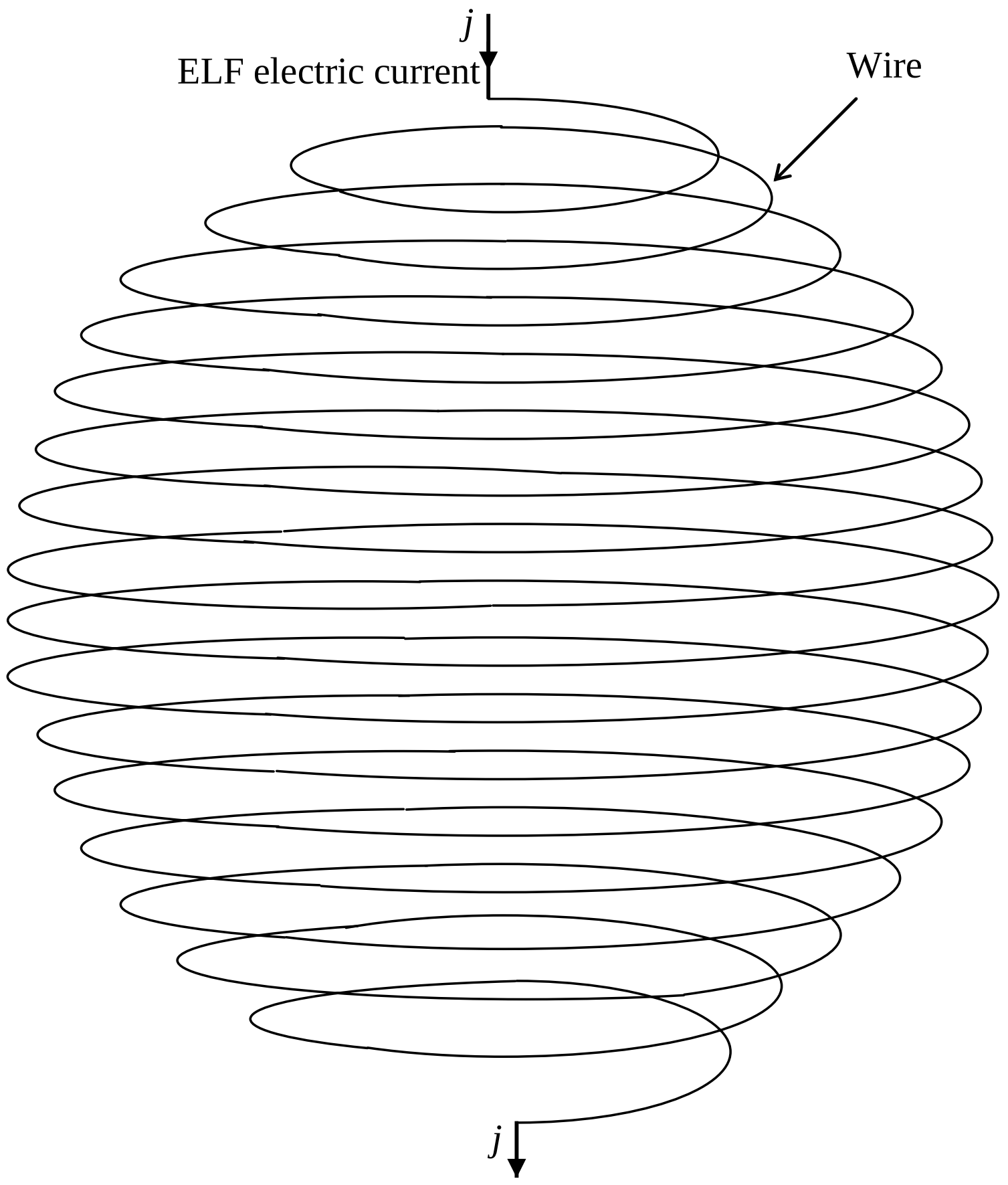

$$m_g = \left\{1 - 2\left[\sqrt{1 + 1.758\times10^{-27}\,\frac{\mu_r j^4}{\sigma\rho^2 f^3}} - 1\right]\right\} m_{i0}$$

Figure A5 – An ELF electric current through a wire, that makes a spherical form as shown above, reduces the gravitational mass of the wire and the gravity inside sphere at the same proportion $\chi = m_g / m_{i0}$ (Gravitational Shielding Effect). Note that this spherical form can be transformed into an ellipsoidal form or a disc in order to coat, for example, a Gravitational Spacecraft. It is also possible to coat with a wire several forms, such as cylinders, cones, cubes, etc. The characteristics of the wire are expressed by: $\mu_r$, $\sigma$, $\rho$; $j$ is the electric current density and $f$ is the frequency.

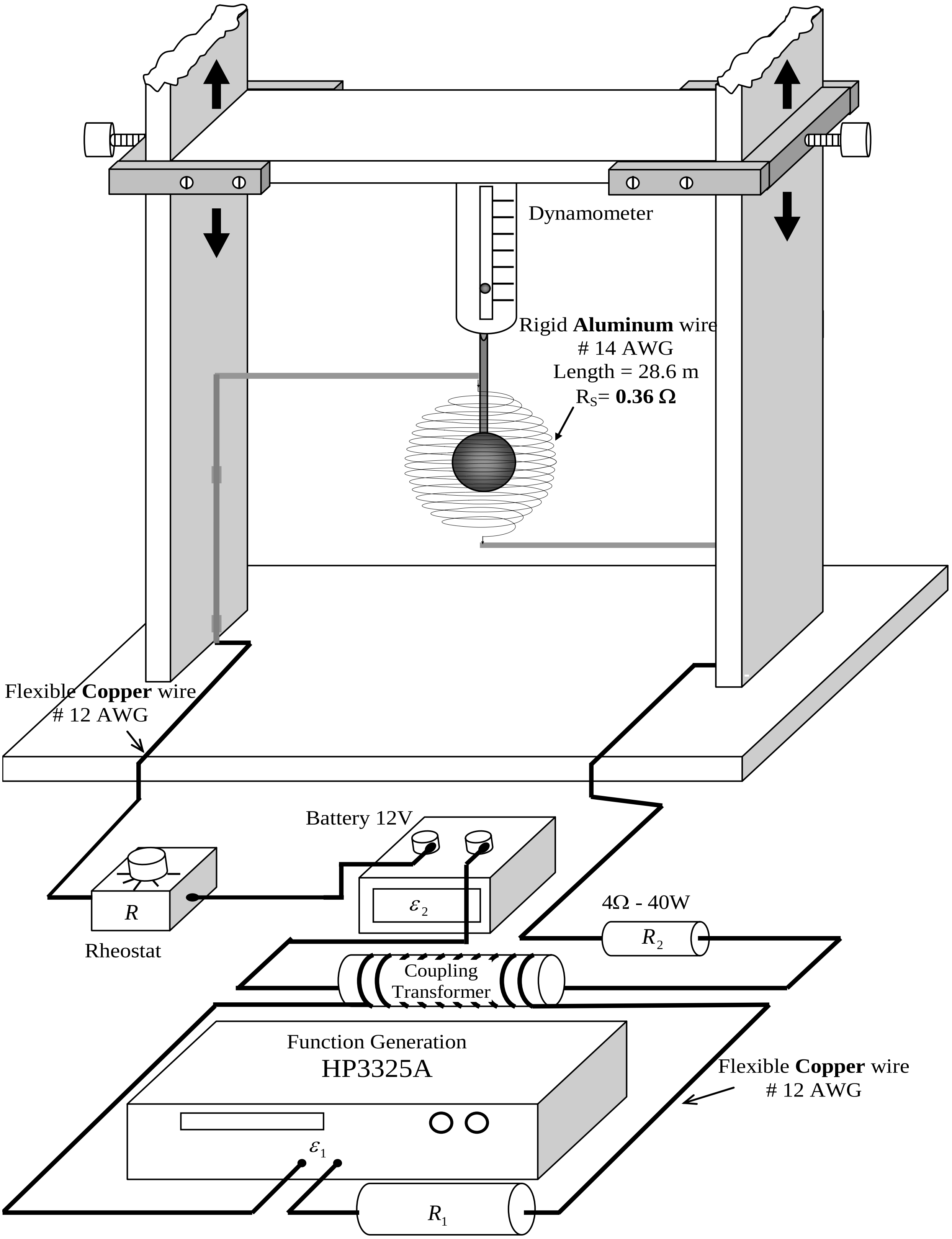

Figure A6 – Experimental set-up 2.

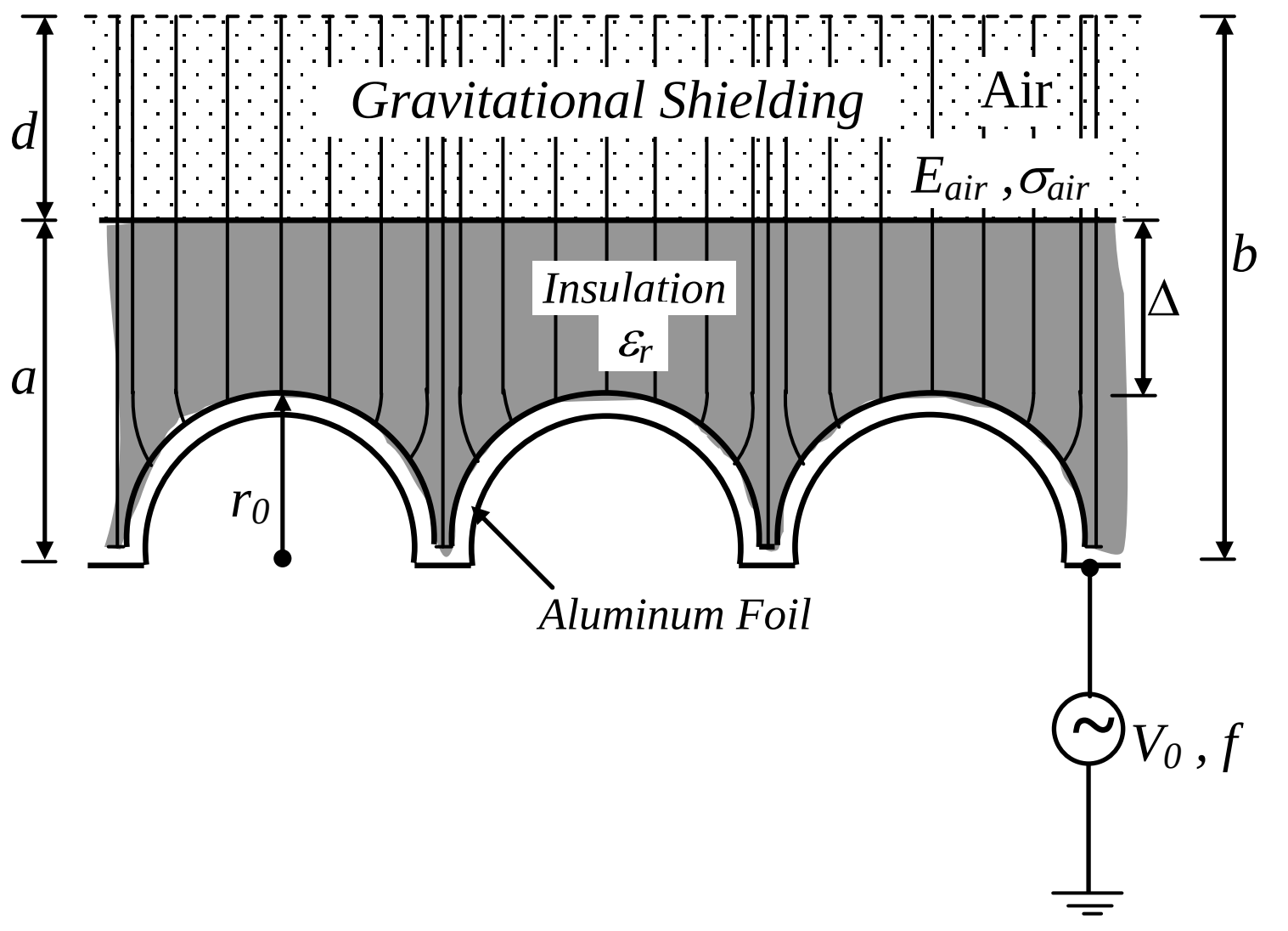

Figure A7 – *Gravitational shielding produced by semi-spheres stamped on the Aluminum foil* - By simply changing the geometry of the surface *of* the Aluminum foil it is possible to increase the working frequency $f$ up to more than *1Hz*.

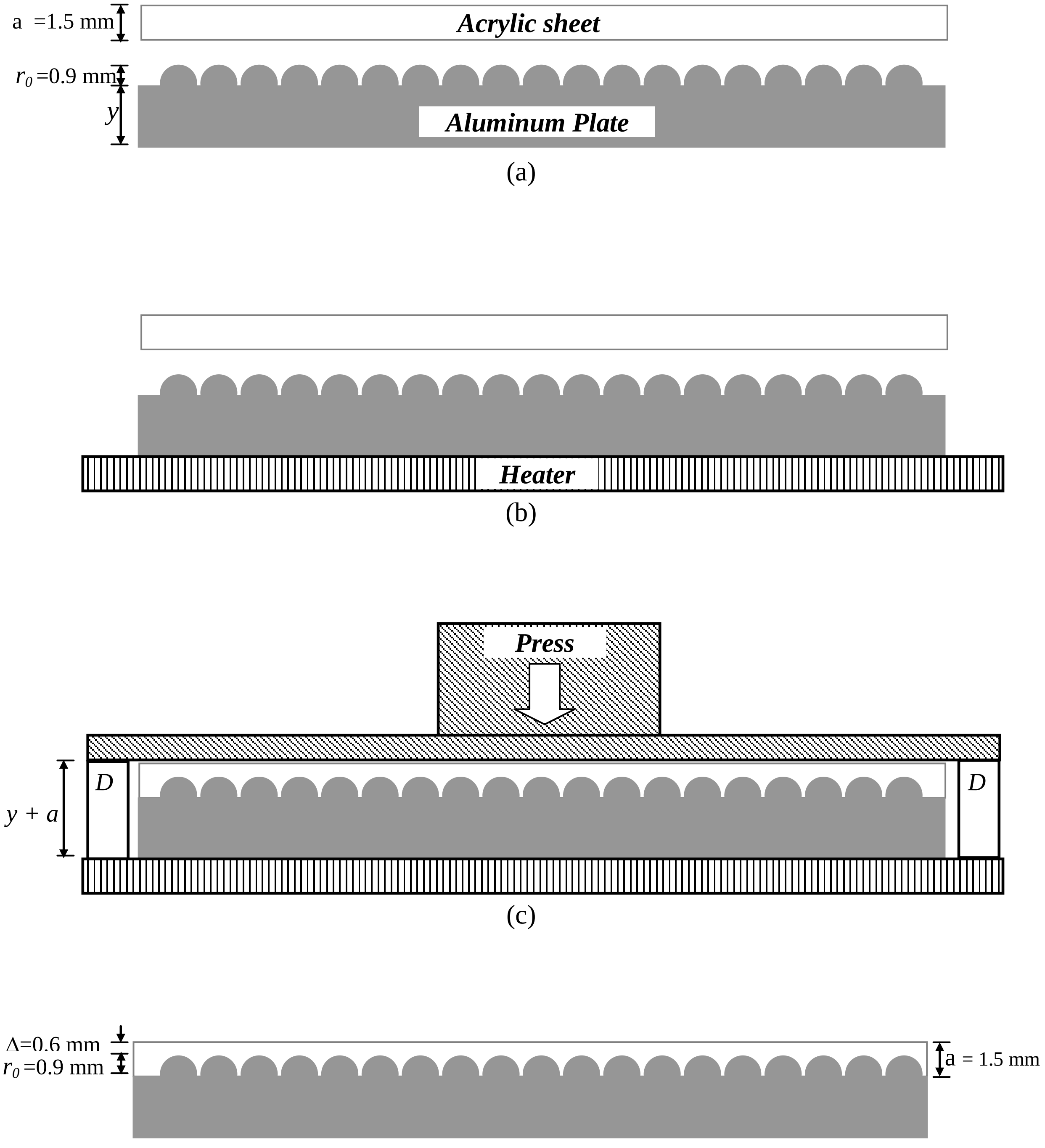

Figure A8 – *Method to coat the Aluminum semi-spheres with acrylic* $(\Delta = a - r_0 = 0.6mm)$. (a)Acrylic sheet (A4 format) with 1.5mm thickness and an Aluminum plate (A4) with several semi-spheres (radius $r_0 = 0.9\ mm$) stamped on its surface. (b)A heater is placed below the Aluminum plate in order to heat the Aluminum. (c)When the Aluminum is sufficiently heated up, the acrylic sheet and the Aluminum plate are pressed, one against the other (The two D devices shown in this figure are used in order to impede that the press compresses the acrylic and the aluminum besides distance $y + a$). (d)After some seconds, the press and the heater are removed, and the device is ready to be used.

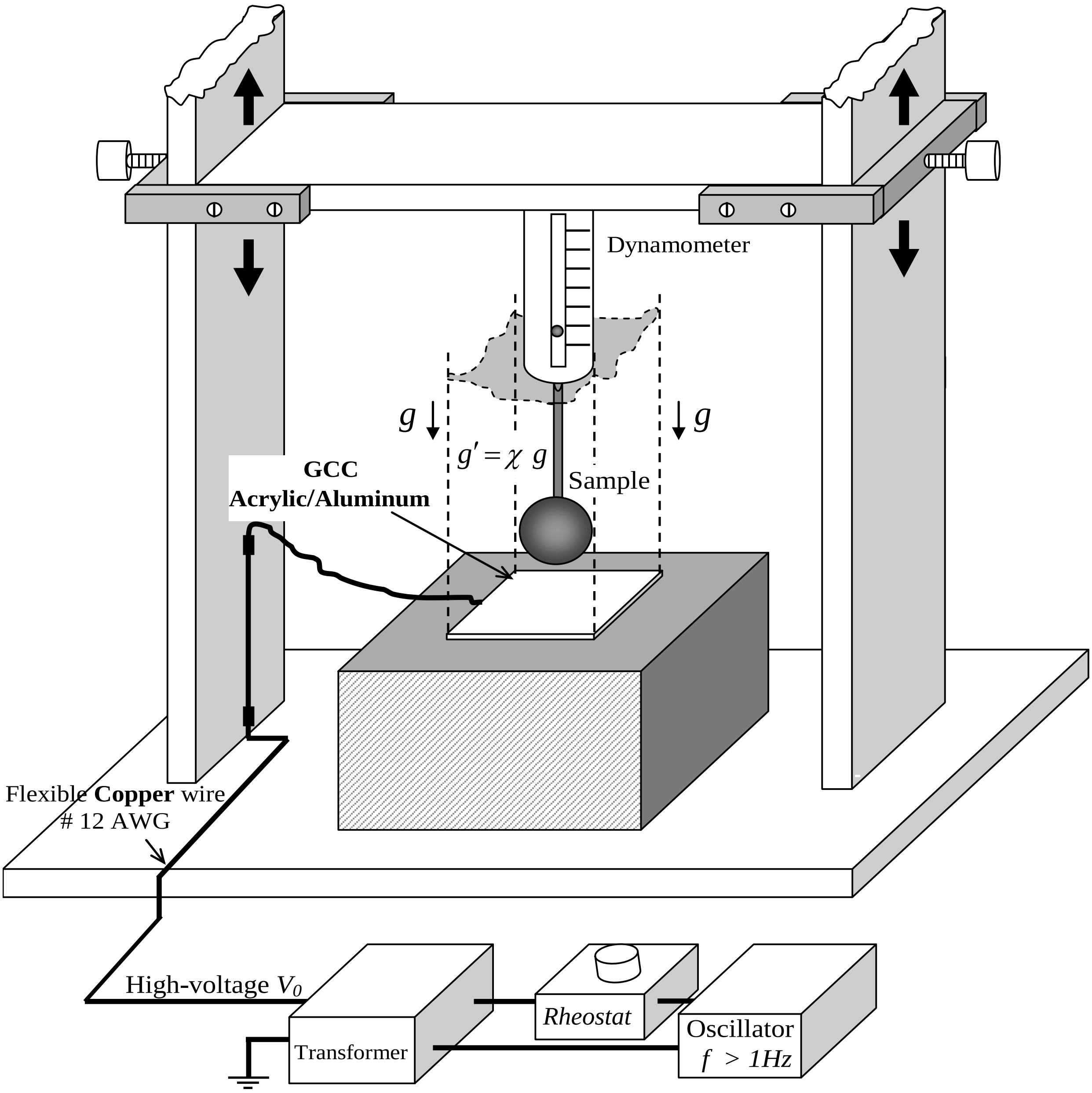

Figure A9 – *Experimental Set-up using a GCC subjected to high-voltage* $V_0$ *with frequency* $f > 1Hz$. Note that in this case, the pan balance is not necessary because the substance of the Gravitational Shielding is an *air layer* with thickness $d$ *above* the acrylic sheet. This is therefore, more a type of Gravity Control Cell (GCC) with *external gravitational shielding.*

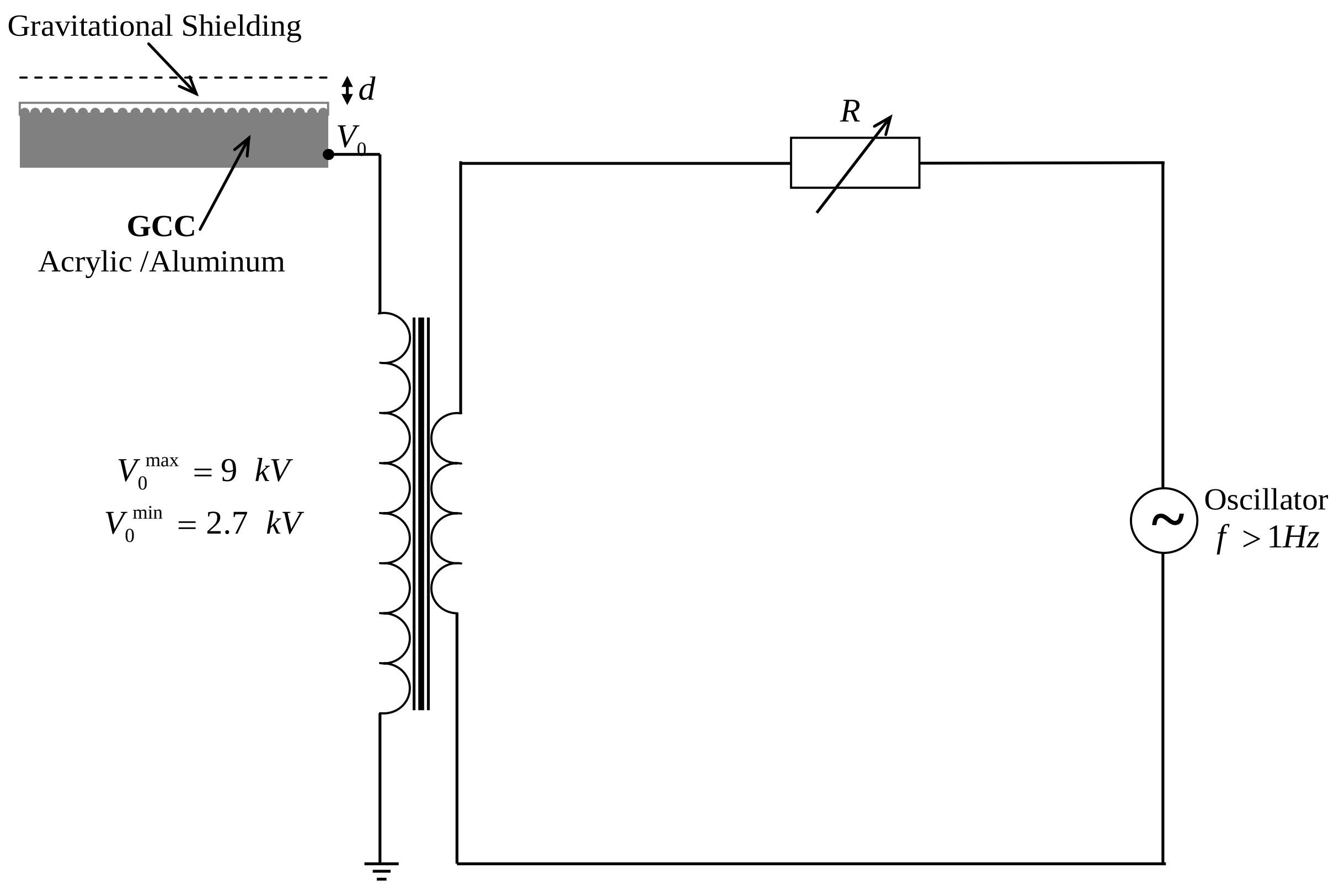

(a)

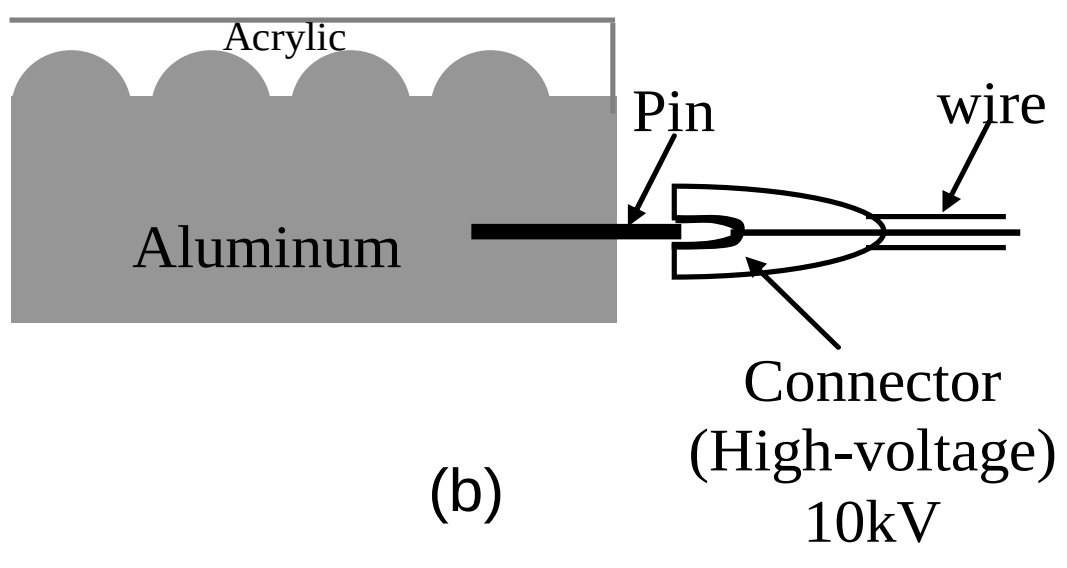

(b)

Figure A10 – (a) *Equivalent Electric Circuit.* (b) Details of the electrical connection with the Aluminum plate. Note that others connection modes (by the top of the device) can produce destructible interference on the electric lines of the $E_{air}$ field.

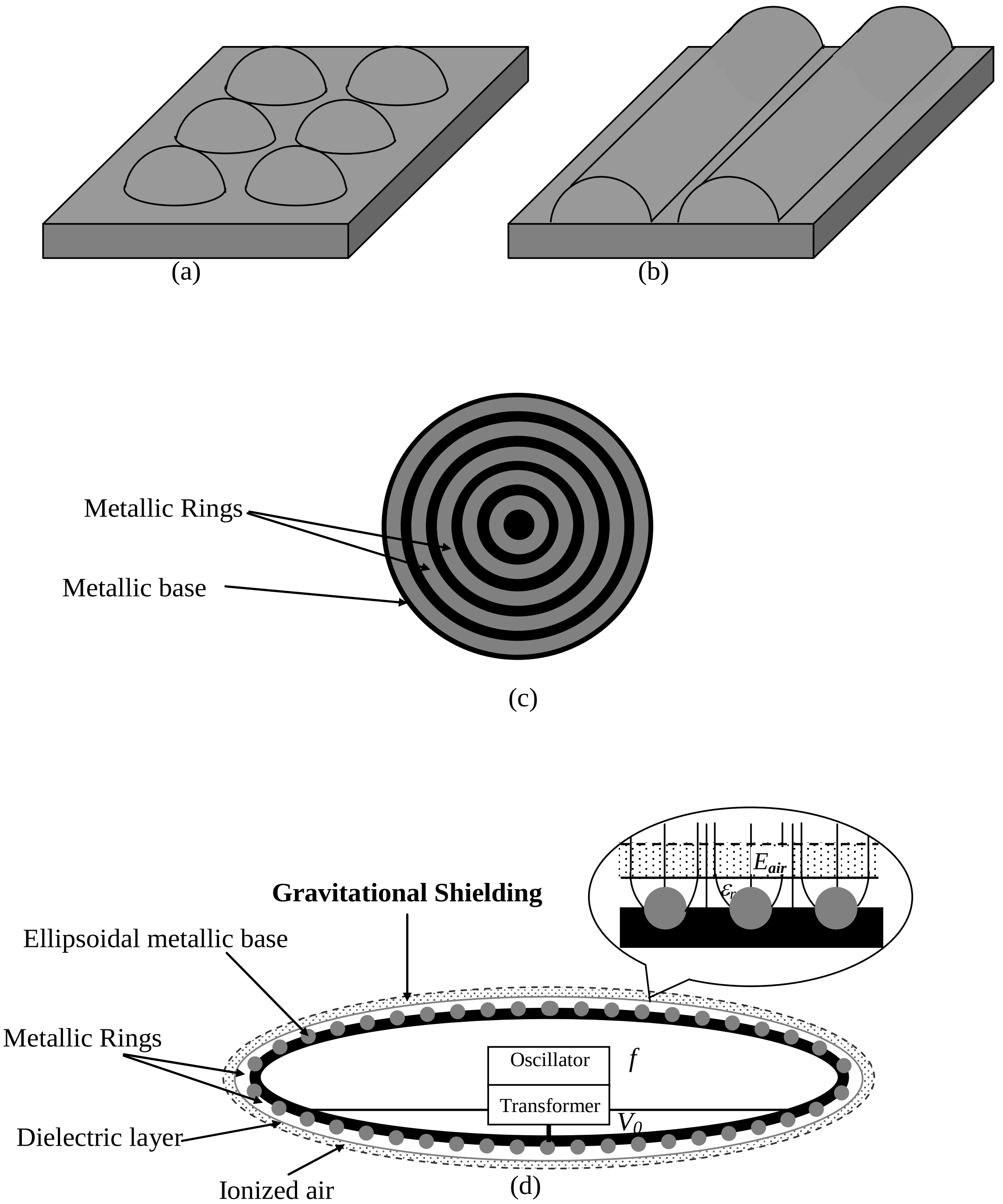

Figure A11 – *Geometrical forms with similar effects as those produced by the semi-spherical form* – (a) shows the semi-spherical form stamped on the metallic surface; (b) shows the *semi-cylindrical* form (an obvious evolution from the semi-spherical form); (c) shows *concentric metallic rings* stamped on the metallic surface, an evolution from semi-cylindrical form. These geometrical forms produce the same effect as that of the semi-spherical form, shown in Fig.A11 (a). By using concentric metallic rings, it is possible to build *Gravitational Shieldings* around bodies or spacecrafts with several formats (spheres, ellipsoids, etc); (d) shows a Gravitational Shielding around a Spacecraft with *ellipsoidal form*.

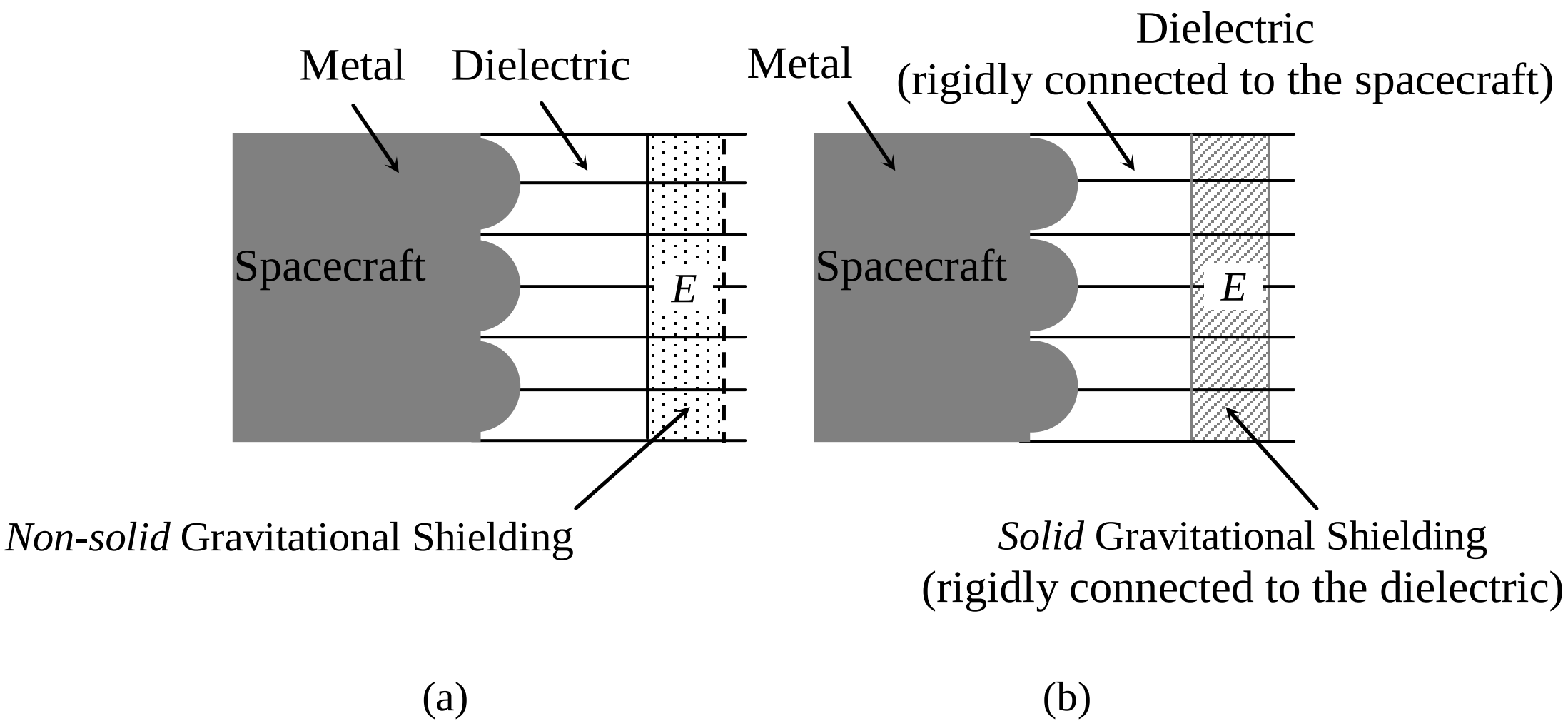

Figure A12 – *Non-solid and Solid Gravitational Shieldings* - In the case of the Gravitational Shielding produced on a *solid substance* (b), when its molecules go to the *imaginary* space-time, *the electric field that produces the effect also goes to the imaginary space-time together with them*, because in this case, the substance of the Gravitational Shielding is *rigidly connected (by means of the dielectric) to the metal* that produces the electric field. This does not occur in the case of *Air* Gravitational Shielding.

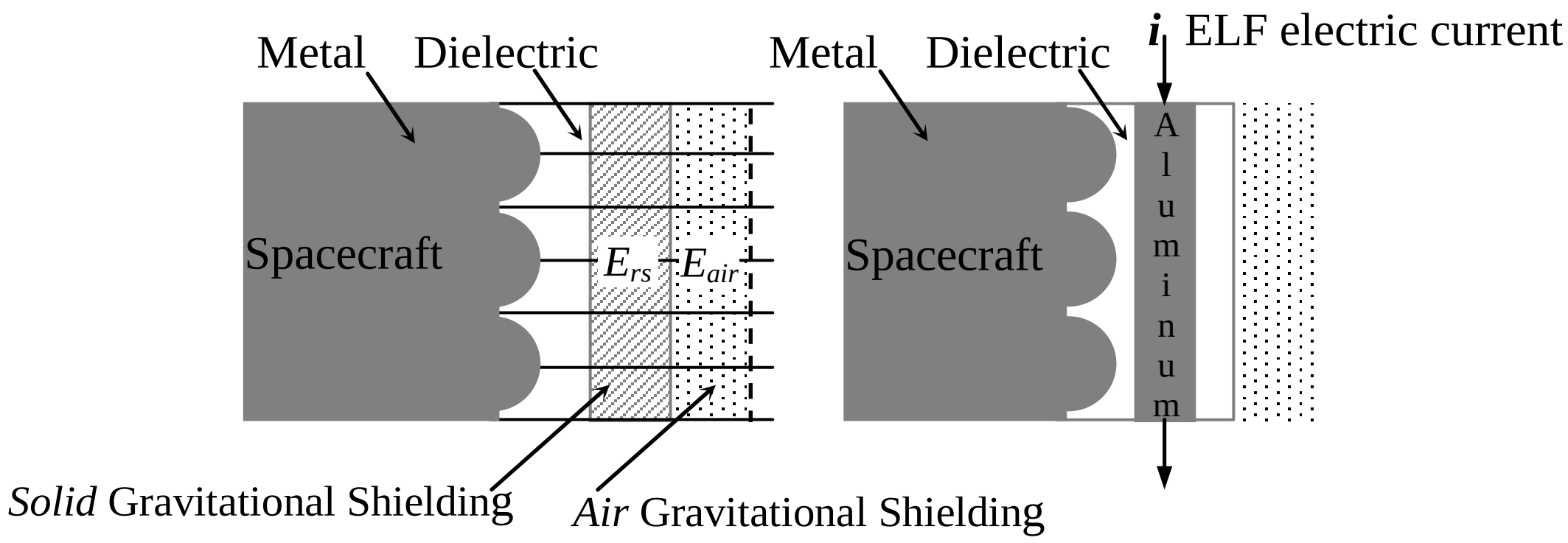

(a)

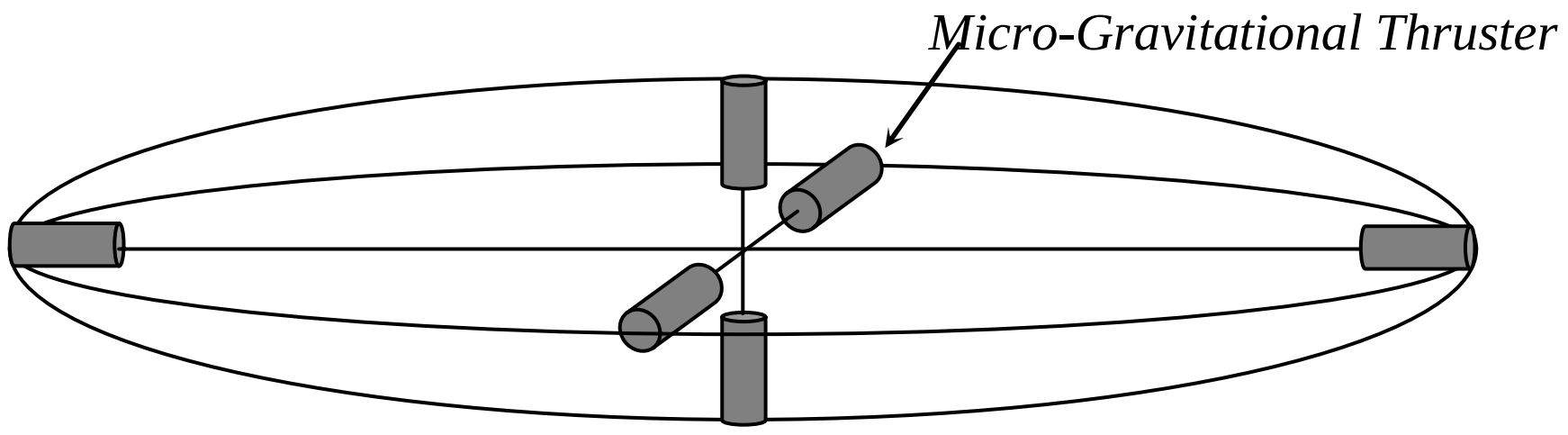

(b)

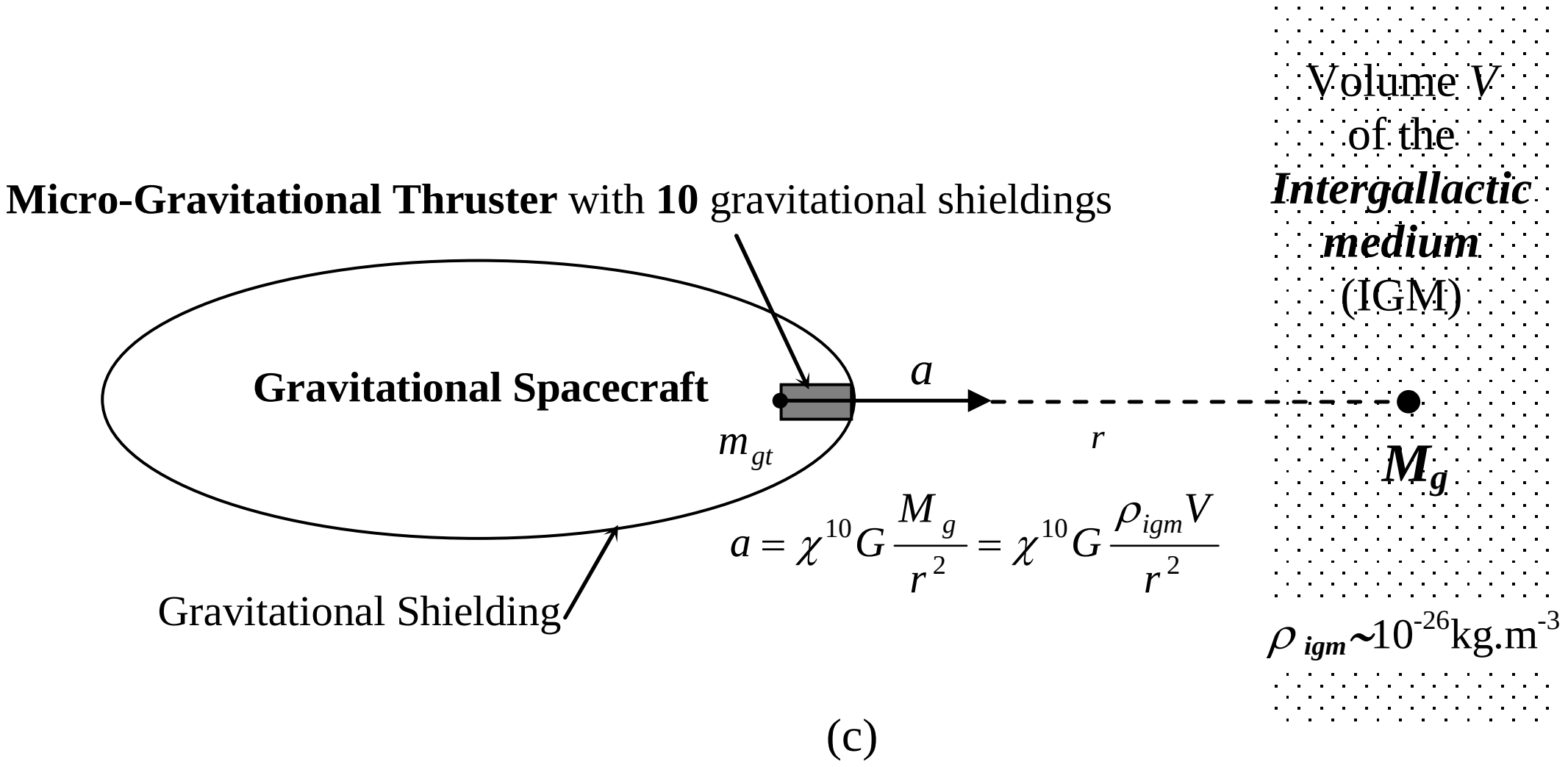

(c)

Figure A13 – *Double Gravitational Shielding and Micro-thrusters* – (a) Shows a double gravitational shielding that makes possible to decrease the *inertial effects* upon the spacecraft when it is traveling both in the *imaginary* space-time and in the *real* space-time. The *solid* Gravitational Shielding also can be obtained by means of *an ELF electric current through a metallic lamina* placed *between the semi-spheres and the Gravitational Shielding of Air* as shown above. (b) Shows 6 *micro-thrusters* placed inside a Gravitational Spacecraft, in order to propel the spacecraft in the directions x, y and z. Note that the Gravitational Thrusters in the spacecraft must have a very small diameter (of the order of *millimeters*) because the hole through the Gravitational Shielding of the spacecraft cannot be large. Thus, these thrusters are in fact *Micro-thrusters*. (c) Shows a micro-thruster inside a spacecraft, and in front of a volume *V* of the intergalactic medium (IGM). Under these conditions, the spacecraft acquires an acceleration *a* in the direction of the volume *V*.

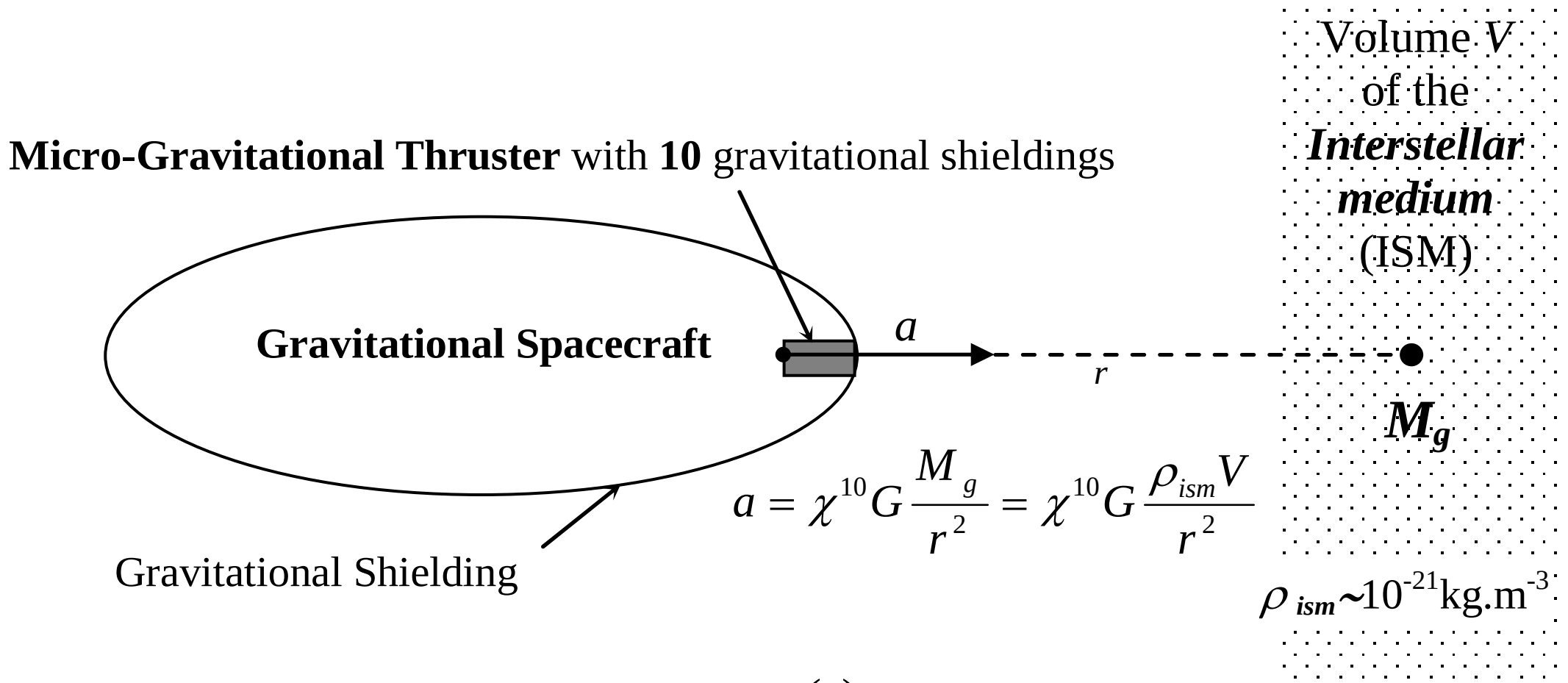

(a)

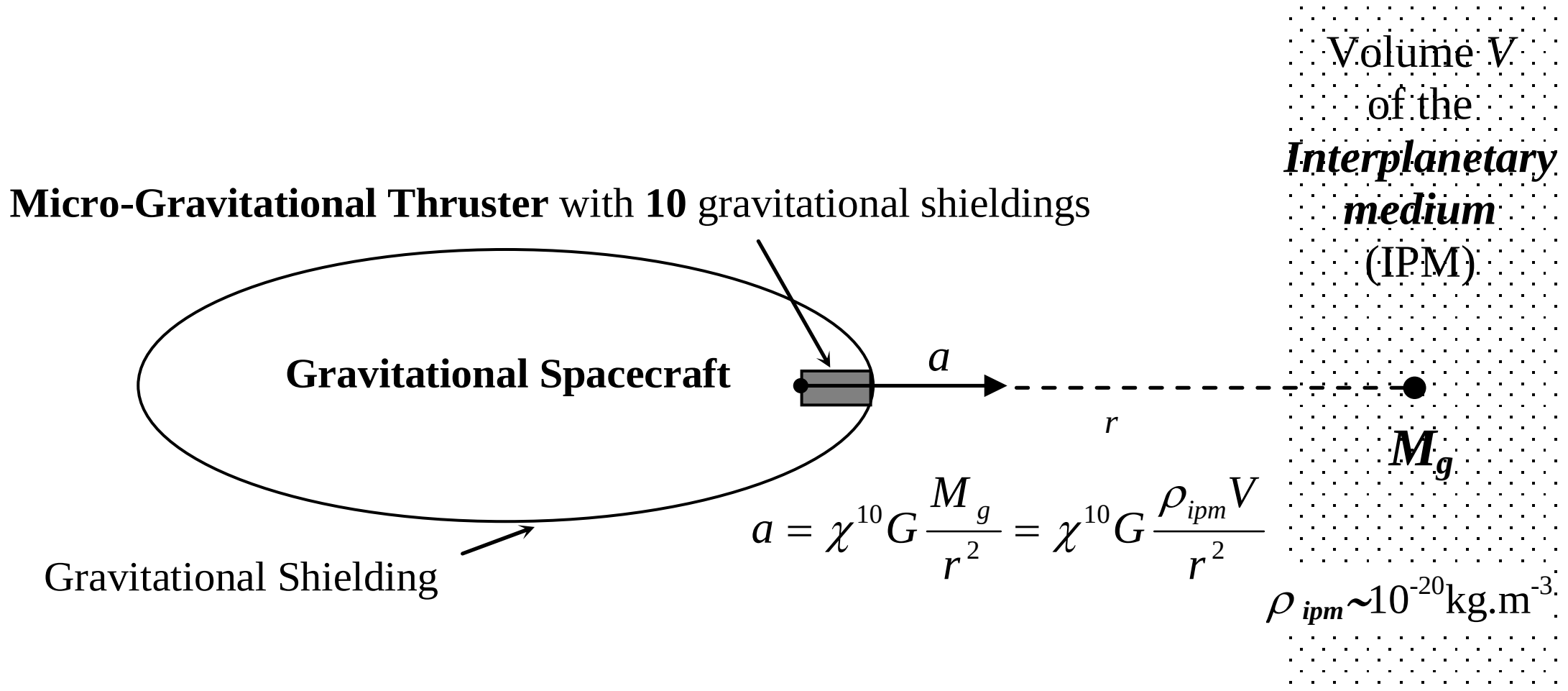

(b)

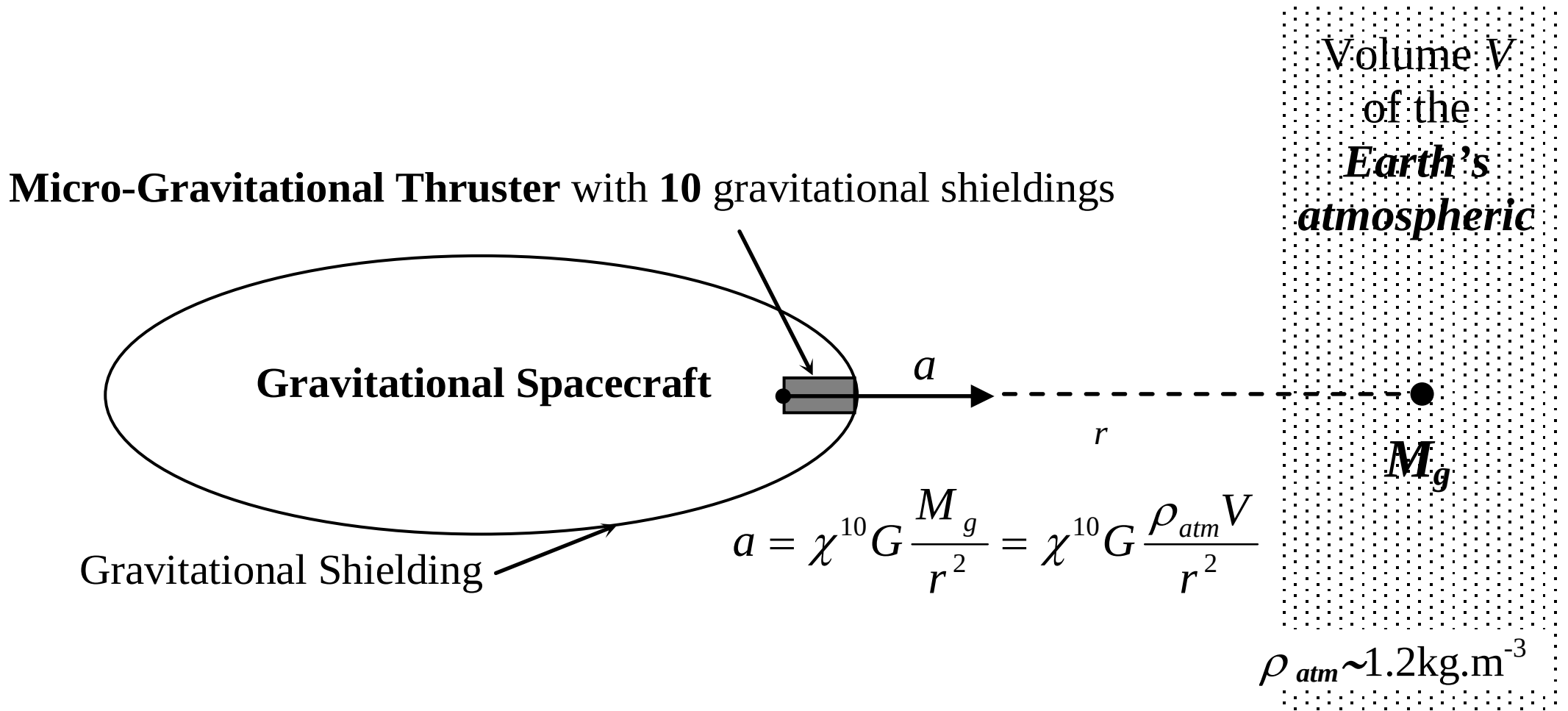

(c)

Figure A14 – *Gravitational Propulsion using Micro-Gravitational Thruster* – (a) Gravitational acceleration produced by a gravitational mass $M_g$ of the *Interstellar Medium*. The density of the Interstellar Medium is about $10^5$ times greater than the density of the *Intergalactic Medium* (b) Gravitational acceleration produced in the *Interplanetary Medium*. (c) Gravitational acceleration produced in the *Earth's atmosphere*. Note that, in this case, $\rho_{atm}$ (*near to the Earth's surface*)is about $10^{26}$ times greater than the density of the *Intergalactic Medium.*

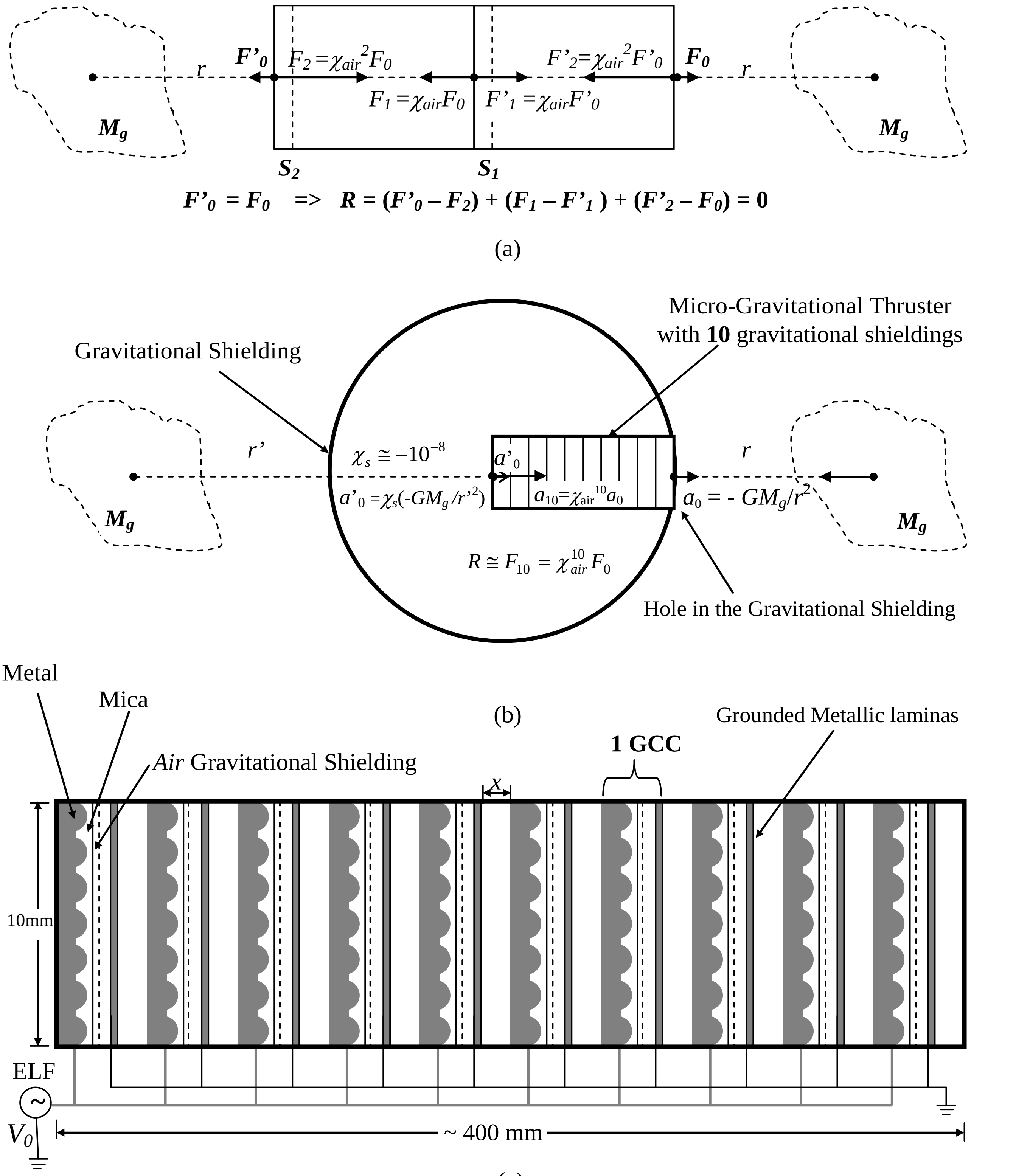

(c)

Figure A15 – *Dynamics and Structure of the Micro-Gravitational Thrusters* - (a) The Micro-Gravitational Thrusters do not work *outside* the Gravitational Shielding, because, in this case, *the resultant upon the thruster is null* due to the symmetry. (b) The Gravitational Shielding $(\chi_s \cong 10^{-8})$ reduces strongly the intensities of the gravitational forces acting on the micro-gravitational thruster, except obviously, through the hole in the gravitational shielding. (c) Micro-Gravitational Thruster with *10 Air Gravitational Shieldings* (10GCCs). The grounded metallic laminas are placed so as to retain the electric field produced by metallic surface behind the semi-spheres.

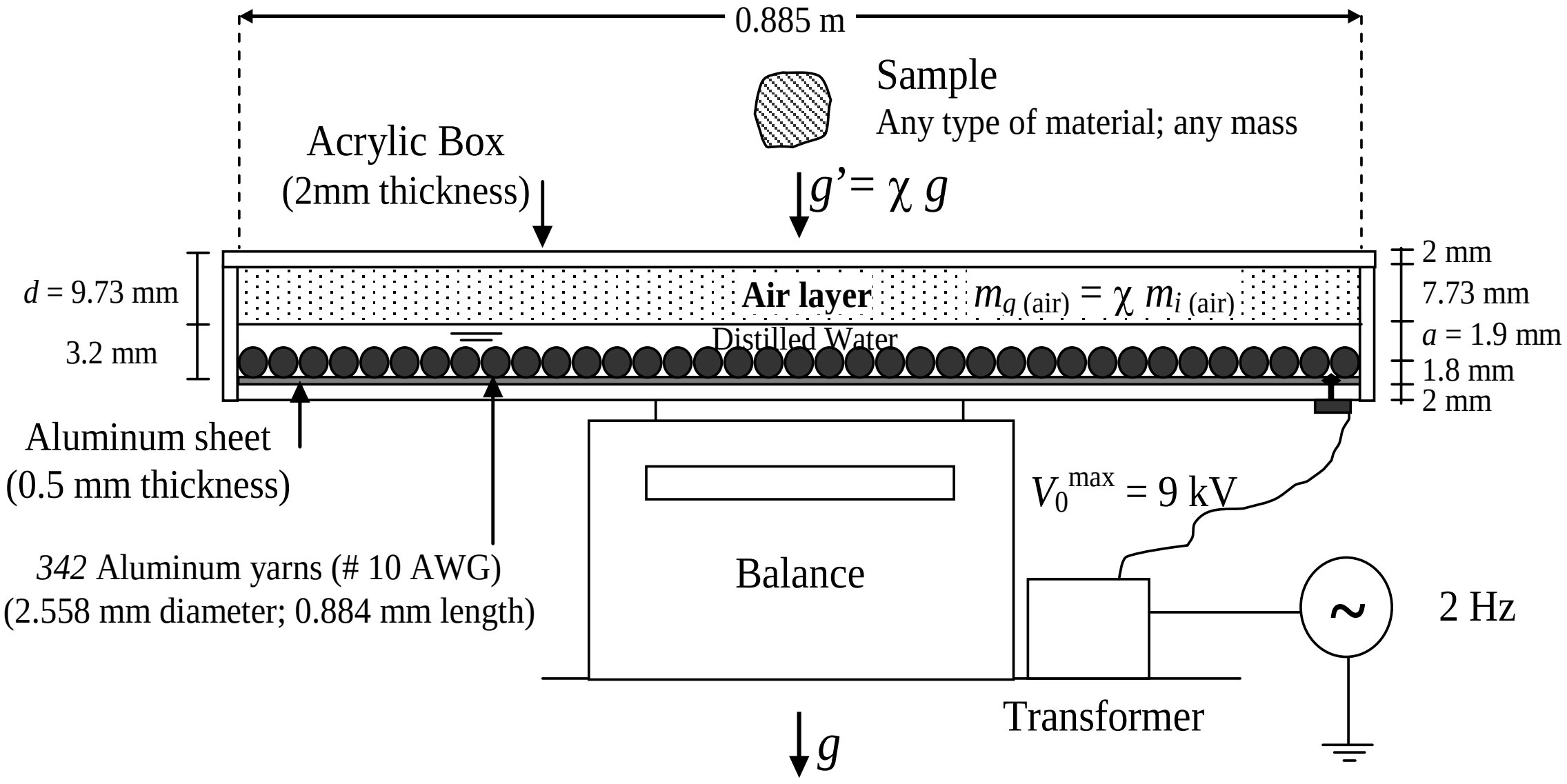

**GCC Cross-section Front view**

(a)

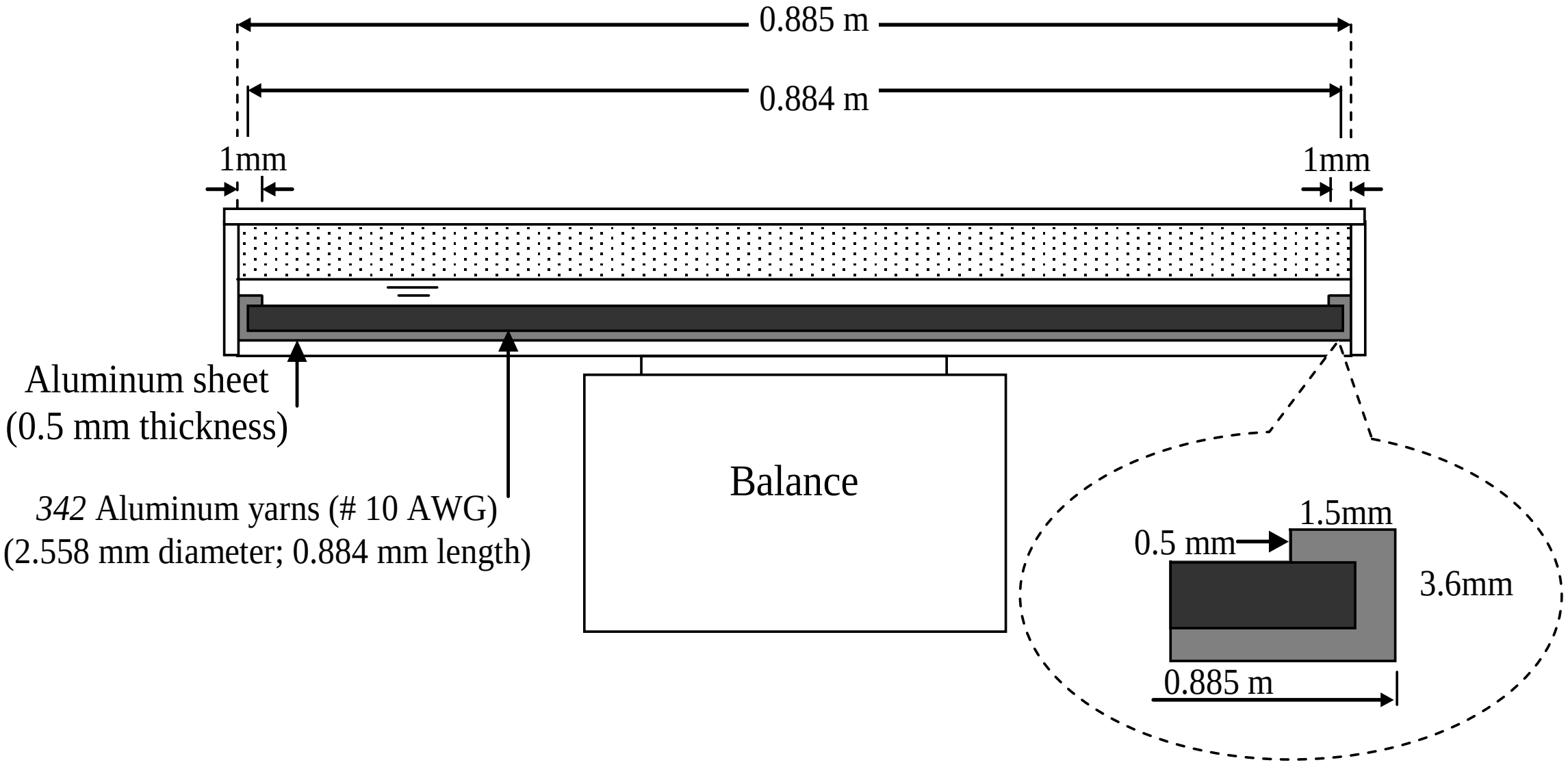

**GCC Cross-section Side View**

(b)

Fig. A16 – *A GCC using distilled Water*.

In total this GCC weighs about 6kg; the air layer 7.3 grams. The balance has the following characteristics: Range 0 – 6kg; readability 0.1g. The yarns are inserted side by side on the Aluminum sheet. Note the detail of fixing of the yarns on the Aluminum sheet.

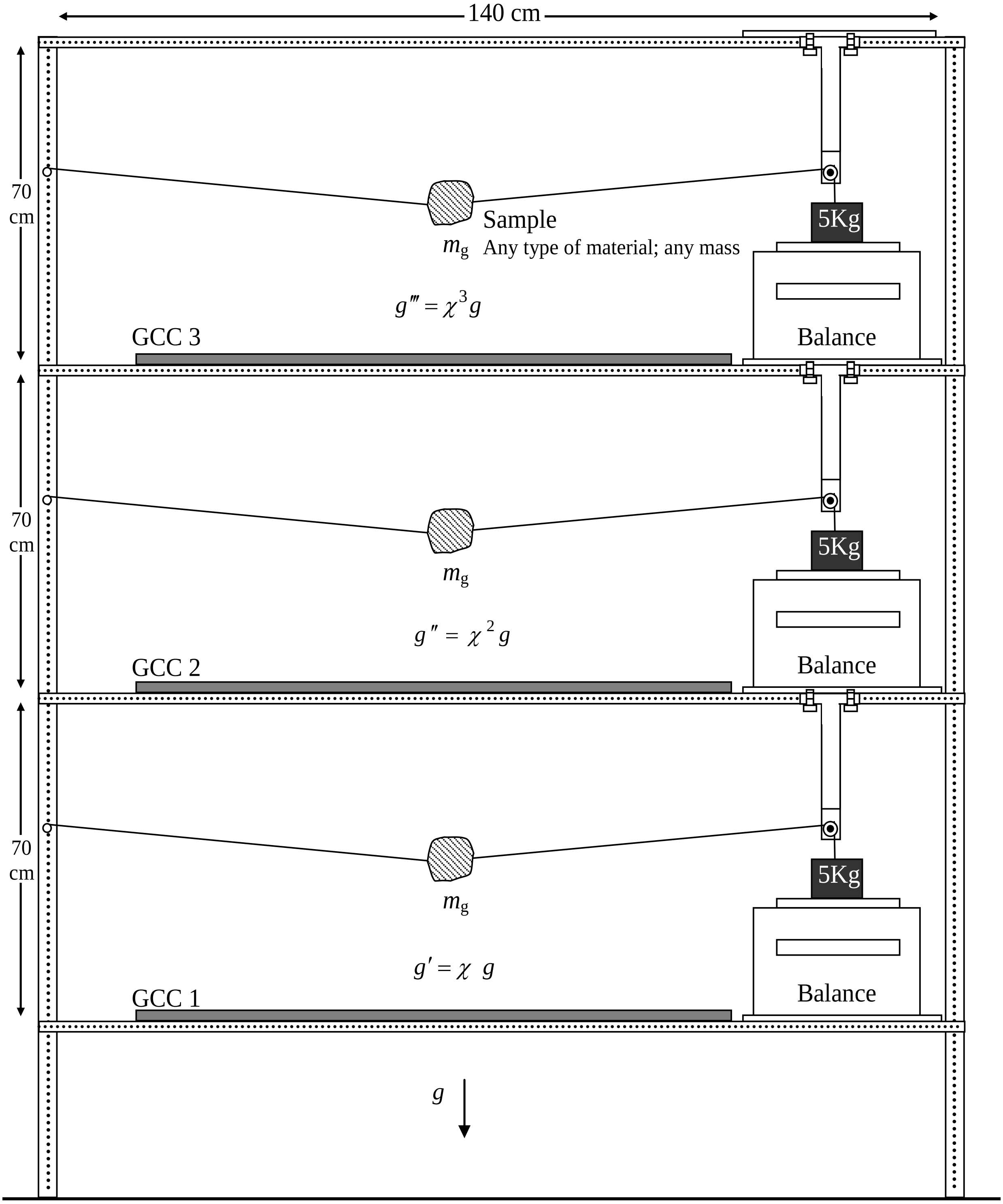

Fig. A17 – *Experimental set-up.* In order to prove *the exponential effect* produced by the superposition of the Gravitational Shieldings, we can take three similar GCCs and put them one above the other, in such way that above the GCC 1 the gravity acceleration will be $g' = \chi\, g$; above the GCC2 $g'' = \chi^2 g$, and above the GCC3 $g''' = \chi^3 g$. Where $\chi$ is given by Eq. (A47). The arrangement above has been designed for values of $m_g < 13g$ and $\chi$ up to -9 or $m_g < 1kg$ and $\chi$ up to -2 .

## APPENDIX B: A DIDACTIC *GCC* USING A BATTERY OF CAPACITORS

Let us now show a new type of GCC - easy to be built with materials and equipments that also can be obtained with easiness.

Consider a battery of $n$ parallel plate capacitors with capacitances $C_1, C_2, C_3, ..., C_n$, connected in parallel. The voltage applied is $V$; $A$ is the area of each plate of the capacitors and $d$ is the distance between the plates; $\varepsilon_{r(water)}$ is the relative permittivity of the dielectric (water). Then the electric charge $q$ on the plates of the capacitors is given by

$$q = (C_1 + C_2 + C_3 + ... + C_n)V = n(\varepsilon_{r(water)}\varepsilon_0)\frac{A}{d}V \qquad (B1)$$

In Fig. I we show a GCC with *two* capacitors connected in parallel. It is easy to see that the electric charge density $\sigma_0$ on each area $A_0 = az$ of the edges B of the thin laminas (*z* is the thickness of the edges B and *a* is the length of them, see Fig.B2) is given by

$$\sigma_0 = \frac{q}{A_0} = n(\varepsilon_{r(water)}\varepsilon_0)\frac{A}{azd}V \qquad (B2)$$

Thus, the electric field $E$ between the edges B is

$$E = \frac{2\sigma_0}{\varepsilon_{r(air)}\varepsilon_0} = 2n\left(\frac{\varepsilon_{r(water)}}{\varepsilon_{r(air)}}\right)\frac{A}{azd}V \qquad (B3)$$

Since $A = L_x L_y$, we can write that

$$E = 2n\left(\frac{\varepsilon_{r(water)}}{\varepsilon_{r(air)}}\right)\frac{L_x L_y}{azd}V \qquad (B4)$$

Assuming $\varepsilon_{r(water)} = 81$ **** (bidistilled water); $\varepsilon_{r(air)} \cong 1$ (vacuum $10^{-4}$ Torr; 300K); $n = 2$; $L_x = L_y = 0.30m$; $a = 0.12m$; $z = 0.1mm$ and $d = 10mm$ we obtain

$$E = 2.43 \times 10^8 V$$

For $V_{max} = 220V$, the electric field is

$$E_{max} = 5.3 \times 10^{10} V/m$$

Therefore, if the frequency of the wave voltage is $f = 60Hz$, $(\omega = 2\pi f)$, we have that $\omega\varepsilon_{air} = 3.3 \times 10^{-9} S.m^{-1}$. It is known that the electric conductivity of the air, $\sigma_{air}$, at $10^{-4}$ Torr and 300K, is much smaller than this value, i.e.,

$$\sigma_{air} << \omega\varepsilon_{air}$$

Under this circumstance $(\sigma << \omega\varepsilon)$, we can substitute Eq. 15 and 34 into Eq. 7. Thus, we get

$$m_{g(air)} = \left\{1 - 2\left[\sqrt{1 + \frac{\mu_{air}\varepsilon_{air}^3}{c^2}\frac{E^4}{\rho_{air}^2}} - 1\right]\right\} m_{i0(air)}$$

$$= \left\{1 - 2\left[\sqrt{1 + 9.68 \times 10^{-57}\frac{E^4}{\rho_{air}^2}} - 1\right]\right\} m_{i0(air)} \qquad (B5)$$

The density of the air at $10^{-4}$ Torr and 300K is

$$\rho_{air} = 1.5 \times 10^{-7} kg.m^{-3}$$

Thus, we can write

$$\chi = \frac{m_{g(air)}}{m_{i(air)}} =$$

$$= \left\{1 - 2\left[\sqrt{1 + 4.3 \times 10^{-43} E^4} - 1\right]\right\} \qquad (B6)$$

Substitution of $E$ for $E_{max} = 5.3 \times 10^{10} V/m$ into this equation gives

$$\chi_{max} \cong -1.2$$

This means that, in this case, the *gravitational shielding* produced in the vacuum between the edges B of the thin laminas can reduce the local gravitational acceleration $g$ down to

$$g_1 \cong -1.2g$$

Under these circumstances, the weight, $P = +m_g g$, of any body just *above* the gravitational shielding becomes

$$P = m_g g_1 = -1.2 m_g g$$

---

****It is easy to see that by substituting the water for Barium Titanate ($BaTiO_3$) the dimensions $L_x, L_y$ of the capacitors can be strongly reduced due to $\varepsilon_{r(BaTiO3)} = 1200$.

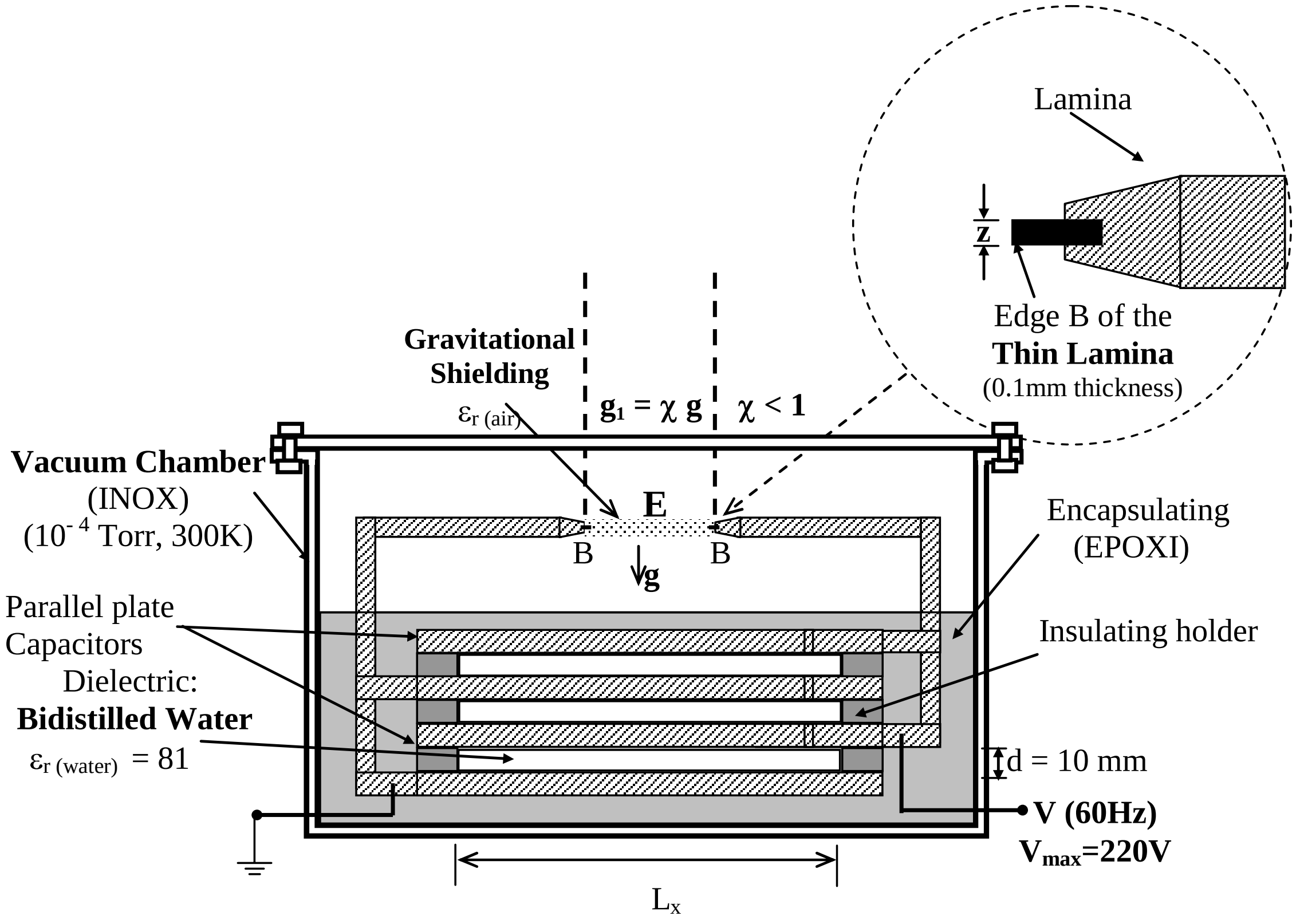

**Vertical Cross Section**

$$q = (C_1 + C_2 + ... + C_n)\,V =$$
$$= n\,[\varepsilon_{r\,(water)} / \varepsilon_{r\,(air)}]\,[A/A_0]\,V/d$$

$$\varepsilon_{r\,(water)} = 81 \;\; ; \;\; \varepsilon_{r\,(air)} \cong 1$$

$$\mathbf{E} = [q/A_0] / \varepsilon_{r\,(air)}\,\varepsilon_0 = n\,[\varepsilon_{r\,(water)} / \varepsilon_{r\,(air)}]\,[A/A_0]\,V/d$$

A is the area of the plates of the capacitors and $A_0$ the cross section area of the edges B of the thin laminas (z is the thickness of the edges).

Figure B1 – **Gravity Control Cell** (GCC) using a *battery of capacitors*. According to Eq. 7 , the electric field, **E**, through the air at $10^{-4}$ Torr; 300K, in the vacuum chamber, produces a gravitational shielding effect. The gravity acceleration above this gravitational shielding is reduced to $\chi \mathbf{g}$ where $\chi < 1$.

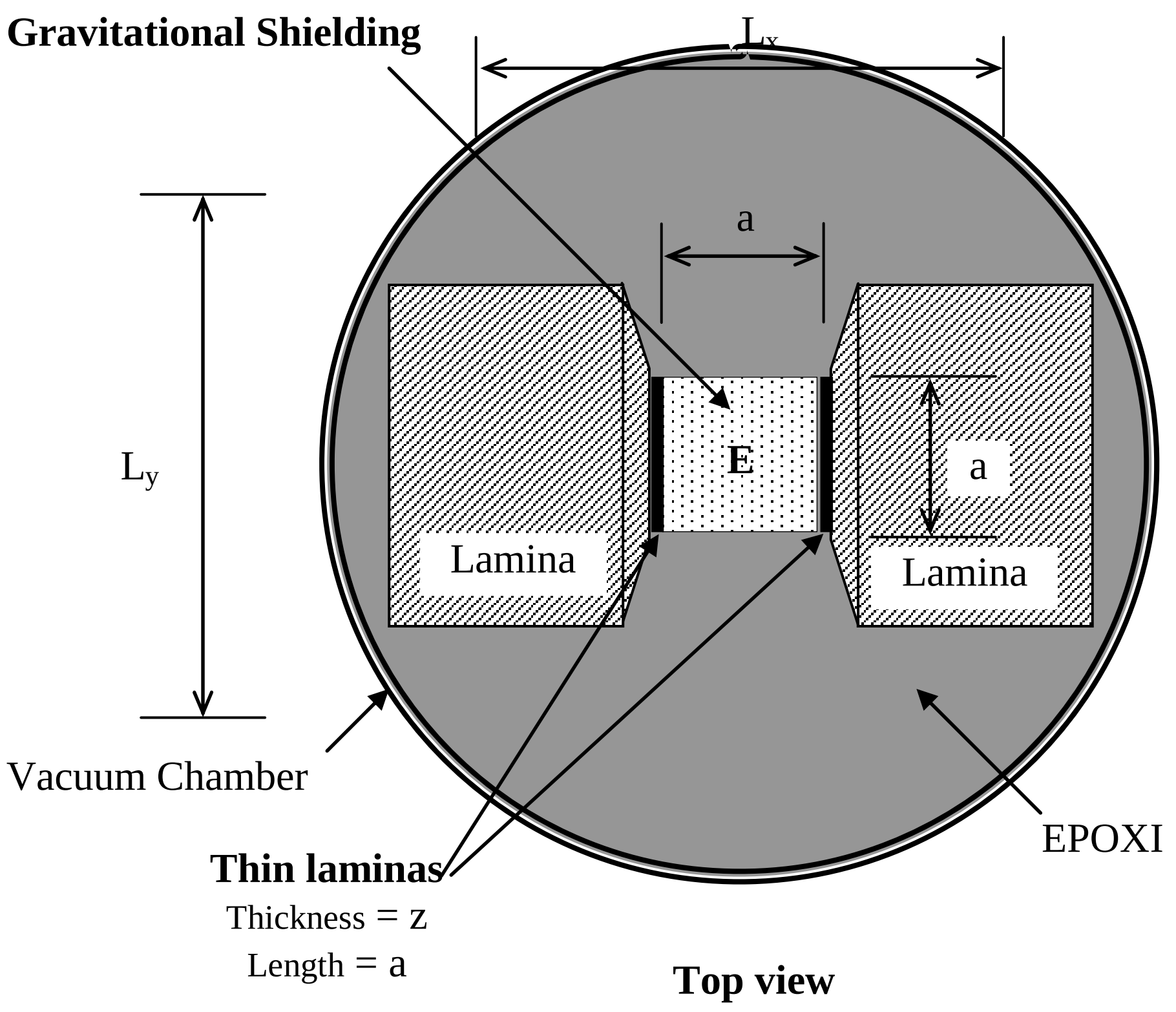

$A_0 = a\ z\ ;\quad A = L_x\ L_y$

Figure B2 – The gravitational shielding produced between the thin laminas.

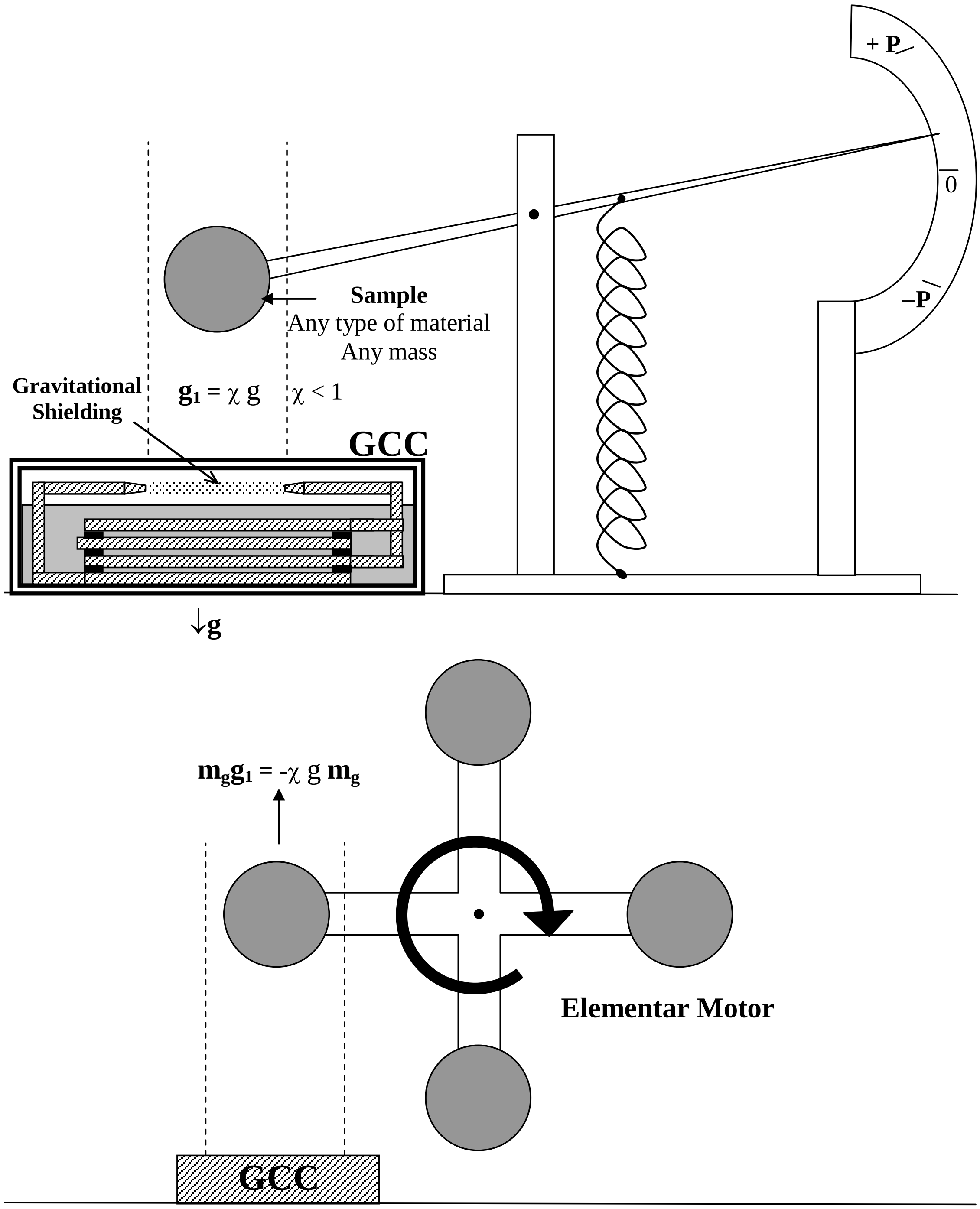

Figure B3 – Experimental arrangement with a GCC using battery of capacitors. By means of this set-up it is possible to check the weight of the sample even when it becomes *negative*.